\documentclass[12pt]{article} 

\usepackage{natbib}
\usepackage[sectionbib]{bibunits}
\defaultbibliographystyle{aer} 
\defaultbibliography{references.bib}

\usepackage[nottoc,notbib]{tocbibind}
\usepackage[dvipsnames,usenames]{color}
\usepackage{hyperref}
\usepackage[table]{xcolor}

\usepackage{adjustbox, algorithm, algpseudocode, amsfonts, amsmath, amssymb, amsthm, appendix, array, bigstrut, bm, bbm, booktabs, caption, commath, datetime, enumerate, enumitem, environ, epsfig, eurosym, footmisc, graphicx, latexsym, longtable, mathtools, mathrsfs, microtype, morefloats, multirow, pifont, pdflscape, rotating, siunitx, subcaption, tikz, threeparttable, tocloft, ulem, verbatim, authblk, tgpagella}

\usepackage[onehalfspacing]{setspace}
\usepackage[right=1in, left=1in, top=1in, bottom=1in]{geometry}
\usepackage[font={bf},justification=centering]{caption}
\usepackage[referable]{threeparttablex}
\usepackage[section]{placeins}

\usdate
\newdateformat{monthyeardate}{%
	\monthname[\THEMONTH], \THEYEAR}

\newcounter{bean}

\newcounter{protocol}
\makeatletter
\newenvironment{protocol}[1][htb]{%
  \let\c@algorithm\c@protocol
  \renewcommand{\ALG@name}{Protocol}
  \begin{algorithm}[#1]%
  }{\end{algorithm}
}
\makeatother

\newcounter{procedure}
\makeatletter
\newenvironment{procedure}[1][htb]{%
  \let\c@algorithm\c@procedure
  \renewcommand{\ALG@name}{Procedure}
  \begin{algorithm}[#1]%
  }{\end{algorithm}
}
\makeatother

\newcommand{\distas}[1]{\mathbin{\overset{#1}{\kern\z@\sim}}}  
\newsavebox{\mybox}\newsavebox{\mysim}

\theoremstyle{plain}

\newtheorem{theorem}{Theorem}

\newtheorem{assumption}{Assumption}
\newtheorem*{assumption*}{Assumption}
\newtheorem{lemma}{Lemma}       
     
\theoremstyle{definition}             
\newtheorem{remark}{Remark}

\renewcommand\theassumption{\arabic{assumption}}         
\AtBeginEnvironment{proof}{\footnotesize}
\makeatother

\DeclareMathOperator{\E}{\mathbb{E}} 					 
\DeclareMathOperator{\V}{\mathbb{V}} 					 
\DeclareMathOperator{\Cov}{\mathbb{C}ov} 					 
\DeclareMathOperator{\R}{\mathbb{R}} 					 
\DeclareMathOperator{\N}{\mathbb{N}} 					 
\DeclareMathOperator*{\argmin}{arg\,min} 				 
\DeclareMathOperator*{\argmax}{arg\,max} 				 

\newcommand{\calA}{\mathcal{A}}

\newcommand{\calE}{\mathcal{E}}
\newcommand{\calF}{\mathcal{F}}
\newcommand{\calG}{\mathcal{G}}

\newcommand{\calX}{\mathcal{X}}
\newcommand{\calY}{\mathcal{Y}}
\newcommand{\indic}{\mathbbm{1}}
\newcommand{\PP}{\mathbb{P}}
\newcommand{\op}{o_{\mathbb{P}}}
\newcommand{\Op}{O_{\mathbb{P}}}

\newcommand{\de}{\,\mathrm{d}}
\newcommand{\iid}{\overset{\mathrm{iid}}{\sim}}
\newcommand{\Be}{\mathsf{Be}}

\newcommand{\sG}{\mathsf{sG}}

\newcommand{\UCB}{\mathsf{UCB}}
\newcommand{\ODR}{\mathsf{ODR}}
\newcommand{\DR}{\mathsf{DR}}

\newcommand{\MIS}{\mathsf{MIS}}
\newcommand{\piucb}{\pi^{\UCB}}
\newcommand{\pidrucb}{\pi^{\DR}}
\newcommand{\piodrucb}{\pi^{\ODR}}

\newcommand{\regr}{\mathsf{Regret}_T}

\newcommand{\acen}{A_{\mathtt{cen}}}
\newcommand{\osigma}{\bar{\sigma}}
\newcommand{\Err}{\mathsf{Err}}

\newcommand{\Kodr}{K_{\mathsf{ODR}}}
\newcommand{\dkl}{D_{\mathsf{KL}}}

\newcommand{\bA}{\mathbf{A}}
\newcommand{\bB}{\mathbf{B}}

\newcommand{\bX}{\mathbf{X}}

\newcommand{\bZ}{\mathbf{Z}}

\newcommand{\ba}{\mathbf{a}}

\newcommand{\bx}{\mathbf{x}}

\newcommand{\bbeta}{\boldsymbol{\beta}}

\newcommand{\bmu}{\boldsymbol{\mu}}

\newcommand{\normal}{\mathsf{N}}

\newcommand\independent{\protect\mathpalette{\protect\independenT}{\perp}}
\def\independenT#1#2{\mathrel{\rlap{$#1#2$}\mkern2mu{#1#2}}}

\newcommand{\xmark}{\ding{55}}%

\DeclareFontFamily{U}{mathx}{\hyphenchar\font45}
\DeclareFontShape{U}{mathx}{m}{n}{
      <5> <6> <7> <8> <9> <10>
      <10.95> <12> <14.4> <17.28> <20.74> <24.88>
      mathx10
      }{}
\DeclareSymbolFont{mathx}{U}{mathx}{m}{n}
\DeclareFontSubstitution{U}{mathx}{m}{n}
\DeclareMathSymbol{\bigtimes}{1}{mathx}{"91}

\let\oldsqrt\sqrt
\def\sqrt{\mathpalette\DHLhksqrt}
\def\DHLhksqrt#1#2{%
\setbox0=\hbox{$#1\oldsqrt{#2\,}$}\dimen0=\ht0
\advance\dimen0-0.2\ht0
\setbox2=\hbox{\vrule height\ht0 depth -\dimen0}%
{\box0\lower0.4pt\box2}}

\definecolor{ptonorange}{RGB}{232,119,34}
\hypersetup{colorlinks=true,linkcolor=ptonorange,filecolor=ptonorange,urlcolor=ptonorange,citecolor=ptonorange}

\allowdisplaybreaks                 

\usepackage{etoolbox}
\makeatletter

\patchcmd{\NAT@citex}
  {\@citea\NAT@hyper@{%
     \NAT@nmfmt{\NAT@nm}%
     \hyper@natlinkbreak{\NAT@aysep\NAT@spacechar}{\@citeb\@extra@b@citeb}%
     \NAT@date}}
  {\@citea\NAT@nmfmt{\NAT@nm}%
   \NAT@aysep\NAT@spacechar\NAT@hyper@{\NAT@date}}{}{}

\patchcmd{\NAT@citex}
  {\@citea\NAT@hyper@{%
     \NAT@nmfmt{\NAT@nm}%
     \hyper@natlinkbreak{\NAT@spacechar\NAT@@open\if*#1*\else#1\NAT@spacechar\fi}%
       {\@citeb\@extra@b@citeb}%
     \NAT@date}}
  {\@citea\NAT@nmfmt{\NAT@nm}%
   \NAT@spacechar\NAT@@open\if*#1*\else#1\NAT@spacechar\fi\NAT@hyper@{\NAT@date}}
  {}{}
\newcommand{\nuisEstLabel}{3}

\begin{document}
\addtocontents{toc}{\protect\setcounter{tocdepth}{2}}

\title{\vspace{-0.5cm} \textbf{Sequential Decision Problems \\ with Missing Feedback}\thanks{I thank my advisor, Matias Cattaneo, for invaluable guidance throughout this project. I also thank Thomas Bearpark, Chiara Motta, Andrea Pugnana, and the participants in the Princeton Econometrics Student Seminar for insightful discussions. All errors and omissions remain my
own. Corresponding author e-mail address: \href{mailto:fpalomba@princeton.edu}{fpalomba@princeton.edu}.}}

\author{Filippo Palomba \\\vspace{-0.3cm} \normalsize Princeton University}

\date{\vspace{-0.15cm} This version: \today.}
\maketitle
\thispagestyle{empty}

\begin{abstract}
\noindent This paper investigates the challenges of optimal online policy learning under missing data. State-of-the-art algorithms implicitly assume that rewards are always observable. I show that when rewards are missing at random, the Upper Confidence Bound ($\UCB$) algorithm maintains optimal regret bounds; however, it selects suboptimal policies with high probability as soon as this assumption is relaxed. To overcome this limitation, I introduce a fully nonparametric algorithm—Doubly-Robust Upper Confidence Bound ($\DR$-$\UCB$)—which explicitly models the form of missingness through observable covariates and achieves a nearly-optimal worst-case regret rate of $\widetilde{O}(\sqrt{T})$. To prove this result, I derive high-probability bounds for a class of doubly-robust estimators that hold under broad dependence structures. Simulation results closely match the theoretical predictions, validating the proposed framework.
\end{abstract}


\onehalfspacing
\setlength\abovedisplayskip{5pt}
\setlength\belowdisplayskip{5pt}
\setlength\parindent{0pt}           
\setlength{\parskip}{6pt}  

\addtocontents{toc}{\protect\setcounter{tocdepth}{0}}

\begin{bibunit}
\section{Introduction}

In recent years, technological advancements and the proliferation of digital platforms have enabled the collection of real-time data describing the interaction between a policy and its targets. Such rich \textit{online} datasets naturally give rise to sequential decision problems in which a decision-maker continuously re-optimizes the implemented policy as more information is gathered. Historically, these decision problems have been modeled as multi‐armed bandits \citep*{thompson1933LikelihoodThatOne, wald1947SequentialAnalysis, robbins1952AspectsSequentialDesign}, in which a decision-maker interactively learns the best policy (\textit{action}) among a set of alternatives by trying out options and observing a signal (\textit{feedback} or \textit{reward}) from the environment.

Naturally, the success of these strategies requires the decision-maker to be able to observe some feedback following each action and, thus, to assess the extent to which the chosen action had the intended impact. However, in many real‐world settings, the environment’s response to a given policy might not always be observable, making it harder to gauge the true efficacy of an intervention. If this particular form of missing data is correlated with the outcome of interest, it introduces sampling bias \citep*{horvitz1952GeneralizationSamplingReplacement} and complicates the identification of optimal decision rules. Despite that, standard bandit algorithms typically rely on the assumption that rewards are always observed upon each action. 

To make the problem more concrete, consider the case of a digital platform that experiments with user engagement strategies—such as personalized push notifications or in‐app promotions—and gauges satisfaction via voluntary review prompts.  While some users may readily submit reviews, a non‐trivial fraction will opt out. Suppose the probability of response itself depends on satisfaction. In that case, reviews are observed only for a particular subpopulation, which may differ from the target one in many aspects, thus potentially inducing a suboptimal policy choice. Other real-life examples include a firm implementing different hiring strategies and a graduate admission committee experimenting with alternative types of offers. 

In this paper, I first show that the popular $\UCB$ algorithm \citep*{auer2002FinitetimeAnalysisMultiarmed} maintains its optimal (up to logarithmic factor) worst-case regret rate of $\widetilde{O}(\sqrt{T})$ whenever the process that causes missing data is independent from rewards. Intuitively, in this case, rewards are missing completely at random \citep*{rubin1976InferenceMissingData}, hence reward-independent missingness only makes the learning process slower, without invalidating it. Nevertheless, this independence assumption is implausible in most practical scenarios where reward observability directly depends on the feedback itself. For example, customer satisfaction is typically surveyed only for extreme types, and different job postings attract different potential employees. 

Whenever the process that causes missing data is reward-dependent, I show that the standard $\UCB$ algorithm may select suboptimal policies with probability approaching one as the number of trials grows. By explicitly modeling the dependence between rewards and the selection process through observable covariates, I develop a theoretically grounded, fully nonparametric procedure for sequential decision problems with missing feedback. The proposed algorithm achieves a worst-case regret rate of order $\widetilde{O}(\sqrt{T})$, which I show to coincide (up to logarithmic factors) with the rate of a novel lower bound on the minimax regret for the class of bandits with reward-dependent missingness. Additionally, the proposed method is doubly robust in the sense that only one among the conditional expectation of rewards and the conditional probability of missingness needs to be correctly specified for the algorithm to function properly. In this spirit, the algorithm's name is Doubly-Robust Upper Confidence Bound ($\DR$-$\UCB$). 

Finally, to the best of my knowledge, this paper provides the first high-probability bounds for a doubly-robust estimator under very general conditions. Specifically, the high-probability bounds derived throughout rely on the theory for martingale difference sequences \citep*{freedman1975TailProbabilitiesMartingales}, and so allow for very general dependence structures in the data. This was necessary in this setting due to the technical challenges brought by the sequential nature of the problem.

\subsection{Related Work}
\label{subsec: related work}

Sequential decision-making problems under uncertainty have been extensively studied in economics, statistics, and computer science after the seminal work of \cite*{wald1947SequentialAnalysis} and have mostly focused on the design of optimal strategies and algorithms. Recent advances have particularly emphasized the role of adaptive algorithms that sequentially learn and adjust to uncertain environments, leading to a proliferation of approaches that efficiently balance exploration and exploitation; for a textbook introduction, see \cite*{bubeck2012RegretAnalysisStochastic}, \cite*{lattimore2020BanditAlgorithms}, and references therein.

This paper builds explicitly on the extensive literature studying the properties of the $\UCB$ algorithm. The idea of being optimistic in the face of uncertainty first appeared in \cite*{lai1985AsymptoticallyEfficientAdaptive}. \cite*{lai1987AdaptiveTreatmentAllocation} provided the first version of the $\UCB$ algorithm, whereas the $\UCB$ algorithm analyzed here is closer to the $\UCB1$ analyzed in \cite*{auer2002FinitetimeAnalysisMultiarmed}. UCB methods have been widely recognized for their effectiveness in handling exploration-exploitation trade-offs, providing provable performance guarantees and optimal regret bounds in multi-armed bandit frameworks.

The literature on multi-armed bandits with delayed feedback is also closely related to this project. Delayed feedback poses significant challenges for standard $\UCB$ algorithms, as the decision-maker is forced to take actions before she can receive any signal from the environment she is interacting with. The closest predecessor of this work is probably \cite*{lancewicki2021StochasticMultiArmedBandits}, where the authors provide problem-specific regret bounds for the Successive Elimination algorithm when rewards are bounded and delays are occasionally infinite and reward-dependent. Different from them, here I propose distribution-free worst-case regret bounds for a novel version of the $\UCB$ and allow rewards to be unbounded.

Finally, this paper intersects with the literature that relies on doubly-robust estimators for different goals, such as handling missing data \citep*{robins1994EstimationRegressionCoefficients,  bang2005DoublyRobustEstimation}, estimating the causal impacts of policies \citep*{cattaneo2010EfficientSemiparametricEstimation, farrell2015RobustInferenceAverage, chernozhukov2018DoubleDebiasedMachine}, or to learn optimal policies in offline \citep*{athey2021PolicyLearningObservational} and online settings \citep*{kallus2022DoublyRobustDistributionally, shen2024DoublyRobustInterval}.

\subsection{Organization of the Paper}
\label{subsec: organization paper}
The paper is organized as follows. Section \ref{sec: problem setup} describes the problem and the algorithms used throughout. Section \ref{sec: MAB} contains the main results and showcases the worst-case regret properties of the classical $\UCB$ algorithm and $\DR$-$\UCB$ in environments with reward-independent (Section \ref{subsec: reward-independent missingness main}) and reward-dependent (Section \ref{subsec: reward-dependent missingness main}) missingness. Section \ref{sec: simulation evidence} illustrates some simulation evidence. Section \ref{sec: conclusion} concludes. The code replicating the simulation study is available at \url{https://github.com/filippopalomba/P_2025_banditMissing}.

\subsection{Notation}
\label{subsec: notation}

For two positive sequences $\left\{a_n\right\}_n,\left\{b_n\right\}_n$, I write $a_n = O(b_n)$ if $\exists\,M\in\R_{++}:a_n\leq M b_n$ for all large $n$, $a_n=o(b_n)$ if $\lim_{n\to\infty} a_n b_n^{-1} = 0$, $a_n=\widetilde{O}(b_n)$ if $\exists\, k\in\N,C\in\R_{++}: a_n=O(b_n\ln^k(Cn))$  $a_n\lesssim b_n$ if there exists a constant $C\in\R_{++}$ such that $a_n\leq C b_n$ for all large $n$, and $a_n \sim b_n$ if $a_n / b_n \rightarrow 1$ as $n \rightarrow \infty$.  For two sequences of random variables $\left\{A_n\right\}_n,\left\{B_n\right\}_n$, I write $A_n=\op(B_n)$ if $\forall\,\varepsilon\in\R_{++}, \lim_{n\to\infty} \PP[|A_n B_n^{-1}|\geq \varepsilon]= 0$ and $A_n =\Op(B_n)$ if $\forall\,\varepsilon\in\R_{++}, \exists \,M,n_0\in\R_{++} : \PP[|A_nB_n^{-1}|>M]<\varepsilon,$ for $n>n_0$. I denote a (possibly multivariate) Gaussian random variable with $\mathsf{N}(\ba, \bB),$ where $\ba$ denotes the mean and $\bB$ the variance-covariance, with $\Be(p)$ a Bernoulli distribution with $p\in(0,1]$ denoting the success probability, with $\sG(\sigma)$ a sub-Gaussian random variable with proxy variance at most $\sigma>0$, and with $\mathcal{SG}(\sigma)$ the space of sub-Gaussian probability distribution with variance proxy at most $\sigma>0$. A random variable $X$ is sub-Gaussian with parameter $\sigma>0$ if $\forall\,\lambda\in\R,\E[\exp(\lambda X)]\leq \exp(\lambda^2\sigma^2/2)$ and $\E[X]=0$ \citep*[Definition 5.1 in][]{lattimore2020BanditAlgorithms}. If $\{X_t\}_{t=1}^\infty$ is an $\calF$-adapted martingale difference sequence with respect to some filtration $\calF=\{\calF_t\}_{t=1}^\infty$, then it is understood that $X_t\sim \sG(\sigma^2)$ requires $\forall\,\lambda\in\R,\E[\exp(\lambda X_t)\mid \calF_t]\leq \exp(\lambda^2\sigma^2/2)$ and $\E[X_t\mid\calF_t]=0$. See also Table \ref{tab: notation} in the supplemental appendix for a summary of the project-specific notation.

\section{Problem Setup and Preliminaries}
\label{sec: problem setup}
I start by describing a generic instance of a stochastic multi-armed bandit (henceforth, MAB) with (possibly) missing rewards and the decision-maker
that interacts with such an environment.

\textbf{Setting.} A decision-maker faces a sequential decision problem over $ T \in \N$ rounds in a stochastic environment. At the beginning of each round $ t \in [T] $, using all the information available at that point, the decision-maker selects an action $ A_t \in \calA:=\{1,\ldots,A\}$. Each action $ a \in \calA $ is associated with a reward $ R_a \in\R, R_a\sim \sG(\sigma_a), \sigma_a>0$, an indicator for \textit{not} being missing $ C_a \in \{0,1\}$, and some covariates $\bX_a \in\mathcal{X}\subseteq \R^k,k\in\N$. All random variables are independent across actions, i.e., $(R_{a},C_{a},\bX_{a})\independent (R_{a'},C_{a'},\bX_{a'})$ for $a\neq a', a, a'\in\calA$. A stochastic MAB problem with missing rewards is defined as a collection of random variables $\{(R_{a,\ell},C_{a,\ell},\bX_{a,\ell})\}_{a\in\calA,\ell\in[T]}$ where each $(R_{a,\ell},C_{a,\ell},\bX_{a,\ell})$ is a random draw from $(R_{a},C_{a},\bX_{a})$. The reward of action $a\in\calA$ in round $t\in[T]$, denoted with $R_{a,t}$, is observed only if $C_{a,t}=1$. At this level of generality, the class of bandits considered is
\[\mathcal{C}:=\left\{(\nu_a)_{a\in\calA}: \nu^{[R]}_a \in \mathcal{SG}(\sigma_a), \: \nu_a^{[C]} = \Be(q_a), q_a\in(0,1]  \right\},\]
where $\nu^{[Y]}_a$ is the marginal distribution of $Y\in\{R,C\}$. Finally, I define $\osigma:= \sqrt{\max_{a\in\calA}\sigma_a^2}$ and $\underline{q} = \min_{a\in\calA}q_a$.

\paragraph{Decision-maker.} Each decision-maker is characterized by a policy that maps \newline $\{(A_\ell, R_{A_\ell,\ell}, C_{A_\ell,\ell}, \bX_{A_\ell,\ell}^\top)\}_{\ell\in[t-1]}$, the history up to the beginning of round $t$, to the space of probability distributions over actions $\Delta(\calA)$. Denote the space of policies as 
$$\Pi:=\left\{\pi: \pi = \{\pi_t\}_{t\in[T]}, \pi_t: (\calA \times \R \times \{0,1\}\times\calX)^{t-1} \to \Delta(\calA)\right\}.$$
I use interchangeably the words ``decision-maker", ``algorithm", and ``policy" when referring to a generic element $\pi\in\Pi$. Protocol \ref{prot: stochastic mab with dependent missingness} describes the interaction between a decision-maker and a MAB with (possibly) missing feedback.

\begin{protocol}
\caption{Multi-Armed Bandit with Missing Rewards}
\label{prot: stochastic mab with dependent missingness}
     Consider a generic bandit $\nu\in\mathcal{C},$ where $\nu=(\nu_a)_{a\in\calA}$
    \begin{algorithmic}
    \For{$\ell=1,2,\ldots,T$}
        \State{Decision-maker chooses $A_\ell=a$ according to some policy $\pi_t$}
        \State{Nature samples $(C_{a,\ell}, R_{a,\ell}, \bX_{a,\ell}) \sim \nu_a$}
        \If{$C_{a,\ell}=1$}
            \State{Decision-maker observes $R_{a,\ell}$}
        \Else
            \State{Decision-maker receives no feedback}
        \EndIf
    \EndFor
    \end{algorithmic}
\end{protocol}

\paragraph{Regret.} The \textit{pseudo-regret} of a decision-maker following a policy $\pi$ in a MAB with missing rewards $\nu\in\mathcal{C}$ is
\[\regr(\pi; \nu) =\sum_{t=1}^T (\max_{a\in\calA}\theta_a - \E_\nu[R_{A_t,t}]) = T\overline{\theta} - \sum_{t=1}^T \theta_{A_t},\] which depends on $\nu$ via the average rewards and it is a random quantity because the $\{A_t\}_{t\in[T]}$ are
random. Note that the latter is true even if the policies considered are deterministic. The reason is that $A_t$ depends on the history, which is random. Furthermore, the regret depends on $R_{A_t,t}$ independently of whether the rewards have been observed. In what follows, I omit the dependence of the regret on $\nu$ and simply write $\regr(\pi)$.

\paragraph{Goal.} The decision-maker's goal is to find a policy $\pi$ with good worst-case regret properties. Formally, the decision-maker seeks to find a policy $\pi$ whose worst-case regret $\overline{\regr}({\pi};\mathcal{C}):=\sup_{\nu\in{\mathcal{C}}} \regr({\pi};\nu)$ has the nearly optimal rate $\widetilde{O}(\sqrt{T})$.
In what follows next, I will not directly optimize the worst case regret over the space of policy $\Pi$ (see \cite{adusumilli2024RiskOptimalPolicies} for an example of such an approach in a standard MAB). Rather, I first derive a lower bound for the minimax regret 
\[\regr^\star(\mathcal{C}) := \inf_{\pi\in\Pi}\overline{\regr}({\pi};\mathcal{C}).\]
Then, I derive an upper bound for the worst-case regret $\overline{\regr}(\widetilde{\pi};\mathcal{C})$ of some specific policy $\widetilde{\pi}$ and, finally, I check that the rates of the two bounds coincide. In what follows, I focus on the popular $\UCB$ algorithm \citep*{auer2002FinitetimeAnalysisMultiarmed} and a novel, doubly-robust modification of it as the algorithms used by the decision-maker to form her policy. 

\subsection{Algorithms}
\label{subsec: algorithms}

The $\UCB$ algorithm selects the action to be played by solving an exploitation-exploration trade-off. Indeed, the algorithm desires to \textit{explore} new actions, but, since playing a sub-optimal action induces regret, it also wants to \textit{exploit} what is already known about the environment.  On the one hand, too little exploration might make a sub-optimal alternative look better than the optimal one because of random fluctuations. On the other hand, too much exploration prevents the algorithm from playing the optimal alternative often enough, which also results in a larger regret. 

Let $\hat{\theta}_{a}(t)$ be an estimator for $\theta_a$ after $t-1$ rounds and denote with $b_{a,t}(\delta)$ a bonus term chosen so that $\theta_a \in [\hat{\theta}_{a}(t) - b_{a,t}(\delta), \hat{\theta}_{a}(t) + b_{a,t}(\delta)]$ with probability at least $1-\delta$. The $\UCB$ algorithm selects the action $a^\star$ at round $t$ that has the highest optimistic mean reward estimate, i.e.
\[a^\star = \argmax_{a\in\calA}\hat{\theta}_{a}(t) + b_{a,t}(\delta).\]
As such, an action $a\in\calA$ can be chosen for two different reasons: because $b_{a,t}(\delta)$ is large, implying that the estimate $\hat{\theta}_{a}(t)$ is noisy (\textit{explorative} choice); or because $\hat{\theta}_{a}(t)$ is large (\textit{exploitative} choice). Since the bonus term $b_{a,t}(\delta)$ is constructed to shrink quickly each time alternative $a$ is selected, exploration becomes less frequent over time. When $b_{a,t}(\delta)$ is sufficiently small, the estimated value $\hat{\theta}_{a}(t)$ closely approximates the true parameter $\theta_a$, assuming that $\hat{\theta}_{a}(t)$ is a ``good" estimator for $\theta_a$. Consequently, $\UCB$ naturally balances between exploration and exploitation.

Before describing in great detail the algorithms, let 
\[P_a(t) := \sum_{\ell=1}^{t-1}\indic[A_\ell=a], \quad\text{and}\quad N_a(t) := \sum_{\ell=1}^{t-1}\indic[A_\ell=a]
    C_{a,\ell}\]
be the number of times an arm $a\in\calA$ has been pulled and the number of times the reward $R_a$ has been observed at the beginning of round $t\in[T]$, respectively.

\subsubsection{Classic $\UCB$ Algorithm}
The (regularized) estimate for $\theta_a, a\in\calA$ at the beginning of round $t\in[T]$ is
\begin{align}\label{eq: mean reward estimator}
    \widehat{R}^{\UCB}_a(t) = \frac{1}{N_a(t) + \lambda}\sum_{\ell=1}^{t-1}\indic[A_\ell=a]C_{a,\ell} R_{a,\ell},
\end{align}
where $\lambda>0$ is a regularization parameter that prevents the estimator from being ill-defined whenever, after initialization, it occurs that  $N_a(t)=0$ for some $a\in\calA$ and $t\geq 1$ (see also Remark \ref{rem: on the regularization of denominators}). The optimistic mean reward estimate of action $a\in\calA$ after $t\in[T]$ rounds is 
\[\widetilde{R}^{\UCB}_a(t,\delta) = \widehat{R}^{\UCB}_a(t) + b^{\UCB}_{a,t}(\delta), \]
where the ``bonus" term $b^{\UCB}_{a,t}(\delta)$ is chosen to make sure that the optimistic mean reward estimate $\widetilde{R}^{\UCB}_a(t,\delta)$ upper bounds the true mean reward $\theta_a$ with high probability. In this specific case, I define
\[b^{\UCB}_{a,t}(\delta) := \frac{\osigma}{\underline{q}_\lambda}\sqrt{\frac{2\ln(2AT/\delta)}{ P_a(t) +\lambda}}+\frac{\lambda\overline{K}}{N_a(t)+\lambda},\] 
where $\overline{K}$ is some constant larger than $\overline{\theta}, \underline{q}_\lambda:=\inf_{a,t}\frac{N_a(t) + \lambda}{P_a(t) + \lambda}$, and $\delta\in(0,1)$. Intuitively, the first term in $b^{\UCB}_{a,t}(\delta)$ governs the probability with which the optimistic estimate overestimates the mean reward, whereas the second term takes into account the bias induced by the regularization term $\lambda > 0$. Under reward-independent missingness (Assumption \ref{assump: independent missingness main} below), Lemma \ref{lemma: probability of failure event} in the supplemental appendix formally justifies the particular choice of $b^{\UCB}_{a,t}(\delta)$ described above by showing that 
\[\forall\,a\in\calA,t\in[T],\quad \theta_a \in \left[\widehat{R}^{\UCB}_a(t) - b^{\UCB}_{a,t}(\delta),\widehat{R}^{\UCB}_a(t) + b^{\UCB}_{a,t}(\delta)\right]\]
holds with probability at least $1-\delta$.

The way the $\UCB$ algorithm works is straightforward: at round $t \in[T]$, it selects the arm $a$ that has the highest optimistic mean reward estimate. Algorithm \ref{alg: UCB algorithm} below summarizes all the steps needed by the classic $\UCB$ algorithm.

\begin{procedure}
\caption{Update Estimators for $\UCB$}
\label{procedure: update estimators}
    \begin{algorithmic}
    \For{$a\in[A]$}
        \State{$N_a(t+1)\leftarrow N_a(t) + \indic[A_t=a]C_{a,t}$}
        \State{$P_a(t+1)\leftarrow P_a(t) + \indic[A_t=a]$}
        \State{$\widehat{R}_a(t+1) \leftarrow \frac{1}{N_a(t+1) + \lambda}\sum_{\ell=1}^{t}\indic[A_\ell=a]C_{a,\ell}R_{a,\ell}$}
        \State{$b^{\UCB}_{a,t+1}(\delta) \leftarrow \frac{\osigma}{\underline{q}_\lambda}\sqrt{\frac{2\ln(2AT/\delta)}{ P_a(t+1)+\lambda}}+\frac{\lambda\overline{K}}{N_a(t+1)+\lambda}$}
        \State{$\widetilde{R}_a(t+1,\delta) \leftarrow  \widehat{R}_a(t+1) + b^{\UCB}_{a,t+1}(\delta)$}
    \EndFor
    \end{algorithmic}
\end{procedure}

\begin{algorithm}[H]
\caption{$\UCB$ algorithm}
    \hspace*{\algorithmicindent} \textbf{Input}: $\lambda>0, \underline{q}_\lambda, \osigma, T, \calA, \delta, \overline{K}$ \\
    \hspace*{\algorithmicindent} \textbf{Initialization}: pull each arm once, get $\widehat{R}^{\UCB}_a(0),$ set $P_a(0)=1, N_a(0)=C_{a,0}, \forall\,a\in\calA$
    \begin{algorithmic}[1]
        \For{$t=1,2,\ldots,T$}
        \State{pull arm $a_t=\argmax_{a\in\calA}\widetilde{R}^{\UCB}_a(t,\delta)$ and set $\pi_t^\UCB= a_t$}
        \State{call \textit{Update Estimators for $\UCB$} (Procedure \ref{procedure: update estimators})}
        \EndFor 
    \end{algorithmic}
    \hspace*{\algorithmicindent} \textbf{Output}: $\piucb = \{\piucb_t\}_{t\in[T]}$
    \label{alg: UCB algorithm}
\end{algorithm}

\begin{remark}\label{rem: on the regularization of denominators} 
I introduce the regularization parameter $\lambda>0$ to take care of those cases in which $N_a(0)=0$, where $t=0$ denotes the initialization period. In the absence of missing data, i.e., when $N_a(t)=P_a(t)$ almost surely, it is common practice to pull each arm once as initialization. If the action set is finite, the initialization just shifts up the regret of a factor not larger than $A \overline{K}$. In the context studied here, pulling each arm once grants that $P_a(0)=1$ for all $a\in\calA$, but does not guarantee that $N_a(0)=1$ for all $a\in\calA$. Rather, the event $\{N_a(0)=1,\forall\,a\in\calA\}$ realizes with probability $\prod_{a\in\calA}q_a\leq 1$, hence the need for regularization.
\end{remark}

\subsubsection{Doubly-Robust $\UCB$ Algorithm}

Let the true conditional mean reward and probability of not being missing for arm $a\in\calA$ as
\[\theta_a(\bX_a) := \E_\nu[R_{a}\mid \bX_a], \qquad q_a(\bX_a) = \E_\nu[C_a\mid \bX_a]\in[\underline{q}, 1]\]
almost surely, and denote with $\hat{\theta}_a(\cdot)$ and $\hat{q}_a(\cdot)$ their estimated counterparts. Throughout, I use interchangeably the terms ``probability of rewards not being missing" and ``probability of missingness". The doubly-robust estimator for mean rewards is defined as
\[ \widehat{R}^{\DR}_a(t) := \frac{1}{P_a(t)} \sum_{\ell=1}^{t-1}\indic[A_\ell=a] \left(\frac{C_{a,\ell}(R_{a,\ell} - \hat{\theta}_a(\bX_{a,\ell}))}{\hat{q}_a(\bX_{a,\ell})} + \hat{\theta}_a(\bX_{a,\ell})\right).\]

The optimistic mean reward estimator for $\DR$-$\UCB$ is defined as 
\begin{align}
    \label{eq: optimistic reward estimator DR-UCB}
    \widetilde{R}_a^{\DR}(t,\delta) = \widehat{R}^{\DR}_a(t) + b_{a,t}^\DR(\delta), \qquad
    b_{a,t}^\DR(\delta) = \Kodr\sqrt{\frac{2\ln(2AT/\delta)}{P_a(t)}} + b_{a,t}^{\mathsf{res}}(\delta),
\end{align}
where $\Kodr := \frac{\osigma}{\underline{q}} + \osigma$ and $b_{a,t}^{\mathsf{res}}(\delta)$ is defined precisely in Section \ref{subsubsec: feasible DR} of the supplemental appendix. Under a reward-dependent process that causes missing data (Assumption \ref{assump: conditional ignorability main} below), Lemma \ref{lemma: probability of failure event with dependent missingness} in the supplemental appendix thoroughly justifies the choice of $b^{\DR}_{a,t}(\delta)$  as a bonus term and proves that
\[\forall\,a\in\calA,t\in[T],\quad \theta_a \in \left[\widehat{R}^{\DR}_a(t) - b^{\DR}_{a,t}(\delta),\widehat{R}^{\DR}_a(t) + b^{\DR}_{a,t}(\delta)\right]\]
holds with probability at least $1-\delta$.

Finally, Algorithm \ref{alg: DR-UCB algorithm} describes in greater detail how
the $\DR$-$\UCB$ algorithm works.

\begin{procedure}
\caption{Update Estimators for $\DR$-$\UCB$}
\label{procedure: update estimators DR-UCB}
    \begin{algorithmic}
    \For{$a\in[A]$}
        \State{$N_a(t+1)\leftarrow N_a(t) + \indic[A_t=a]C_{a,\ell}$}
        \State{$P_a(t+1)\leftarrow P_a(t) + \indic[A_t=a]$}
        \State{Update $\hat{q}_a$ and $\hat{\theta}_a$ if required}
        \State{$\widehat{R}^{\DR}_a(t+1) := \frac{1}{P_a(t+1)} \sum_{\ell=1}^{t}\indic[A_\ell=a] \left(\frac{C_{a,\ell}(R_{a,\ell} - \hat{\theta}_a(\bX_{a,\ell}))}{\hat{q}_a(\bX_{a,\ell})} + \hat{\theta}_a(\bX_{a,\ell})\right)$}
        \State{$\widetilde{R}^\DR_a(t+1,\delta) \leftarrow
      \widehat{R}^\DR_a(t+1) + b_{a,t}^\DR(\delta)$}
    \EndFor
    \end{algorithmic}
\end{procedure}

\begin{algorithm}[H]
\caption{$\DR$-$\UCB$ algorithm}
    \hspace*{\algorithmicindent} \textbf{Input}: $\lambda>0, T, \calA, \delta, \{\hat{q}_a(\cdot),\hat{\theta}_a(\cdot)\}_{a\in\calA}$ \\
    \hspace*{\algorithmicindent} \textbf{Initialization}: pull each arm once, get $\widehat{R}^{\DR}_a(0)$ and set $P_a(0)=1, N_a(0)=C_{a,0}, \forall\,a\in\calA$\\
    \hspace*{\algorithmicindent} \textbf{Nuisances}: get estimates $\{\hat{q}_a(\bX_{a,0}),\hat{\theta}_a(\bX_{a,0})\}_{a\in\calA}\}$ according to Assumption \ref{assump: nuisance estimation main}\textcolor{ptonorange}{(iii)}
    \begin{algorithmic}[1]
        \For{$t=1,2,\ldots,T$}
        \State{pull arm $a_t=\argmax_{a\in\calA}\widetilde{R}^\DR_a(t,\delta)$ and set $\pidrucb_t= a_t$}
        \State{call \textit{Update Estimators for $\DR$-$\UCB$} (Procedure \ref{procedure: update estimators DR-UCB})}
        \EndFor        
    \end{algorithmic}
    \hspace*{\algorithmicindent} \textbf{Output}: $\pidrucb = \{\pidrucb_t\}_{t\in[T]}$
    \label{alg: DR-UCB algorithm}
\end{algorithm}

\section{Multi-armed Bandits with Missing Data}
\label{sec: MAB}
In Section \ref{subsec: reward-independent missingness main}, I begin by analyzing the performance of the $\UCB$ algorithm under the assumption that the process that causes missing data does not depend on rewards (Assumption \ref{assump: independent missingness main}). In Section \ref{subsec: reward-dependent missingness main}, I then show that, when the previous assumption is relaxed to allow for reward-dependent missingness (Assumption \ref{assump: conditional ignorability main}), the $\UCB$ algorithm can suffer from linear regret, whereas $\DR$-$\UCB$ achieves sub-linear worst-case regret. In Section \ref{subsec: lower bound}, I present a lower bound for the minimax regret. Finally, in Section \ref{subsec: nuisance estimation}, I conclude by giving some practical advice on how to implement $\DR$-$\UCB$.

\subsection{Reward-independent Missingness}
\label{subsec: reward-independent missingness main}
I begin by considering the scenario where rewards are missing at random, as formalized in Assumption \ref{assump: independent missingness main}. Additionally, rewards are assumed to be sub-Gaussian, a common condition that constrains their tail behavior and enables the use of standard concentration inequalities; see \cite*{vershynin2018HighDimensionalProbabilityIntroduction} and \cite*{wainwright2019HighDimensionalStatisticsNonAsymptotic} for an introduction.

\begin{assumption}\label{assump: independent missingness main}
    For each action $a\in\calA,C_a\independent R_a$.
\end{assumption}

Under Assumption \ref{assump: independent missingness main}, the class of bandits of interest is restricted to
\[\mathcal{C}_1:=\left\{(\nu_a)_{a\in\calA}: \nu^{[R]}_a\in\mathcal{SG}(\sigma_a), \:\:\nu_a^{[C]}= \Be(q_a), q_a\in(0,1],\:\: \nu_a = \nu^{[R]}_a\cdot \nu^{[C]}_a  \right\}\subset \mathcal{C}.\]

The paper’s first main result establishes that the $\UCB$ algorithm achieves a near-optimal worst-case regret rate over the class of bandits $\mathcal{C}_1$. The proof of such a result follows using standard arguments. Namely, I consider a ``good" event and show that it occurs with high probability in the setting considered throughout. In this spirit, define such an event as
$$\calG(\delta_1,\delta_2) = \overline{\calF^{\UCB}(\delta_1)} \cap \overline{\calF^{\MIS}(\delta_2)},$$ 
for some $\delta_1,\delta_2\in(0,1)$, where
\begin{align*}
    \calF^\UCB(\delta) &:= \{\exists\,a\in\calA,t\in[T],\left|\widehat{R}^{\UCB}_a(t) - \theta_a\right|\geq b^{\UCB}_{a,t}(\delta)\}, \\
    \calF^{\MIS}(\delta):=&\{\exists\,a\in\calA,t\in[T]:N_a(t)\leq (1-\delta)q_aP_a(t), P_a(t) \geq \underline{T}_a\},
\end{align*}
where $\underline{T}_a:= 1+\frac{24\ln(T)}{q_a}$. Under the good event, which realizes with probability at least $1-\delta_1-\delta_2$, it holds that: (\textit{i}) for each action, the optimistic reward estimator always covers the true mean; (\textit{ii}) after a minimum amount of pulls $\underline{T}_a$, the missing data mechanism is not too extreme in terms of percentage deviation from its mean.

In the supplemental appendix, Lemma \ref{lemma: probability of failure event} and Lemma \ref{lemma: probability of missingness event} show that $\calF^\UCB(\delta_1)$ and $\calF^\MIS(\delta_2)$ occur with arbitrarily small probability. Then, under the good event $\calG(\delta_1,\delta_2)$, it is possible to bound the worst case regret of $\UCB$ uniformly over bandits in $\mathcal{C}_1$ and horizons $T\in\N$. This result is presented formally in the next theorem and proven in the supplemental appendix.

\begin{theorem}
\label{thm: regret UCB in MAB - main text}
Let Assumption \ref{assump: independent missingness main} hold, $\lambda=o(T^{1/2}),\delta_1\in(0,1) ,\delta_2=\sqrt{\frac{1+\kappa}{12}},$ and $\kappa>0.$ Then, for any $T\in\N$ and bandit $\nu\in\mathcal{C}_1$
\[ \regr(\piucb) \lesssim \frac{4\osigma}{\underline{q}_\lambda}\sqrt{2AT\ln(2AT/\delta_1)},\]
with probability at least $1-\delta_1 - O(T^{-\kappa})$.
\end{theorem}

\subsection{Reward-dependent Missingness}
\label{subsec: reward-dependent missingness main}

The previous section showed that, under the assumption that the process that causes missing data does not depend on rewards, the $\UCB$ algorithm has nearly-optimal regret. However, this independence assumption is hard to defend in practical applications. Borrowing from the program evaluation literature, I relax Assumption \ref{assump: independent missingness main} and, instead, rely on a \textit{conditional ignorability assumption} (CIA), which states that independence holds only conditional on the vector of covariates $\bX_a$ \citep*{rosenbaum1983CentralRolePropensity}.\footnote{This is a standard assumption in the causal inference literature. I refer the interested reader to \cite*{imbens2004NonparametricEstimationAverage} and Chapter 21 in \cite*{wooldridge2010EconometricAnalysisCross} for thorough discussions of the plausibility of such an assumption in various contexts.} Put differently, the CIA imposes that knowledge of $\bX_a$ is sufficient to break the dependence between rewards and the missing data mechanism in each arm. 

The CIA can be expressed in two ways: by imposing some structure on the conditional expectation of $C_a$ (so-called \textit{design-based} approach); by imposing some structure on the conditional expectation of $R_a$ (so-called \textit{model-based} approach). Depending on the specific application, either version of the CIA might be more appealing.  The next assumption formalizes this idea.

\begin{assumption}[Model- and Design-based Ignorability]
\label{assump: conditional ignorability main}
For each $a\in\calA$, either
\[\E_\nu[R_a\mid \bX_a, C_a] = \E_\nu[R_a\mid \bX_a]=:\theta_a(\bX_a)\qquad \text{a.s.}\tag{\textbf{MB}}\]
or    
\[\E_\nu[C_a\mid \bX_a, R_a] = \E_\nu[C_a\mid \bX_a]=:q_a(\bX_a)\qquad \text{a.s.}\tag{\textbf{DB}}\]
holds. Moreover, $R_a \mid \bX_a,  \sim \sG(\sigma_a) $ and  $q_a(\bx) \in[\underline{q},1]$,
for some constants $0\leq \overline{K}_\theta<\infty$ and $\underline{q}\in(0,1]$.
\end{assumption}

On top of the CIA, Assumption \ref{assump: conditional ignorability main} requires sub-Gaussianity of rewards only conditional on $\bX_a$ and has two other mild requirements: \textit{(i)} the conditional expectation of each $R_a$ is uniformly bounded over $\calX$; \textit{(ii)} the probability of observing rewards is non-zero ($q_a>0$). Hence, the class of bandits considered throughout is
\[\mathcal{C}_2:=\left\{\nu=(\nu_a)_{a\in\calA}: \nu_a^{[C]} = \Be(q_a), q_a\in(0,1], \text{ and Assumption \ref{assump: conditional ignorability main} holds} \right\}\subset \mathcal{C}.\]

In this framework, it is well-known that the mean-reward estimator defined in \eqref{eq: mean reward estimator} is not consistent anymore for $\theta_a$ \citep*{horvitz1952GeneralizationSamplingReplacement}. Even more worryingly, the estimators $\widehat{R}^{\UCB}_a(t), a\in\calA$ might have probability limits $\widetilde{\theta}_a$ such that
\[\argmax_{a\in\calA}\widetilde{\theta}_a \neq \argmax_{a\in\calA}{\theta}_a,\]
so that even after a large number of rounds $T$, the $\UCB$ algorithm will not learn the best arm. Section \ref{subsubsec: classic ucb} of the supplemental appendix shows an example of a bandit in $\mathcal{C}_2$ such that this realizes.

Assumption \ref{assump: conditional ignorability main} ensures that $\theta_a$ can be identified from the data had the nuisance functions $\{\theta_a(\cdot),q_a(\cdot), a\in\calA\}$ been known. However, in practice, these nuisances need to be estimated, and extra care is required in doing so. Before elucidating how nuisance estimation should be conducted, define the following $\ell_2$-estimation errors for each $a\in\calA$
\begin{align*}
    \Err_{t}(\hat{\theta}_a)&:= \sqrt{\frac{1}{P_a(t)}\sum_{\ell=1}^{t-1}\indic[A_\ell=a](\hat{\theta}_a(\bX_{a,\ell}) -\widetilde{\theta}_a(\bX_{a,\ell}))^2},
\end{align*}
and
\begin{align*}
    \Err_{t}(\hat{q}_a)&:= \sqrt{\frac{1}{P_a(t)}\sum_{\ell=1}^{t-1}\indic[A_\ell=a](\hat{q}_a(\bX_{a,\ell}) -\widetilde{q}_a(\bX_{a,\ell}))^2} 
\end{align*}
for some known $\widetilde{q}_a(\bx)$ and $\widetilde{\theta}_a(\bx)$. 

Assumption \ref{assump: nuisance estimation main} precisely states all the requirements the nuisance estimators have to satisfy; see also Section \ref{subsec: nuisance estimation} for two examples of practical procedures that satisfy Assumption \ref{assump: nuisance estimation main}. 

\begin{assumption}[Nuisance Estimation]\label{assump: nuisance estimation main}
For each $a\in\calA,$ the following are true:
\begin{enumerate}[label=(\alph*), ref={\nuisEstLabel(\alph*)}]
    \item (double robustness) either $\widetilde{q}_a(\bx)=q_a(\bx)$ or $\widetilde{\theta}_a(\bx)=\theta_a(\bx)$;
    \label{assump: nuisance estimation - DR main}
    \item (truncation) $\forall\,\bx\in\calX,\widehat{q}_a(\bx) \in [\underline{q}, 1], \underline{q}\in(0,1]$; \label{assump: nuisance estimation - truncation main}
    \item (independence)  $(\hat{q}_a(\bX_{a}),\hat{\theta}_a(\bX_{a}))\independent (R_{a},C_{a})\mid \bX_{a}$; \label{assump: nuisance estimation - independence main}
    \item ($\ell_2$-error rate) there exist rates $\alpha> 1/2, \alpha_q>0,$ and $\alpha_\theta>0$ such that 
    \[\Err_t(\hat{q}_a) \lesssim \frac{1}{P_a(t)^{\alpha_q}}, \qquad \Err_t(\hat{\theta}_a) \lesssim \frac{1}{P_a(t)^{\alpha_\theta}}, \qquad \Err_t(\hat{q}_a)\Err_t(\hat{\theta}_a)\lesssim \frac{1}{P_a(t)^{\alpha}}\]
    with probability $1-\delta_\mathfrak{c}, \delta_\mathfrak{c}\in(0,1)$.
    \label{assump: nuisance estimation - l2 error main}
\end{enumerate}
\end{assumption}

First, Assumption \ref{assump: nuisance estimation - DR main} requires one between the true conditional probability of missingness, $q_a(\cdot)$, and the true conditional expectation of rewards, $\theta_a(\cdot)$, to be the probability limit of one of the nuisance estimators, $\hat{\theta}_a(\cdot)$ and $\hat{q}_a(\cdot)$. In other words, it suffices to have well-specified only one of the two conditional expectations. Second, Assumption \ref{assump: nuisance estimation - truncation main} bounds the estimated probability of (not) being missing away from 0, a typical regularity condition in such problems. Third, to avoid over-fitting bias, Assumption \ref{assump: nuisance estimation - independence main} asks the nuisance functions to be estimated in an independent (conditional on $\bX_a$) sample. Fourth, \ref{assump: nuisance estimation - l2 error main} controls the estimation error of the nuisance estimators in two ways: \textit{(i)} the estimation error of each nuisance need to be shrinking in $P_a(t)$; \textit{(ii)} the product of the estimation errors must decay faster than $1/\sqrt{P_a(t)}$. These conditions make the sampling error dominate the estimation error induced by the fact that $\{(\theta_a(\cdot),q_a(\cdot)),a\in\calA\}$ are estimated. 

Assumption \ref{assump: nuisance estimation main} is fundamental to make the term $b_{a,t}^{\mathsf{res}}(\delta)$ of higher order in the bonus term $b_{a,t}^\DR(\delta)$ defined in \eqref{eq: optimistic reward estimator DR-UCB}. Formally, under such an assumption, it follows that $b_{a,t}^{\mathsf{res}}(\delta)=\op(1/\sqrt{P_a(t)})$ and the next lemma follows.
\begin{lemma}
    \label{lemma: concentration of estimated DR rewards under dependent missingness - main text}
    Let Assumptions \ref{assump: conditional ignorability main} and \ref{assump: nuisance estimation main} hold, $\delta\in(0,1), a\in\calA,$ and $t\in[T]$. Then, 
    \[\left| \widehat{R}^{\DR}_a(t) - \theta_a\right|\geq  b_{a,t}^\DR(\delta AT)\]
    with probability at most $\delta$.
\end{lemma}
The above result shows that the bonus term for the $\DR$-$\UCB$ algorithm has been chosen appropriately in the sense that it controls the probability with which $\widehat{R}^{\DR}_a(t)$ deviates from $\theta_a$. This result is of independent interest as it is the first one that provides high-probability bounds for a doubly-robust estimator under mild assumptions; see Lemma \ref{lemma: probability of failure event with dependent missingness} in the supplemental appendix for a formal statement, a proof of the result, and some heuristics on the logic behind the strategy used in the proof.

To provide a bound on the worst-case regret over the class $\mathcal{C}_2$, a similar strategy to the one used in Section \ref{subsec: reward-independent missingness main} is adopted. Define the failure event
\[\calF^{\DR}(\delta) := \left\{\exists\,a\in\calA,t\in[T],\left| \widehat{R}^{\DR}_a(t) - \theta_a\right|\geq b_{a,t}^\DR(\delta)\right\}.\]
When $\calF^{\DR}(\delta)$ occurs the optimistic doubly-robust reward estimator $\widetilde{R}_a^{\DR}(t) = \widehat{R}_a^{\DR}(t) + b_{a,t}^{\DR}(t)$ does not cover the true mean reward $\theta_a$. Using Lemma \ref{lemma: concentration of estimated DR rewards under dependent missingness - main text}, it is immediate to see that $\PP[\calF^{\DR}(\delta)]\leq \delta$ for some $\delta\in(0,1)$. The next theorem shows that the regret of the $\DR$-$\UCB$ algorithm is nearly optimal (up to logarithmic factors) and provides an upper bound that holds with high probability uniformly over the horizon $T$ and the class of bandits $\mathcal{C}_2$.
\begin{theorem}
\label{thm: regret UCB in MAB with dependent missingness - main text}
Let Assumptions \ref{assump: conditional ignorability main} and \ref{assump: nuisance estimation main} hold with $\delta \in(0,1)$ and  $\delta_{\mathfrak{c}} \in(0,1)$. Then, for any horizon $T\in\N$ and bandit $\nu\in\mathcal{C}_2$ 
\begin{align*}
\regr(\pidrucb) \lesssim \frac{4\osigma}{\underline{q}}\sqrt{AT\ln(2AT/\delta)}
\end{align*}
with probability $1-\delta-\delta_\mathfrak{c}$.
\end{theorem}

Comparing Theorem \ref{thm: regret UCB in MAB with dependent missingness - main text} with Theorem \ref{thm: regret UCB in MAB - main text}, it is possible to see that the leading order terms of the worst case regret bound are identical. Indeed, Assumption \ref{assump: nuisance estimation - l2 error main} is crucial in granting that the uncertainty due to the estimation of the nuisance functions is dominated by the one due to sampling error; see the supplemental appendix for a formal proof and additional heuristics to foster intuition for this result.

\subsection{Lower Bound}
\label{subsec: lower bound}
In this section, I show that the minimax regret
\[\regr^\star(\mathcal{C}_j) := \inf_{\pi\in\Pi}\sup_{\nu\in\mathcal{C}_j} \regr(\pi;\nu), \:j\in\{1,2\}\]
is lower bounded by a constant times a factor of $\sqrt{T}$. 

The logic of the proof is similar to that behind classic lower bound results obtained via Le Cam's two-point lemma. In particular, a generic policy $\tilde{\pi}\in\Pi$ is considered and its regret on two particular instances $\nu,\nu'\in\mathcal{C}$ is lower-bounded, where $\mathcal{C}$ is some class of bandits. Then, it follows that 
\[\sup_{\tilde{\nu}\in\mathcal{C}}\regr(\tilde{\pi};\tilde{\nu}) \geq \max\{\regr(\tilde{\pi},\nu), \regr(\tilde{\pi},\nu')\}\geq f(T),\]
for some function $f(\cdot)$. Because the policy $\tilde{\pi}$ was generic, then the bound above holds for any policy in $\Pi$, hence
\[\regr^\star(\mathcal{C}) = \inf_{\pi\in\Pi}\sup_{{\nu}\in\mathcal{C}}\regr({\pi};{\nu})\geq f(T).\]

In particular, to get a lower bound for the minimax regret over the classes of bandits $\mathcal{C}_1$ and $\mathcal{C}_2$, I derive a lower bound for the minimax regret for the classes of Gaussian bandits 
\[\mathcal{C}_1^{\mathtt{gau}} := \{(\nu_a)_{a\in\calA}: \nu_a^{[R]}=\normal(\theta_a, 1), \nu_a^{[C]}=\Be(q_a),q_a\in(0,1]\}\subset \mathcal{C}_1\]
and
\begin{align*}
    \mathcal{C}^{\mathtt{gau}}_2:= \big\{\nu=(\nu_a)_{a\in\calA}: \nu_a^{[R|X]}=&\normal(\theta_a(X), 1), \nu_a^{[C]} = \Be(q_a), q_a\in(0,1], \\
    &\text{ and Assumption \ref{assump: conditional ignorability main} holds} \big\}\subset \mathcal{C}_2.
\end{align*}
Because the ``sup" of the minimax is taken over a smaller subset, the minimax bound for Gaussian bandits extends immediately to the classes of sub-Gaussian bandits $\mathcal{C}_1$ and $\mathcal{C}_2$. A textbook example can be found in \cite*{lattimore2020BanditAlgorithms}, Chapter 15, and the proof of Theorem \ref{thm: lower bound minimax regret main} is almost identical to that of Theorem 15.2 in $\mathsf{LS}$ and, indeed, the bound is identical.

\begin{theorem}
    \label{thm: lower bound minimax regret main}
    Let $T\in\N, T\geq A-1$ and consider the classes of bandits $\mathcal{C}_1$ and $\mathcal{C}_2$. Then,
    \[\regr^\star(\mathcal{C}_j) = \inf_{\pi\in\Pi}\sup_{{\nu}\in\mathcal{C}_j}\regr(\pi;{\nu})\geq \frac{\sqrt{T(A-1)}}{16\sqrt{e}}.\] 
\end{theorem}

\subsection{Nuisance Estimation}
\label{subsec: nuisance estimation}
The near-optimality of $\DR$-$\UCB$ hinges on having nuisance estimators $\hat{\theta}_a(\cdot)$ and $\hat{q}_a(\cdot)$ that satisfy Assumption \ref{assump: nuisance estimation main}. In what follows, I give two examples of classes of estimators that satisfy the conditions in Assumption \ref{assump: nuisance estimation main} and give practical advice for implementation.

First, I start by highlighting a classical trade-off between Assumptions \ref{assump: nuisance estimation - DR main} and \ref{assump: nuisance estimation - l2 error main}: relying on flexible nonparametric estimators makes Assumption \ref{assump: nuisance estimation - DR main} more likely to be satisfied than when parametric estimators are used; however, nonparametric estimators typically have slower convergence rates than their parametric competitors, thus making \ref{assump: nuisance estimation - l2 error main} harder to be satisfied. As an instance, machine learning methods --such as lasso, ridge, random forests, neural networks-- can be used to estimate the nuisances $\{(\theta_a(\cdot), q_a(\cdot)),a\in\calA\}$, as long as their convergence rate (or $\ell_2$ error bounds) decay at a faster rate than $1/\sqrt{P_a(T)}$. For some real-life applications and discussions of the feasibility of these methods, see \cite*{ahrens2025IntroductionDoubleDebiased}.

I now turn to Assumption \ref{assump: nuisance estimation - independence main}, but before discussing it, it is necessary to introduce some notation. Define the main estimation dataset as
\[\mathcal{D}_{\mathtt{main}}^T:=\left\{\bZ_\ell, \ell \in[T]\right\},\qquad \bZ_j:=(R_{A_j,j}, C_{A_j,j}, \bX_{A_j,j}^\top)^\top.\] 
Appropriate nuisance estimators $\hat{\theta}_a$ and $\hat{q}_a$ that satisfy Assumption \ref{assump: nuisance estimation - independence main} can be constructed in (at least) two ways:
\begin{enumerate}
    \item[\texttt{M1}] \textit{Different batch}. Use samples from a dataset $\mathcal{D}_{\mathtt{M1}}$ such that
    \[\mathcal{D}_{\mathtt{M1}}:=\left\{\bZ_\ell, \ell \not\in[T]\right\}.\]
    For example, suppose two similar decision-makers are interacting with two copies of the same bandit with missing data. Let the datasets generated by their interactions with the MAB be denoted with $\mathcal{D}_{\mathtt{main}, j}^{T_j}, j=1,2$. Then, $\hat{\theta}_a$ and $\hat{q}_a$ can be estimated for the first decision-maker using $\mathcal{D}_{\mathtt{main}, 2}^{T_2}$ and for the second decision-maker using $\mathcal{D}_{\mathtt{main}, 1}^{T_1}$. 
    \item[\texttt{M2}] \textit{Leave-one-out}. When updating $\widehat{R}^{\DR}_a(t)$ at the end of round $t-1$, one can use estimators $\hat{\theta}_a$ and $\hat{q}_a$ that use the dataset $\mathcal{D}_{\mathtt{main}}^{t-2}$, i.e.,
    \[\mathcal{D}_{\mathtt{main}}^{t-2}:=\left\{\bZ_\ell, \ell=1,\ldots,t-2\right\}.\] 
    In words, $\hat{\theta}_a$ and $\hat{q}_a$ are constructed online using the ``leave-one-out" principle, where the sample left out corresponds to the last observed sample of data.
\end{enumerate}

As a final remark, I stress that in virtue of approaches such as \texttt{M2}, I should be writing $\hat{\theta}^{(\ell)}_a$ and $\hat{q}^{(\ell)}_a$ for each round $\ell\in[T]$. I avoid doing so to save notation, but it is maintained that the nuisance estimators are allowed to be updated with the rounds.

Third, the estimated conditional probability of missingness needs to be bounded between $\underline{q}$ and 1 to avoid dealing with ``small denominators", i.e., when some of the $\hat{q}_a$'s are close to 0.  Various alternatives exist in the literature: \cite*{crump2009DealingLimitedOverlap} proposes a data-driven trimming procedure that minimizes the variance of the estimator, and it is optimal under homoskedasticity (see \cite*{khan2025DoublyRobustHeteroscedasticityaware} for a recent extension to the heteroskedastic case); \citep{ma2020RobustInferenceUsing} warn against ad hoc trimming and propose a data-driven procedure that minimizes the asymptotic mean squared error of the resulting estimator. I stress that, while these procedures have been extensively validated via simulations, they rely on asymptotic guarantees and should therefore be applied with caution in finite-sample settings such as the one analyzed here.

\section{Simulation Evidence}
\label{sec: simulation evidence}
In this section, I present simulation results for the performance of the algorithms discussed. In the simulation, for each $a\in\calA$, I model $X_{a,j} \iid \mathsf{N}(0,1), j=1,\ldots, d, u_{a,j}\iid \mathsf{N}(0,\sigma_j^2),j\in\{C,R\},$ and define 
\[C_a = \indic\left[\sum_{\ell=1}^dX_{a,\ell} \beta_\ell+u_{a,C}>\tau(q_a)\right], \quad R_a = \theta_a + \sum_{\ell=1}^d X_{a,\ell}\beta_\ell + u_{a,R},\]
where $\bbeta:=(\beta_1,\ldots,\beta_d)^\top\in\R_{++}$ and $\tau(q_a): \PP[\sum_{\ell=1}^dX_{a,\ell} \beta_\ell+u_{a,C}>\tau(q_a)] = q_a.$ Note that
\[R_a\sim\mathsf{N}(\theta,d+\sigma_R^2), \qquad C_a\sim \mathsf{Be}(q_a),\qquad C_a\independent R_a \mid \bX_a.\]

I then set $A=2, T=5000, d=1,$ and consider three different scenarios: no missing data; reward-independent missingness; and reward-dependent missingness. Table \ref{tab: simulation scenarios main} reports the most important characteristics of these scenarios, which are the true mean rewards $(\theta_1,\theta_2)$, the probability limit of the mean reward estimator that uses only observed rewards $(\widetilde{\theta}_1,\widetilde{\theta}_2)$, and the probability of missingness. More details about the simulation, estimation of nuisance functions, and parametrization can be found in the supplemental appendix Section \ref{sec: main results}.

\begin{table}[!ht]
\caption{Details of the three simulated scenarios.}
\centering
\begin{tabular}{ccccc}
\toprule\toprule
   MAB    & $(\theta_1,\theta_2)$ & $(\widetilde{\theta}_1,\widetilde{\theta}_2)$ & $(q_1,q_2)$ \\
\midrule
\textit{Standard}     & $(0.5,1)$ & $(0.5,1)$ & $(1,1)$ \\
\textit{w/ independent missingness (Section \ref{subsec: reward-independent missingness main})}      & $(0.5,1)$ & $(0.5,1)$ & $(0.25,0.9)$ \\
\textit{w/ dependent missingness (Section \ref{subsec: reward-dependent missingness main})}        & $(0.5,1)$ & $(1.16, 1.08)$ & $(0.25,0.9)$ \\
\bottomrule
\end{tabular}%
\label{tab: simulation scenarios main}
\end{table}

\begin{figure}[h]
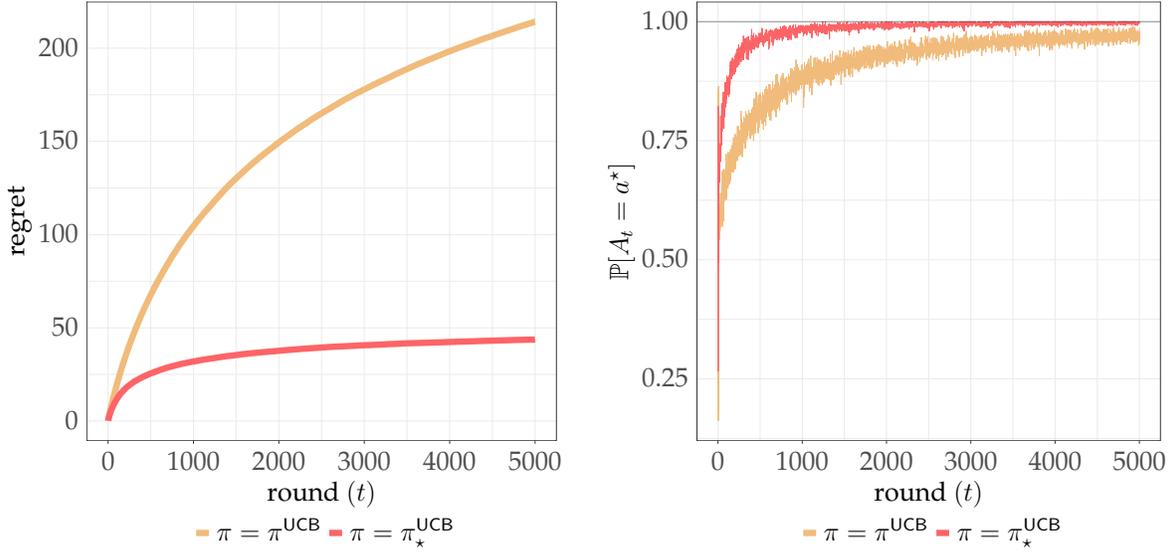

    \centering
    \caption{Regret and optimal arm selection - No missingness}
    \label{fig: no missingness}
     \begin{subfigure}[b]{0.48\textwidth}
         \centering
         \resizebox{0.95\textwidth}{!}{%
         \input{figures/cumulRegret_noCen.tex}
         }
     \end{subfigure}
    \begin{subfigure}[b]{0.48\textwidth}
         \centering
         \resizebox{0.95\textwidth}{!}{%
         \input{figures/probBestAction_noCen.tex}
         }
     \end{subfigure} 
 \par
	\begin{center}
		\parbox[1]{\textwidth}{\footnotesize \textit{Notes:} the left panel shows the cumulative regret averaged over $S=500$ draws of a bandit parametrized according to the specifics of scenario 1. A similar exercise has been conducted in the right panel to plot the probability with which each policy selects the optimal arm. The algorithm behind the policy $\pi^{\UCB}$ is described in Algorithm \ref{alg: UCB algorithm}, whereas the one behind $\pi_\star^{\UCB}$ is described in Section \ref{subsec: reward-dependent missingness supp} of the supplemental appendix.}
	\end{center}     
\end{figure}

In the first two scenarios, there is no wedge between the probability limits $(\widetilde{\theta}_1,\widetilde{\theta}_2)$ and the true mean reward, suggesting that the $\UCB$ algorithm would work fine. This is exactly what can be deduced from Figure \ref{fig: no missingness} and Figure \ref{fig: independent missingness}. 

The left panel of those figures shows the instance-specific regret of the $\UCB$ algorithm, which exhibits the classical logarithmic shape in the number of rounds. As a benchmark, the performance of $\UCB$ is compared with an oracle version of the algorithm, which does not rely on the optimistic mean reward estimator $\widetilde{R}^{\UCB}_a(t;\delta)$, but rather on a confidence interval constructed as if the data-generating process was known. The right panel of Figure \ref{fig: no missingness} and Figure \ref{fig: independent missingness} portrays the probability with which each algorithm selects the optimal arm $a^\star=2$ and shows that it approaches one as the number of rounds grows large. Notably, there is not much of a difference between scenario 1 and scenario 2. The only exception is that $\UCB$ needs more rounds to discover the optimal arm, consistent with the fact that rewards are not observed at all rounds.

\begin{figure}[h]
    \centering
    \caption{Regret and optimal arm selection - reward-independent missingness}
    \label{fig: independent missingness}
     \begin{subfigure}[b]{0.48\textwidth}
         \centering
         \resizebox{0.95\textwidth}{!}{%
         \input{figures/cumulRegret_indepCen.tex}
         }
     \end{subfigure}
    \begin{subfigure}[b]{0.48\textwidth}
         \centering
         \resizebox{0.95\textwidth}{!}{%
         \input{figures/probBestAction_indepCen.tex}
         }
     \end{subfigure} 
 \par
	\begin{center}
		\parbox[1]{\textwidth}{\footnotesize \textit{Notes:} the left panel shows the cumulative regret averaged over $S=500$ draws of a bandit parametrized according to the specifics of scenario 2. A similar exercise has been conducted in the right panel to plot the probability with which each policy selects the optimal arm. The algorithm behind the policy $\pi^{\UCB}$ is described in Algorithm \ref{alg: UCB algorithm}, whereas the one behind $\pi_\star^{\UCB}$ is described in Section \ref{subsec: reward-dependent missingness supp} of the supplemental appendix.}
	\end{center}     
\end{figure}

Sensible differences emerge under the last scenario. Indeed, the parametrization under the third scenario has been chosen so that 
\[\theta_1 < \theta_2\quad \text{but} \quad \widetilde{\theta}_1>\widetilde{\theta}_2,\]
which implies that $\UCB$ will select with high probability $a=1$, which is not the best action. On the contrary, $\DR$-$\UCB$ should still be able to pick the best arm $a^\star=2$. Figure \ref{fig: dependent missingness} demonstrates that this intuition is indeed correct, together with the theoretical results presented in Section \ref{sec: MAB}. 

Most importantly, the one under scenario 3 is an instance of a bandit in $\mathcal{C}_2$ that makes the regret of $\pi^\UCB$ grow linearly with the number of rounds. Indeed, the vanilla $\UCB$ algorithm selects the correct action only a quarter of the time after 5,000 rounds. If the number of rounds were to increase, this probability would slowly approach 0. Intuitively, it takes time for the $\UCB$ algorithm to stick to the suboptimal arm $a^\star=1$ because the probability limits $\widetilde{\theta}_1$ and $\widetilde{\theta}_2$ are very close to each other.

On the other hand, the $\DR$-$\UCB$ shows logarithmic regret, and its probability of selecting the optimal arm quickly approaches one. Even in Figure \ref{fig: dependent missingness}, I also display the performance of an oracle algorithm that does not use the bonus term $b_{a,t}^{\DR}(\delta),$ but instead leverages knowledge of the underlying data-generating process.

\begin{figure}[h]
    \centering
    \caption{Regret and optimal arm selection - Reward-dependent missingness}
    \label{fig: dependent missingness}
     \begin{subfigure}[b]{0.48\textwidth}
         \centering
         \resizebox{0.95\textwidth}{!}{%
         \input{figures/cumulRegret_depCen.tex}
         }
     \end{subfigure}
    \begin{subfigure}[b]{0.48\textwidth}
         \centering
         \resizebox{0.95\textwidth}{!}{%
         \input{figures/probBestAction_depCen.tex}
         }
     \end{subfigure} 
    \par
    \begin{center}
        \parbox[1]{\textwidth}{\footnotesize \textit{Notes:} the left panel shows the cumulative regret averaged over $S=500$ draws of a bandit parametrized according to the specifics of scenario 3. A similar exercise has been conducted in the right panel to plot the probability with which each policy selects the optimal arm. The algorithm behind the policy $\pi^{\UCB}$ is described in Algorithm \ref{alg: UCB algorithm}, the one to implement the policy $\pi^{\DR}$ in Algorithm \ref{alg: DR-UCB algorithm}, whereas the one behind $\pi_\star^{\DR}$ is described in Section \ref{sec: main results} of the supplemental appendix.}
    \end{center}     
\end{figure}

\begin{figure}
    \centering
    \caption{Mean reward estimators behavior}
    \label{fig: estimators worms}
    \resizebox{0.95\textwidth}{!}{%
    \input{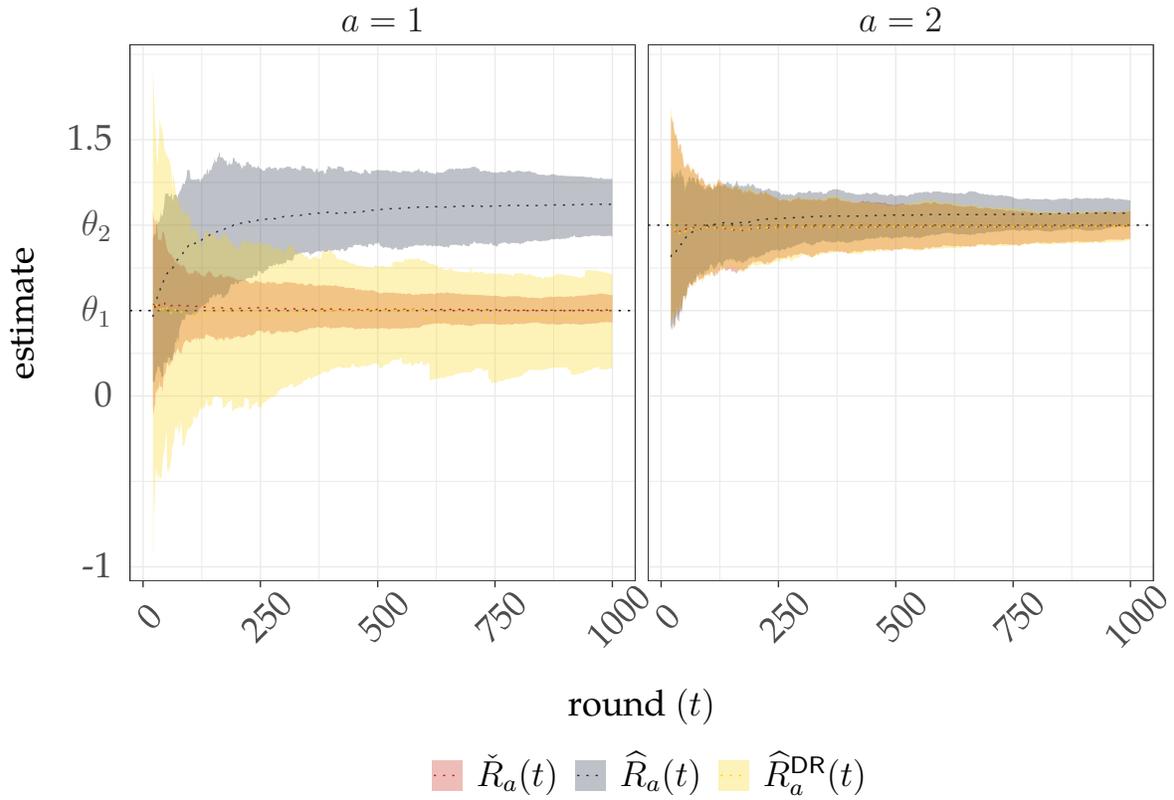}
    }
    \par
    \begin{center}
        \parbox[1]{\textwidth}{\footnotesize \textit{Notes:} dotted lines are the average value taken by an estimator across 500 draws of a bandit parametrized as under scenario 3; shaded areas are bounded between the 2.5th and 97.5th percentiles. Horizontal black lines indicate the values of the true mean rewards $\theta_a$.}
    \end{center}  
\end{figure}

Finally, Figure \ref{fig: estimators worms} illustrates why $\DR$-$\UCB$ works, whereas $\UCB$ does not under reward-dependent missingness. The graph plots the mean value, along with the 2.5th and 97.5th percentiles of the values obtained by three estimators across 1,000 draws of a bandit parametrized as in scenario 3. The three estimators are: the estimator that uses only observed rewards, $\widehat{R}_a(t)$; the doubly-robust estimator, $\widehat{R}^{\DR}_a(t)$; and an oracle estimator that always observes rewards, $\check{R}_a(t)$. 

The right panel showcases the results for action $a=2$. Under this arm, the difference between the true mean reward $\theta_2=1$ and the probability limit of $\widehat{R}_2(t)$, $\widetilde{\theta}_2=1.08$ is minimal, but still detectable from the graph. As expected, the doubly-robust and oracle estimators both quickly converge to the true mean reward $\theta_2$. The left panel portrays the differences between the estimators in a neater way. The na{\"i}ve estimator $\widehat{R}_1(t)$ rapidly approaches $\widetilde{\theta}_1=1.16$. The other two estimators converge to $\theta_1=0.5$. The larger uncertainty of the doubly-robust estimator, when compared to the oracle, comes from the fact that it estimates the nuisance functions, and because the incidence of missing data is strong under this arm, $q_1=0.2$. 

Comparing the two panels also shows that the $\UCB$ algorithm flips the true ordering of mean rewards. Put differently, $\UCB$ picks with increasing probability $a=1$, because $\widetilde{\theta}_1>\widetilde{\theta}_2$. This is nothing more than a standard sample selection problem, where the selection occurs on dimensions that are (possibly directly) related to the outcomes of interest.

\section{Conclusion}
\label{sec: conclusion}
This paper examined a sequential decision‑making problem in which feedback may be missing when the decision-maker interacts with the environment. I showed that standard methods—most notably the popular $\UCB$ algorithm—incur linear minimax regret across a wide range of such problems. In contrast, the proposed $\DR$-$\UCB$ algorithm matches the optimal minimax regret rate (up to logarithmic factors), as established by a new lower bound for this problem class. I also provide practical guidance for implementing $\DR$-$\UCB$. Extending this framework to more general settings—such as contextual bandits or models with time‑dependent feedback—constitutes a promising avenue for future work.

\singlespacing
\putbib

\end{bibunit}

\clearpage
\addtocontents{toc}{\protect\setcounter{tocdepth}{2}}


\renewcommand{\nuisEstLabel}{SA5}

\renewcommand{\thesection}{SA\arabic{section}}
\makeatletter
\renewcommand{\numberline}[1]{%
  \@cftbsnum #1\@cftasnum~\@cftasnumb%
}
\makeatother

\setcounter{section}{0}
\setcounter{equation}{0}
\setcounter{page}{0}
\setcounter{figure}{0}
\setcounter{theorem}{0}
\setcounter{lemma}{0}
\setcounter{remark}{0}
\setcounter{assumption}{0}
\setcounter{protocol}{0}
\setcounter{procedure}{0}
\setcounter{algorithm}{0}

\renewcommand\thefigure{SA-\arabic{figure}} 
\setcounter{table}{0}
\renewcommand{\thetable}{SA-\arabic{table}}
\renewcommand{\thelemma}{SA-\arabic{lemma}}
\renewcommand{\thecorollary}{SA-\arabic{corollary}}
\renewcommand{\theexample}{SA-\arabic{example}}
\renewcommand{\theremark}{SA-\arabic{remark}}
\renewcommand{\thetheorem}{SA-\arabic{theorem}}
\renewcommand{\theassumption}{SA\arabic{assumption}}
\renewcommand{\theprotocol}{SA\arabic{protocol}}
\renewcommand{\theprocedure}{SA\arabic{procedure}}
\renewcommand{\thealgorithm}{SA\arabic{algorithm}}

\renewcommand{\refname}{Supplement References}

\begin{center}
{\LARGE
\vspace{-0.75in} Supplement to ``Sequential Decision Problems with Missing Feedback''
}
\end{center}

\bigskip

\small 

This supplement contains all proofs, additional results, and other technical details. Section \ref{sec: introduction} describes the setup and notation, states the assumptions we rely on, and introduces some auxiliary results. Section \ref{sec: main results} illustrates the main technical results. Section \ref{sec: simulation} describes in detail the setting used for the simulation study. Section \ref{sec: proofs} contains all the proofs.

\setcounter{tocdepth}{2}

\singlespacing
\tableofcontents
\onehalfspacing
\thispagestyle{empty}
\clearpage

\begin{table}[ht]
    \centering
    \caption{Summary of Notation}
    \label{tab: notation}
    \doublespacing
    \resizebox{.85\textwidth}{!}{
    \setlength\extrarowheight{-6pt}
    \begin{tabular}{p{0.08\linewidth}  p{0.92\linewidth}}
        \toprule\toprule
        \multicolumn{1}{l}{\textbf{Quantity}} & \multicolumn{1}{l}{\textbf{Description}} \\
        \midrule
        \multicolumn{2}{l}{\uline{Environment}} \\
        $\calA$ & set of actions \\ 
        $A$ & number of actions \\
        $R_{a}$ & reward for action $a\in\calA$\\
        $C_{a}$ & censoring mechanism for action $a\in\calA$ \\
        $\bX_{a}$ & observed covariates for action $a\in\calA$ \\
        $\mathcal{R}$ & support of rewards \\
        $\mathcal{X}$ & support of covariates \\
        $\mathcal{Z}$ & support of $(R_a,C_a,\bX_a)$ \\
        $\mathcal{C}_j$ & class of bandits, $j\in\{0,1,2\}$ \\
        $\nu_a$ & probability measure of $(R_a,C_a,\bX_a)$ for action $a\in\calA$\\
        $\theta_a$ & unconditional mean reward for action $a\in\calA$ \\
        $\overline{\theta}$ & best mean reward \\
        $a^\star$ & action associated with the best mean reward \\
        $\theta_a(\bX_a)$ & conditional (on $\bX_a$) mean reward for action $a\in\calA$ \\
        $q_a$ & probability of censoring for action $a\in\calA$ \\
        $\acen$ & sum of inverse of censoring probabilities \\
        $\underline{q}$ & minimum probability of censoring across actions \\
        $q_a(\bX_a)$ & conditional probability of censoring \\ 
        $\sigma_a$ & variance proxy of the reward for action $a\in\calA$ \\
        $\osigma$ & largest variance proxy across all actions \\
        $\overline{K}$ & uniform upper bound for all $\theta_a$s \\ \\
        \multicolumn{2}{l}{\uline{Algorithms}} \\
        $T$ & number of rounds \\
        $\Pi$ & space of policy functions \\
        $\pi_t^{\mathsf{x}}$ & policy function at round $t\in[T]$, $\mathsf{x}\in \{\UCB,\DR,\ODR\}$\\
        $A_t$ & action chosen at round $t\in[T]$ \\
        $\regr^{\mathsf{x}}$ & pseudo-regret of algorithm $\mathsf{x}\in \{\UCB,\DR,\ODR\}$\\
        $\regr^\star(\mathcal{C})$ & minimax regret over class of bandits $\mathcal{C}$ and policies $\Pi$\\
        $\delta$ & probability with which high-probability bounds do not hold \\
        $P_a(t)$ & times action $a\in\calA$ has been played at the beginning of round $t\in[T]$ \\
        $N_a(t)$ & times the reward of action $a\in\calA$ has been observed at the beg. of round $t\in[T]$ \\
        $\widehat{R}_a^\mathsf{x}(t)$ & mean reward estimator, $\mathsf{x}\in \{\UCB,\DR,\ODR\}$ \\
        $\widetilde{R}_a^\mathsf{x}(t,\delta)$ & optimistic mean reward estimator, $\mathsf{x}\in \{\UCB,\DR,\ODR\}$ \\
        $b_a^\mathsf{x}(\delta)$ & bonus term, $\mathsf{x}\in \{\UCB,\DR,\ODR\}$ \\
        $\underline{q}_\lambda$ & minimum regularized probability of censoring across actions \\
        $\hat{q}_a(\cdot)$ & estimator of the conditional probability of censoring for action $a\in\calA$ \\
        $\hat{\theta}_a(\cdot)$ & estimator of the conditional expected reward for action $a\in\calA$ \\
        $\Err_t(\mathsf{x})$ & $\ell_2$-error rate for estimator $\mathsf{x}\in\{(\hat{q}_a,\hat{\theta}_a), a\in\calA\}$\\
        $\lambda$ & regularization parameter for $\UCB$ algorithm \\
        $\Kodr$ & regret constant \\
        \bottomrule\bottomrule
    \end{tabular} 
    }
\end{table}

\clearpage
\normalsize
\setcounter{page}{1}

\begin{bibunit}
\section{Introduction}
\label{sec: introduction}
This section introduces the notation used in this project (Section \ref{subsec: notation supp}), describes the setup and the algorithms analyzed (Sections \ref{subsec: setup} and \ref{subsec: algorithms supp}), outlines the assumptions I rely on (Section \ref{subsec: assumptions}), and presents some auxiliary lemmas that will prove useful throughout (Section \ref{subsec: aux lemmas}).

\subsection{Notation}
\label{subsec: notation supp}

\textbf{Sets.} In general, blackboard bold uppercase letters $(e.g., \N,\R)$ are used to denote standard sets of numbers. All the other sets are denoted with uppercase calligraphic letters $(e.g., \calF,\calG)$. The set of natural numbers is denoted with $\N$, the set of real numbers with $\R$, the set of non-negative real numbers with $\R_+$, and the set of positive real numbers with $\R_{++}$. I write $\R^d$ for $d\in\N$ to denote $\R^d = \bigtimes_{j=1}^d\R$, where $\bigtimes$ denotes the Cartesian product between sets. I denote an ordered set $\{1,\ldots, n\}, n\in\N$ with $[n]$. The complement of a set $\calF$ is denoted as $\overline{\calF}$. I denote the sigma-algebra generated by a random variable $X$ as $\sigma(X)$ and with $\mathscr{B}(\mathcal{S})$ the Borel $\sigma$-algebra on the topological space $\mathcal{S}$.

\textbf{Linear algebra.} Throughout the text, 
$\mathbf{0}_k$ and $\mathbf{1}_k$ denote the $k$-dimensional zero and one vectors, respectively. For a $k$ by $m$ matrix $\bA$ I use $\bA^\top$ to denote the transpose of $\bA$ and, if $k=m$ and $\bA$ is non-singular I use $\bA^{-1}$ to denote the inverse of $\bA$. For $x\in\R^k$, I write $\bx\succeq \mathbf{0}_k$ to denote the component-wise inequality in $\R^k$. Let $\bx\in\R^d$ I use $|\bx|:=\left(\sum_{i\in[d]}x_i^2\right)^{1/2}$ to the denote the Euclidean norm and $\|\bx\|_\infty:= \sup_{i\in[d]}|x_i| $ to denote the sup-norm.

\textbf{Asymptotic statements.} For two positive sequences $\left\{a_n\right\}_n,\left\{b_n\right\}_n$, I write $a_n = O(b_n)$ if $\exists\,M\in\R_{++}:a_n\leq M b_n$ for all large $n$, $a_n=o(b_n)$ if $\lim_{n\to\infty} a_n b_n^{-1} = 0$, $a_n=\widetilde{O}(b_n)$ if $\exists\, k\in\N,C\in\R_{++}: a_n=O(b_n\ln^k(Cn))$  $a_n\lesssim b_n$ if there exists a constant $C\in\R_{++}$ such that $a_n\leq C b_n$ for all large $n$, and $a_n \sim b_n$ if $a_n / b_n \rightarrow 1$ as $n \rightarrow \infty$.  For two sequences of random variables $\left\{A_n\right\}_n,\left\{B_n\right\}_n$, I write $A_n=\op(B_n)$ if $\forall\,\varepsilon\in\R_{++}, \lim_{n\to\infty} \PP[|A_n B_n^{-1}|\geq \varepsilon]= 0$ and $A_n =\Op(B_n)$ if $\forall\,\varepsilon\in\R_{++}, \exists \,M,n_0\in\R_{++} : \PP[|A_nB_n^{-1}|>M]<\varepsilon,$ for $n>n_0$. 

\textbf{Statistical Distributions.} I denote a (possibly multivariate) Gaussian random variable with $\mathsf{N}(\ba, \bB),$ where $\ba$ denotes the mean and $\bB$ the variance-covariance, with $\Be(p)$ a Bernoulli distribution with $p\in(0,1]$ denoting the success probability, and with $\sG(\sigma)$ a sub-Gaussian random variable. A random variable $X$ is sub-Gaussian with variance proxy $\sigma>0$ if $\forall\,\lambda\in\R,\E[\exp(\lambda X)]\leq \exp(\lambda^2\sigma^2/2)$ and $\E[X]=0$. With a slight abuse of terminology, I say that a random variable $Y$ with non-zero mean is sub-Gaussian when $(Y-\E[Y])\sim \sG(\sigma)$. If $\{X_t\}_{t=1}^\infty$ is an $\calF$-adapted martingale difference sequence with respect to some filtration $\calF=\{\calF_t\}_{t=1}^\infty$, then it is understood that $X_t\sim \sG(\sigma)$ requires $\forall\,\lambda\in\R,\E[\exp(\lambda X_t)\mid \calF_t]\leq \exp(\lambda^2\sigma^2/2)$ and $\E[X_t\mid\calF_t]=0$.  All probability measures are assumed to belong to the set of all Borel probability measures on an appropriately defined topological space $\mathcal{S}$.

Table \ref{tab: notation} summarizes the notation specific to this project. Notation that is only used in the proofs is omitted from the table and defined throughout.

\subsection{Setup}
\label{subsec: setup}

I start by describing a generic instance of a stochastic MAB with possibly missing rewards and the decision-maker that interacts with such an environment.

\textbf{Setting.} A decision-maker faces a sequential decision problem over $ T \in \N$ rounds in a stochastic environment. At the beginning of each round $ t \in [T] $, using all the information available at that point, the decision-maker selects an action $ A_t \in \calA:=\{1,\ldots,A\}$. Each action $ a \in \calA $ is associated with a tuple of random variables: a reward $ R_a \in \mathcal{R}\subseteq\R$, an indicator for not being missing $ C_a \in\{0, 1\} $, and some covariates $\bX_a \in\mathcal{X}\subseteq \R^k,k\in\N$. A stochastic MAB problem with missing rewards is defined as a collection of random variables $\{(R_{a,\ell},C_{a,\ell},\bX_{a,\ell})\}_{a\in\calA,\ell\in[T]}$ with the first index running over the set of actions $\calA$, the second index running over rounds, and satisfying the following three conditions for fixed actions $a,a'\in\calA:a\neq a'$:

\begin{enumerate}
    \item $\{(R_{a,t},C_{a,t},\bX_{a,t})\}_{t\in[T]}$ are $T$ independent draws from
    $(R_a, C_a, \bX_a)$, which is distributed according to some (unknown) probability measure $\nu_a$
    defined on the measurable space $(\mathcal{R}\times\{0,1\}\times
    \mathcal{X},\sigma(R_a,C_a,\bX_a))$;
    \item 
    $(R_{a},C_{a},\bX_{a})\independent (R_{a'},C_{a'},\bX_{a'})$, so that the
    unknown joint distribution of $\{(R_{a},C_{a},\bX_{a})\}_{a\in\calA}$ can be
    defined as $\nu=\prod_{a\in\calA}\nu_a$;
    \item the reward $R_{a,t}$ is observed by the decision-maker only if $C_{a,t}=1$.
\end{enumerate}

Each action $a\in\calA$ has an associated mean reward
\[\theta_a \equiv \theta_a(\nu) = \E_{\nu}[R_a],\] which I assume to be finite. I define the best (in hindsight) action, the associated best mean reward, and the sub-optimality gap as
\[a^\star:= \argmax_{a\in\calA} \theta_a, \qquad
\qquad\overline{\theta}:=\max_{a\in\calA} \theta_a, \qquad \qquad \Delta_a:=
\overline{\theta} - \theta_a,\: a\in\calA.\]

It follows from the description above that all that is needed to characterize a
MAB with possibly missing rewards is the collection of probability measures $\{\nu_a\}_{a\in\calA}$.
In this work, I focus on the following class of bandits
\[\mathcal{C}:=\left\{(\nu_a)_{a\in\calA}: \nu^{[R]}_a\in\mathcal{SG}(\sigma_a), \: \nu_a^{[C]} = \Be(q_a), q_a\in(0,1]  \right\},\] where $\nu_a^{[Y]}$ denotes the marginal of $\nu_a$ with respect to $Y\in\{C,R\}$ and $\mathcal{SG}(\sigma)$ denotes the space of sub-Gaussian probability distribution with variance proxy at most $\sigma>0$, and $\Be(p)$ denotes the probability distribution of a Bernoulli random variable with parameter $p$. The class of bandits $\mathcal{C}$ is very general, as it only restricts the mean reward to be finite, the tails to be sub-Gaussian, and rules out the trivial case in which the reward of action $a\in\calA$ is not observable (which would occur when $q_a =0$). Finally, I define $\osigma:= \sqrt{\max_{a\in\calA}\sigma_a^2}$. I conjecture that many of the results proved in this appendix can be extended to more general families of random variables, such as the class sub-Exponential or sub-Weibull random variables. For example, Lemma \ref{lemma: freedmans inequality} can be shown to hold for sub-Exponential random variables using an identical strategy to the one used throughout.

\paragraph{Decision-maker.} The interaction between the decision-maker and the environment
produces the following collection of random variables 
$$\{(A_t, \bZ_t)\}_{t\in[T]}, \qquad \bZ_j:=(R_{A_j,j}, C_{A_j,j}, \bX_{A_j,j}^\top)^\top. $$

Each decision-maker is characterized by a policy that maps the history up to round $t$,\newline  $\{(A_\ell, R_{A_\ell,\ell}, C_{A_\ell,\ell}, \bX_{A_\ell,\ell}^\top)\}_{\ell\in[t-1]}$, to the space of probability distributions over actions $\Delta(\calA)$. Denote the space of policies as 
$$\Pi:=\left\{\pi: \pi = \{\pi_t\}_{t\in[T]}, \pi_t: (\calA \times \mathcal{Z})^{t-1} \to \Delta(\calA)\right\}, \qquad \mathcal{Z}:= \mathcal{R} \times \{0,1\}\times\calX.$$
I use interchangeably the words ``decision-maker", ``algorithm", and ``policy"
when referring to $\pi\in\Pi$.

Protocol \ref{prot: stochastic mab with dependent missingness} below describes the interaction between the decision-maker and the MAB with censoring.

\begin{protocol}
\caption{Multi-Armed Bandit with Missing Rewards}
     Consider a generic bandit $\nu\in\mathcal{C},$ where $\nu=(\nu_a)_{a\in\calA}$
    \begin{algorithmic}
    \For{$\ell=1,2,\ldots,T$}
        \State{Decision-maker chooses $A_\ell=a$ according to some policy $\pi_t$}
        \State{Nature samples $(C_{a,\ell}, R_{a,\ell}, \bX_{a,\ell}) \sim \nu_a$}
        \If{$C_{a,\ell}=1$}
            \State{Decision-maker observes $R_{a,\ell}$}
        \Else
            \State{Decision-maker receives no feedback}
        \EndIf
    \EndFor
    \end{algorithmic}
\end{protocol}

\paragraph{Regret.} The \textit{pseudo-regret} of a decision-maker following a
policy $\pi$ in a MAB with missing rewards $\nu\in\mathcal{C}$ is
\[\regr(\pi; \nu) =\sum_{t=1}^T (\max_{a\in\calA}\theta_a - \E_\nu[R_{A_t,t}]) =
T\overline{\theta} - \sum_{t=1}^T \theta_{A_t},\] which depends on $\nu$ via the
average rewards, and it is a random quantity because the $\{A_t\}_{t\in[T]}$ are
random. Note that the latter is true even if the policies considered are
deterministic. The reason is that $A_t$ depends on $\{\bZ_\ell\}_{\ell=1}^{t-1}$
which are random. In what follows, I omit the dependence of the regret on $\nu$
and simply write $\regr(\pi)$.

\subsection{Algorithms}
\label{subsec: algorithms supp}

In what follows, I focus on the popular $\UCB$ algorithm
\citep{auer2002FinitetimeAnalysisMultiarmed} and modifications thereof as the algorithm used by the decision-maker to obtain a policy $\{\piucb_t\}_{t\in[T]}$. The $\UCB$ algorithm selects the optimal policy using optimistic estimates of previous rewards. Hereafter, I first describe the classic $\UCB$ algorithm and then showcase the novel doubly-robust version proposed in this project, first in its unfeasible (oracle) version and then in its feasible form.

\subsubsection{Classic $\UCB$ Algorithm}

Before formalizing the algorithm, it is necessary to introduce some notation:
\begin{itemize}
    \item The number of times an arm $a\in\calA$ has been pulled at the beginning of round $t\in[T]$ $$P_a(t) := \sum_{\ell=1}^{t-1}\indic[A_\ell=a].$$
    \item The number of times the reward $R_a$ has been observed at the beginning of round $t\in[T]$ $$N_a(t) := \sum_{\ell=1}^{t-1}\indic[A_\ell=a]
    C_{a,\ell}.$$
    \item The (regularized) estimate for the mean reward of action $a\in\calA$ at the beginning of round $t\in[T]$ is
    \begin{align*}
        \widehat{R}^{\UCB}_a(t) = \frac{1}{N_a(t) + \lambda}\sum_{\ell=1}^{t-1}\indic[A_\ell=a]C_{a,\ell} R_{a,\ell}= \frac{1}{P_a(t) + \lambda}\sum_{\ell=1}^{t-1}\indic[A_\ell=a]\frac{C_{a,\ell}R_{a,\ell}}{\hat{q}_a(\ell)},
    \end{align*}
    where $\hat{q}_a(\ell):= \frac{N_a(\ell) + \lambda}{P_a(\ell) + \lambda}$ and $\lambda>0$ is a regularization parameter that prevents the estimator from being ill-defined whenever, after initialization, it occurs that  $N_a(t)=0$ for some $a\in\calA$ and $t\geq 1$.
    \item The optimistic mean reward estimate of action $a\in\calA$ after $t\in[T]$ rounds is
    \[\widetilde{R}^{\UCB}_a(t,\delta) = \widehat{R}^{\UCB}_a(t) + b^{\UCB}_{a,t}(\delta), \]
    where the ``bonus" term $b^{\UCB}_{a,t}(\delta)$ is chosen to make sure that the optimistic mean reward estimate $\widetilde{R}^{\UCB}_a(t,\delta)$ upper bounds the true mean reward $\theta_a$ with high probability. In this specific case, I define
    \[b^{\UCB}_{a,t}(\delta) := \frac{\osigma}{\underline{q}}\sqrt{\frac{2\ln(2AT/\delta)}{ P_a(t) +\lambda}}+\frac{\lambda\overline{K}}{N_a(t)+\lambda},\] 
    where $\overline{K}$ is some constant larger than $\overline{\theta}, \underline{q}:=\inf_{a,t}\hat{q}_a(t)$, and $\delta\in(0,1)$. Intuitively, the first term in $b^{\UCB}_{a,t}(\delta)$ governs the probability with which we want the optimistic estimate to overestimate the mean reward, whereas the second term takes into account the bias induced by the regularization term $\lambda > 0$. Lemma \ref{lemma: concentration of estimated rewards} formally justifies the particular choice of $b^{\UCB}_{a,t}(\delta)$ described above. Finally, note that the introduction of the regularization parameter $\lambda>0$ makes
    $\underline{q}$ bounded away from 0.
\end{itemize}

The way the $\UCB$ algorithm works is straightforward: at round $t \in[T]$, it
selects the arm $a$ that has the highest optimistic mean reward estimate.
Algorithm \ref{alg: UCB algorithm supp} below summarizes all the steps needed by the 
classic $\UCB$ algorithm.

\begin{procedure}
\caption{Update Estimators for $\UCB$}
\label{procedure: update estimators supp}
    \begin{algorithmic}
    \For{$a\in[A]$}
        \State{$N_a(t+1)\leftarrow N_a(t) + \indic[A_t=a]C_{a,t}$}
        \State{$P_a(t+1)\leftarrow P_a(t) + \indic[A_t=a]$}
        \State{$\widehat{R}_a(t+1) \leftarrow \frac{1}{N_a(t+1) + \lambda}\sum_{\ell=1}^{t}\indic[A_\ell=a]C_{a,\ell}R_{a,\ell}$}
        \State{$b^{\UCB}_{a,t+1}(\delta) \leftarrow \frac{\osigma}{\underline{q}_\lambda}\sqrt{\frac{2\ln(2AT/\delta)}{ P_a(t+1)+\lambda}}+\frac{\lambda\overline{K}}{N_a(t+1)+\lambda}$}
        \State{$\widetilde{R}_a(t+1,\delta) \leftarrow  \widehat{R}_a(t+1) + b^{\UCB}_{a,t+1}(\delta)$}
    \EndFor
    \end{algorithmic}
\end{procedure}

\begin{algorithm}[H]
\caption{$\UCB$ algorithm}
    \hspace*{\algorithmicindent} \textbf{Input}: $\lambda>0, \underline{q}_\lambda, \osigma, T, \calA, \delta, \overline{K}$ \\
    \hspace*{\algorithmicindent} \textbf{Initialization}: pull each arm once, get $\widehat{R}^{\UCB}_a(0),$ set $P_a(0)=1, N_a(0)=C_{a,0}, \forall\,a\in\calA$
    \begin{algorithmic}[1]
        \For{$t=1,2,\ldots,T$}
        \State{pull arm $a_t=\argmax_{a\in\calA}\widetilde{R}^{\UCB}_a(t,\delta)$ and set $\pi_t^\UCB= a_t$}
        \State{call \textit{Update Estimators for $\UCB$} (Procedure \ref{procedure: update estimators supp})}
        \EndFor 
    \end{algorithmic}
    \hspace*{\algorithmicindent} \textbf{Output}: $\piucb = \{\piucb_t\}_{t\in[T]}$
    \label{alg: UCB algorithm supp}
\end{algorithm}

\subsubsection{Oracle Doubly-Robust $\UCB$ Algorithm}

Let the true conditional mean reward and probability of rewards not being missing for  arm $a\in\calA$ as
\[\theta_a^\star(\bX_a) := \E_\nu[R_{a}\mid \bX_a], \qquad q_a^\star(\bX_a) = \E_\nu[C_a\mid \bX_a]\in[\underline{q}_a, 1]\]
almost surely. Throughout, I use interchangeably the terms ``probability of rewards not being missing" and ``probability of missingness".

The two doubly-robust versions of the classic $\UCB$ algorithm -- one feasible, one unfeasible -- described here differ from the standard one because they rely on alternative mean reward estimators and bonus terms.

The oracle doubly-robust $\UCB$ algorithm ($\ODR$-$\UCB$) uses the following mean reward estimator
 \begin{align}
    \widehat{R}^{\ODR}_a(t) :&= \frac{1}{P_a(t)} \sum_{\ell=1}^{t-1}\indic[A_\ell=a] \left(\frac{R_{a,\ell}C_{a,\ell}}{q_a(\bX_{a,\ell})} - \frac{\theta_a(\bX_{a,\ell})}{q_a(\bX_{a,\ell})}\left(C_{a,\ell}-q_a(\bX_{a,\ell})\right)\right)
    \label{eq: mean reward estimator ODR} \\
    &=\frac{1}{P_a(t)} \sum_{\ell=1}^{t-1}\indic[A_\ell=a] \left(\frac{C_{a,\ell}(R_{a,\ell} - \theta_a(\bX_{a,\ell}))}{q_a(\bX_{a,\ell})} + \theta_a(\bX_{a,\ell})\right), \nonumber
\end{align}
for some known functions $\{\theta_a(\cdot), q_a(\cdot)\}_{a\in\calA}$. The ``Oracle'' part comes from the requirement that $q_a(\cdot)$ and $\theta_a(\cdot)$ being known functions, whilst the ``Double-Robust'' follows from the fact that only one among $q_a(\cdot)=q_a^\star(\cdot)$ and $\theta_a(\cdot)=\theta_a^\star(\cdot)$ needs to be true to make $\widehat{R}^{\ODR}_a(t)$ a consistent estimator of $\theta_a$ (see Lemma \ref{lemma: concentration of estimated rewards under dependent missingness}). The optimistic mean reward estimator is defined accordingly as
    \[\widetilde{R}^\ODR_a(t) = 
      \widehat{R}^\ODR_a(t) + b_{a,t}^{\ODR}(\delta),\qquad b_{a,t}^{\ODR}(\delta):= \Kodr\sqrt{\frac{2\ln(2AT/\delta)}{P_a(t)}},
    \]
where $\Kodr := \frac{\osigma}{\underline{q}} + \osigma$ and $\underline{q}:=\min_a\underline{q}_a$. Algorithm \ref{alg: ODR-UCB algorithm} details all the steps needed by $\ODR$-$\UCB$.

\begin{procedure}
\caption{Update Estimators for $\ODR$-$\UCB$}
\label{procedure: update estimators ODR-UCB supp}
    \begin{algorithmic}
    \For{$a\in[A]$}
        \State{$N_a(t+1)\leftarrow N_a(t) + \indic[A_t=a]C_{a,\ell}$}
        \State{$P_a(t+1)\leftarrow P_a(t) + \indic[A_t=a]$}
        \State{Update ${q}_a$ and ${\theta}_a$ if required}
        \State{$\widehat{R}^{\ODR}_a(t+1) := \frac{1}{P_a(t+1)} \sum_{\ell=1}^{t}\indic[A_\ell=a] \left(\frac{C_{a,\ell}(R_{a,\ell} - {\theta}_a(\bX_{a,\ell}))}{{q}_a(\bX_{a,\ell})} + {\theta}_a(\bX_{a,\ell})\right)$}
        \State{$\widetilde{R}^\ODR_a(t+1,\delta) \leftarrow
      \widehat{R}^\ODR_a(t+1) + b_{a,t}^\ODR(\delta)$}
    \EndFor
    \end{algorithmic}
\end{procedure}

\begin{algorithm}[H]
\caption{$\ODR$-$\UCB$ algorithm}
    \hspace*{\algorithmicindent} \textbf{Input}: $\lambda>0, T, \calA, \{{q}_a(\cdot),{\theta}_a(\cdot)\}_{a\in\calA}$ \\
    \hspace*{\algorithmicindent} \textbf{Initialization}: pull each arm once, get $\widehat{R}^{\ODR}_a(0)$ and set $P_a(0)=1, N_a(0)=C_{a,0}, \forall\,a\in\calA$\\
    \begin{algorithmic}[1]
        \For{$t=1,2,\ldots,T$}
        \State{pull arm $a_t=\argmax_{a\in\calA}\widetilde{R}^\ODR_a(t,\delta)$ and set $\piodrucb_t= a_t$ \Comment{ties are broken randomly}}
        \State{call \textit{Update Estimators for $\ODR$-$\UCB$} (Procedure \ref{procedure: update estimators ODR-UCB supp})}
        \EndFor        
    \end{algorithmic}
    \hspace*{\algorithmicindent} \textbf{Output}: $\piodrucb = \{\piodrucb_t\}_{t\in[T]}$
    \label{alg: ODR-UCB algorithm}
\end{algorithm}

\subsubsection{Feasible Doubly-Robust $\UCB$ Algorithm}
\label{subsubsec: feasible DR}

The feasible doubly-robust $\UCB$ algorithm ($\DR$-$\UCB$) differs from $\ODR$-$\UCB$
because it attempts to estimate the true conditional mean reward and probability of
missingness for each arm. To grant good properties in terms of regret, such estimation
needs to be conducted in appropriate ways, which is formalized in Assumptions \ref{assump: nuisance estimation - independence}-\ref{assump: nuisance estimation - l2 error}.

Once $\hat{\theta}_a$ and $\hat{q}_a$ have been constructed, the following estimator for mean rewards can be obtained as follows:
\[ \widehat{R}^{\DR}_a(t) := \frac{1}{P_a(t)} \sum_{\ell=1}^{t-1}\indic[A_\ell=a] \left(\frac{C_{a,\ell}(R_{a,\ell} - \hat{\theta}_a(\bX_{a,\ell}))}{\hat{q}_a(\bX_{a,\ell})} + \hat{\theta}_a(\bX_{a,\ell})\right).\]

The optimistic mean reward estimator for $\DR$-$\UCB$ is defined as 
\begin{alignat*}{3}
    \widetilde{R}_a^{\DR}(t,\delta) &= \widehat{R}^{\DR}_a(t) + b_{a,t}^\DR(\delta), \qquad
    &b_{a,t}^\DR(\delta) &= b_{a,t}^\ODR(\delta) + {b}_{a,t}^{[1]}(\delta) + {b}_{a,t}^{[2]}(\delta) + {b}_{a,t}^{[3]}(\delta), \\
   {b}_{a,t}^\ODR(\delta) &= \Kodr\sqrt{\frac{2\ln(2AT/\delta)}{P_a(t)}}, \qquad
   &{b}_{a,t}^{[1]}(\delta) &:= \frac{\osigma}{\underline{q}^2}\sqrt{\frac{2\ln(2AT/\delta)}{P_a(t)} }\Err_{t}(\hat{q}_a), \\
   {b}_{a,t}^{[2]}(\delta) &:=  \frac{1}{\underline{q}}\sqrt{\frac{2\ln(2AT/\delta)}{P_a(t)} }\Err_{t}(\hat{\theta}_a),\qquad & {b}_{a,t}^{[3]}(\delta)&=\Err_{t}(\hat{\theta}_a)\Err_{t}(\hat{q}_a), 
\end{alignat*}
where
\begin{align*}
    \Err_{t}(\hat{\theta}_a)&:= \sqrt{\frac{1}{P_a(t)}\sum_{\ell=1}^{t-1}\indic[A_\ell=a](\hat{\theta}_a(\bX_{a,\ell}) -\theta_a(\bX_{a,\ell}))^2},
\end{align*}
and
\begin{align*}
    \Err_{t}(\hat{q}_a)&:= \sqrt{\frac{1}{P_a(t)}\sum_{\ell=1}^{t-1}\indic[A_\ell=a](\hat{q}_a(\bX_{a,\ell}) -q_a(\bX_{a,\ell}))^2} 
\end{align*}
are the sample $\ell_2$ estimation errors the mean reward estimator incurs as it relies on the estimated counterparts of $\theta_a(\cdot)$ and $q_a(\cdot)$.

Finally, Algorithm \ref{alg: DR-UCB algorithm supp} describes in greater detail how
the $\DR$-$\UCB$ algorithm works.

\begin{procedure}
\caption{Update Estimators for $\DR$-$\UCB$}
\label{procedure: update estimators DR-UCB supp}
    \begin{algorithmic}
    \For{$a\in[A]$}
        \State{$N_a(t+1)\leftarrow N_a(t) + \indic[A_t=a]C_{a,\ell}$}
        \State{$P_a(t+1)\leftarrow P_a(t) + \indic[A_t=a]$}
        \State{Update $\hat{q}_a$ and $\hat{\theta}_a$ if required}
        \State{$\widehat{R}^{\DR}_a(t+1) := \frac{1}{P_a(t+1)} \sum_{\ell=1}^{t}\indic[A_\ell=a] \left(\frac{C_{a,\ell}(R_{a,\ell} - \hat{\theta}_a(\bX_{a,\ell}))}{\hat{q}_a(\bX_{a,\ell})} + \hat{\theta}_a(\bX_{a,\ell})\right)$}
        \State{$\widetilde{R}^\DR_a(t+1,\delta) \leftarrow
      \widehat{R}^\DR_a(t+1) + b_{a,t}^\DR(\delta)$}
    \EndFor
    \end{algorithmic}
\end{procedure}

\begin{algorithm}[H]
\caption{$\DR$-$\UCB$ algorithm}
    \hspace*{\algorithmicindent} \textbf{Input}: $\lambda>0, T, \calA, \delta, \{\hat{q}_a(\cdot),\hat{\theta}_a(\cdot)\}_{a\in\calA}$ \\
    \hspace*{\algorithmicindent} \textbf{Initialization}: pull each arm once, get $\widehat{R}^{\DR}_a(0)$ and set $P_a(0)=1, N_a(0)=C_{a,0}, \forall\,a\in\calA$\\
    \hspace*{\algorithmicindent} \textbf{Nuisances}: get estimates $\{\hat{q}_a(\bX_{a,0}),\hat{\theta}_a(\bX_{a,0})\}_{a\in\calA}\}$ according to Assumption \ref{assump: nuisance estimation}\textcolor{ptonorange}{(iii)}
    \begin{algorithmic}[1]
        \For{$t=1,2,\ldots,T$}
        \State{pull arm $a_t=\argmax_{a\in\calA}\widetilde{R}^\DR_a(t,\delta)$ and set $\pidrucb_t= a_t$}
        \State{call \textit{Update Estimators for $\DR$-$\UCB$} (Procedure \ref{procedure: update estimators DR-UCB supp})}
        \EndFor        
    \end{algorithmic}
    \hspace*{\algorithmicindent} \textbf{Output}: $\pidrucb = \{\pidrucb_t\}_{t\in[T]}$
    \label{alg: DR-UCB algorithm supp}
\end{algorithm}

\subsection{Assumptions}
\label{subsec: assumptions}

Assumption \ref{assump: independent missingness} is a strong assumption that implies that rewards are missing at random. Sub-gaussianity allows to control the tail behavior of rewards.

\begin{assumption}[Reward-Independent Missingness]\label{assump: independent missingness}
    For each action $a\in\calA,C_a\independent R_a$ and $R_a\sim\sG(\sigma_a)$.
\end{assumption}

Assumption \ref{assump: conditional ignorability} relaxes the previous assumption and instead assumes the process that causes missing data does not depend on rewards only conditional on a vector of observable variables $\bX_a$. On top of that, Assumption \ref{assump: conditional ignorability} has two other mild requirements: \textit{(i)} the conditional expectation of each $R_a$ is uniformly bounded over $\calX$; \textit{(ii)} the probability of observing rewards is non-zero ($q_a>0$).

\begin{assumption}[Model- and Design-based Ignorability]
\label{assump: conditional ignorability}
For each $a\in\calA$, either
\[\E_\nu[R_a\mid \bX_a, C_a] = \E_\nu[R_a\mid \bX_a]=:\theta_a(\bX_a)\qquad \text{a.s.}\tag{\textbf{MB}}\]
or    
\[\E_\nu[C_a\mid \bX_a, R_a] = \E_\nu[C_a\mid \bX_a]=:q_a(\bX_a)\qquad \text{a.s.}\tag{\textbf{DB}}\]
holds. Moreover, $R_a \mid \bX_a,  \sim \sG(\sigma_a) $ and  $q_a(\bx) \in[\underline{q},1]$,
for some constants $0\leq \overline{K}_\theta<\infty$ and $\underline{q}\in(0,1]$.
\end{assumption}

Assumptions \ref{assump: oracle DR} and \ref{assump: non-oracle DR} are equivalent in all aspects and have been differentiated only for illustration and interpretation purposes. Assumption \ref{assump: oracle DR} requires one between the true conditional probability of missingness, $q^\star_a(\bx)$, and the true conditional expectation of rewards, $\theta^\star_a(\bx)$, to be a known function. As mentioned, Assumption \ref{assump: non-oracle DR} is equivalent, but requires such known functions to be the probability limit of an estimator.

\begin{assumption}[Oracle Double Robustness]
\label{assump: oracle DR}
For each $a\in\calA,\bx\in\calX$, either
    \begin{align*}
        q_a(\bx) = q_a^\star(\bx), \tag{a}
    \end{align*}
    or
    \begin{align*}
        \theta_a(\bx)=\theta_a^\star(\bx),\tag{b}
    \end{align*}
holds for some known functions $q_a:\calX\to[\underline{q},1]$ and $\theta_a:\calX\to \mathcal{R}$.
\end{assumption}

\begin{assumption}[Double Robustness]
\label{assump: non-oracle DR}
For each $a\in\calA,\bx\in\calX,$ either
\begin{align*}
    q_a(\bx) = q_a^\star(\bx), \tag{a}
\end{align*}
holds or
\begin{align*}
    \theta_a(\bx)=\theta_a^\star(\bx),\tag{b}
\end{align*}
where $q_a(\bx)$ and $\theta_a(\bx)$ are the probability limits as $T\to\infty$ of the estimators $\hat{q}_a(\cdot)$ and $\hat{\theta}_a(\cdot)$, respectively.    
\end{assumption}

Assumption \ref{assump: nuisance estimation} disciplines the estimation of the nuisance functions $\{(q_a(\cdot),\theta_a(\cdot),a\in\calA\}$. Assumption \ref{assump: nuisance estimation - truncation} requires the estimator to bound the estimated probability of missingness away from 0, a typical regularity condition in such problems. To avoid over-fitting biases, Assumption \ref{assump: nuisance estimation - independence} asks the nuisance functions to be estimated in an independent (conditional on $\bX_a$) sample. Lastly, \ref{assump: nuisance estimation - l2 error} controls the estimation error of the nuisance estimators in two ways: \textit{(i)} the estimation error of each nuisance need to be shrinking in $P_a(t)$; \textit{(ii)} the product of the estimation errors must decay faster than $1/\sqrt{P_a(t)}$. These conditions make the sampling error dominate the estimation error induced by the fact that $\{(\theta_a(\cdot),q_a(\cdot)),a\in\calA\}$ are estimated.
\begin{assumption}[Nuisance Estimation]\label{assump: nuisance estimation}
For each $a\in\calA,$ the following are true:
\begin{enumerate}[label=(\alph*), ref={\nuisEstLabel(\alph*)}]
    \item (double robustness) either $\widetilde{q}_a(\bx)=q_a(\bx)$ or $\widetilde{\theta}_a(\bx)=\theta_a(\bx)$;
    \label{assump: nuisance estimation - DR}
    \item (truncation) $\forall\,\bx\in\calX,\widehat{q}_a(\bx) \in [\underline{q}, 1], \underline{q}\in(0,1]$; \label{assump: nuisance estimation - truncation}
    \item (independence)  $(\hat{q}_a(\bX_{a}),\hat{\theta}_a(\bX_{a}))\independent (R_{a},C_{a})\mid \bX_{a}$; \label{assump: nuisance estimation - independence}
    \item ($\ell_2$-error rate) there exist rates $\alpha> 1/2, \alpha_q>0,$ and $\alpha_\theta>0$ such that 
    \[\Err_t(\hat{q}_a) \lesssim \frac{1}{P_a(t)^{\alpha_q}}, \qquad \Err_t(\hat{\theta}_a) \lesssim \frac{1}{P_a(t)^{\alpha_\theta}}, \qquad \Err_t(\hat{q}_a)\Err_t(\hat{\theta}_a)\lesssim \frac{1}{P_a(t)^{\alpha}}\]
    with probability $1-\delta_\mathfrak{c}, \delta_\mathfrak{c}\in(0,1)$.
    \label{assump: nuisance estimation - l2 error}
\end{enumerate}
\end{assumption}

\subsection{Auxiliary Lemmas}
\label{subsec: aux lemmas}
The following lemma helps bound sums involving dependent random variables.
\begin{lemma}[Freedman's Inequality]
\label{lemma: freedmans inequality}
Let $\left\{\left(D_k, \mathcal{F}^k\right)\right\}_{k \geq 1}$ be a martingale
difference sequence and let $\{\nu_k\}_{k=1}^n$ be random variables such that
$\nu_k$ is $\mathcal{F}^{k-1}$-measurable. If $\forall\, \kappa\in\R\setminus\{0\}$
    \[\mathbb{E}\left[e^{\kappa D_k} \mid \mathcal{F}_{k-1}\right] \leq
    e^{\kappa^2 \nu_k^2 / 2}\quad \text {a.s.},\]
    then 
\[\left|\sum_{k=1}^n D_k\right| \leq \sqrt{2\log(2/\delta)\sum_{k=1}^n\nu_k^2}\]
with probability at least $1-\delta$. Furthermore, if $\sum_{k=1}^n\nu_k^2\leq V$ a.s., then
\[\left|\sum_{k=1}^n D_k\right| \leq\sqrt{2V\log(2/\delta)}\] with probability at least $1-\delta$.
\end{lemma}
\begin{flushright}
[\hyperref[proof: freedmans inequality]{Proof}]
\end{flushright}

The next lemma shows that the product of a sub-Gaussian random variable and a Bernoulli random variable is still sub-Gaussian despite imposing no structure on the joint behavior of the two variates.
\begin{lemma}
    \label{lemma: bernoulli and sg product}
    Let $X \sim\sG(\sigma), \sigma>0,Y\sim \Be(p),p\in(0,1],$ and $Z:=X\cdot Y$. Then $Z\sim \sG(\sigma)$.
\end{lemma}
\begin{flushright}
[\hyperref[proof: bernoulli and sg product]{Proof}]
\end{flushright}

I had a much more slack result showing that $Z\sim \mathsf{sE}(2\sigma, 2\sqrt{2}\sigma)$, where $\mathsf{sE}$ denotes a sub-Exponential random variable. This tighter result was suggested by the user VHarisop in the following StackExchange \href{https://math.stackexchange.com/questions/4969890/product-of-sub-gaussian-and-bernoulli-random-variable/4970742#4970742}{post}. Note that nothing is assumed on the joint of $X$ and $Y$. 

The next lemma shows that sub-Gaussianity of $Y\mid X$ is inherited by the fluctuations of $\E[Y\mid X]$ around $\E[Y]$.
\begin{lemma}\label{lemma: deviation of conditional expectation}
Let $Y\mid X \sim\sG(\sigma), \sigma>0,$ and define $W:=\E[Y\mid X] -\E[Y]$. Then, $W\sim\sG(\sigma)$.
\end{lemma}
\begin{flushright}
[\hyperref[proof: deviation of conditional expectation]{Proof}]
\end{flushright}

Finally, the next lemma shows some useful properties for the Kullback-Leibler divergence, $\dkl$, between two probability distributions.

\begin{lemma}\label{lemma: kl properties}
    Let $P$ and $Q$ be two probability distributions on $\calX\times\calY$ that admit densities $p$ and $q$, with respect to the Lebesgue measure. Then,
    \[\dkl(P,Q) = \dkl(P_X, Q_X) + \E_{X\sim P_X}[\dkl(P_{Y\mid X}, Q_{Y\mid X})].\]
    Furthermore, if $X \independent Y$, then
     \[\dkl(P,Q) = \dkl(P_X, Q_X) + \dkl(P_Y, Q_Y).\]   
\end{lemma}
\begin{flushright}
[\hyperref[proof: kl properties]{Proof}]
\end{flushright}

\clearpage
\section{Main Results}
\label{sec: main results}

In this section, I analyze how the algorithms described in Section \ref{subsec: algorithms supp} perform in the two variations of the MAB with missing rewards problem described in Section \ref{subsec: setup}. In Section \ref{subsec: reward-independent missingness supp}, I assume that rewards and censoring mechanisms are independent and show that the standard $\UCB$ algorithm still achieves nearly-optimal regret. In Section \ref{subsec: reward-dependent missingness supp}, I relax the independence assumption and show that the $\UCB$ algorithm has linear regret, whereas $\ODR$-$\UCB$ and $\DR$-$\UCB$ both possess nearly-optimal regret rates. Specifically, I provide high-probability bounds that hold uniformly over the number of rounds $T$ and a subclass of bandits contained in $\mathcal{C}$.

The probability measure underlying all these statements is a probability measure induced by the interaction between the policy $\pi\in\Pi$ and some bandit $\nu\in\mathcal{C}$. Formally, let $R_t=R_{A_t,t}, C_t=C_{A_t,t},$ and $\bX_t=\bX_{A_t,t}$. The probability measure considered throughout is the probability distribution associated to the tuple $(A_1,R_1,C_1, \bX_1, \ldots, A_T, R_T,C_T,\bX_T)$ on the measurable space $(\Omega_T, \calG_T),$ where $\Omega_T:=(\calA\times \R \times \{0,1\}\times \mathcal{X})^{A\cdot T}$ and $\calG_T:=\mathscr{B}(\Omega_T)$. For more technical details on the construction of the appropriate measures and the underlying probability space, I refer the reader to Chapter 4.4 in \cite{lattimore2020BanditAlgorithms} and references therein.

\subsection{Reward-independent Missingness}
\label{subsec: reward-independent missingness supp}

In what follows, I provide an upper bound for the regret of $\UCB$ using standard arguments. Namely, I consider a ``good" event and show that it occurs with high probability in the setting considered throughout. In this spirit, define such an event as
$$\calG(\delta_1,\delta_2) = \overline{\calF^{\UCB}(\delta_1)} \cap \overline{\calF^{\MIS}(\delta_2)},$$ 
for some $\delta_1,\delta_2\in(0,1)$, where
\begin{align*}
    \calF^\UCB(\delta) &:= \left\{\exists\,a\in\calA,t\in[T],\left|\widehat{R}^{\UCB}_a(t) - \theta_a\right|\geq b^{\UCB}_{a,t}(\delta)\right\}, \\
    \calF^{\MIS}(\delta)&:=\left\{\exists\,a\in\calA,t\in[T]:N_a(t)\leq (1-\delta)q_aP_a(t), P_a(t) \geq \underline{T}_a\right\},
\end{align*}
where $\underline{T}_a:= 1+\frac{24\ln(T)}{q_a}$. Under the good event: (\textit{i}) for each action, the optimistic reward estimator always covers the true mean; (\textit{ii}) the censoring mechanism is not too extreme in terms of percentage deviation from its mean.

The next lemma justifies the particular choice of bonus term for the $\UCB$ algorithm, that is
\[b^{\UCB}_{a,t}(\delta) := \frac{\osigma}{\underline{q}_\lambda}\sqrt{\frac{2\ln(2AT/\delta)}{ P_a(t) +\lambda}}+\frac{\lambda\overline{K}}{N_a(t)+\lambda}.\]
The following result shows that the absolute deviation between the mean reward estimator $\widehat{R}^{\UCB}_a(t)$ and the true mean reward $\theta_a$ is larger than $b^{\UCB}_{a,t}(\delta AT)$ with probability at most $\delta$. 
\begin{lemma}
    \label{lemma: concentration of estimated rewards}
    Let Assumption \ref{assump: independent missingness} hold, $\delta\in(0,1), a\in\calA,$ and $ t\in[T]$. Then 
    \[\left|\widehat{R}^{\UCB}_a(t) - \theta_a\right|\geq b^{\UCB}_{a,t}(\delta AT)\]
    with probability at most $\delta$.
\end{lemma}
\begin{flushright}
[\hyperref[proof: concentration of estimated rewards]{Proof}]
\end{flushright}

Lemma \ref{lemma: probability of failure event}, using Lemma \ref{lemma: concentration of estimated rewards} as a building block, quantifies the probability with which $\calF^\UCB(\delta)$ realizes. Lemma \ref{lemma: probability of missingness event} serves a similar purpose for $\calF^{\mathsf{CEN}}(\delta)$. Lemma \ref{lemma: UCB property} shows an implication of the event $\overline{\calF}^\UCB(\delta)$, which turns out to be useful when bounding the regret.

\begin{lemma}
\label{lemma: probability of failure event}
    Let Assumption \ref{assump: independent missingness} hold and $\delta \in(0,1)$. Then $$\PP[\calF^\UCB(\delta)]\leq\delta.$$
\end{lemma}
\begin{flushright}
[\hyperref[proof: probability of failure event]{Proof}]
\end{flushright}

\begin{lemma}
\label{lemma: probability of missingness event}
Let Assumption \ref{assump: independent missingness} hold, $\delta\in(0,1),$ and $a\in\calA$. Then 
$$ \PP[\calF^{\MIS}(\delta)]\leq  \frac{2}{\delta^2}T^{1-12\delta^2} \acen,$$ 
where $\acen:=\sum_{a\in\calA}q_a^{-1}$.
\end{lemma}
\begin{flushright}
[\hyperref[proof: probability of missingness event]{Proof}]
\end{flushright}

\begin{lemma}\label{lemma: UCB property}
    Let Assumption \ref{assump: independent missingness} hold and $\delta\in(0,1)$. With probability at least $1-\delta$ or, equivalently, under the event $\overline{\calF^\UCB(\delta)}$, it holds that
    \[\forall\,a\in\calA,t\in[T],\qquad \widetilde{R}^{\UCB}_a(t,\delta) > \theta_a.\]
\end{lemma}
\begin{flushright}
[\hyperref[proof: UCB property]{Proof}]
\end{flushright}

Before stating the first theorem, I present Lemma \ref{lemma: bound on inverse of number of observed}. This is a technical lemma that plays a crucial role in bounding the reciprocal of the (random) number of times an arm has been pulled and its feedback observed. This quantity needs to be handled because of the second term in $b_{a,t}^\UCB(\delta)$ that addresses the presence of regularization bias.

\begin{lemma}
\label{lemma: bound on inverse of number of observed}
Let Assumption \ref{assump: independent missingness} hold and $\delta\in(0,1)$. With probability at least $1-\delta$  or, equivalently, under the event $\overline{\calF^{\MIS}(\delta)}$, it holds that
\[\sum_{t=1}^T\frac{1}{N_a(t)+\lambda}\leq \frac{\acen}{1-\delta}\ln\left(\frac{T}{\acen} +\frac{\lambda}{1-\delta}\right),\]
and
\[\sum_{t=1}^T\frac{1}{\sqrt{N_a(t)+\lambda}}\leq \frac{\acen}{\sqrt{1-\delta}}\sqrt{\frac{T}{\acen} +\frac{\lambda}{1-\delta}}.\]
\end{lemma}
\begin{flushright}
[\hyperref[proof: bound on inverse of number of observed]{Proof}]
\end{flushright}

I am now able to state and prove the first main result: the standard $\UCB$ algorithm has optimal (up to logarithmic factors) regret when the process that causes missing data does not depend on rewards.
\begin{theorem}
\label{thm: regret UCB in MAB}
Let Assumption \ref{assump: independent missingness} hold, $\lambda=o(T^{1/2}),\delta_1\in(0,1) ,\delta_2=\sqrt{\frac{1+\kappa}{12}},$ and $\kappa>0.$ Then, for any $T\in\N$ and bandit $\nu\in\mathcal{C}_1$
\[ \regr(\piucb) \leq \frac{4\osigma}{\underline{q}_\lambda}\sqrt{2AT\ln(2AT/\delta_1)} + o(\sqrt{T}) + A\overline{K},\]
with probability at least $1-\delta_1 - O(T^{-\kappa})$.
\end{theorem}
\begin{flushright}
[\hyperref[proof: regret UCB in MAB]{Proof}]
\end{flushright}

\subsection{Reward-dependent Missingness}
\label{subsec: reward-dependent missingness supp}
Now, I relax Assumption \ref{assump: independent missingness} and instead assume that rewards and the censoring mechanisms are independent only conditional on a vector of covariates $\bX_a$. Accordingly, in this section, the class of bandit considered is 
\[\mathcal{C}_2:=\left\{\nu=(\nu_a)_{a\in\calA}: \nu_a^{[C]} = \Be(q_a), q_a\in(0,1], \text{ and Assumption \ref{assump: conditional ignorability} holds} \right\}.\]
In this more complex setting, the sample average of the observed rewards $\widehat{R}_a^{\UCB}(t)$, is not a consistent estimator of $\theta_a$ anymore. On the contrary, $\widehat{R}_2^{\ODR}(t)$ and $\widehat{R}_2^{\DR}(t)$ are consistent estimators for $\theta_a$ under Assumption \ref{assump: oracle DR} and Assumptions \ref{assump: non-oracle DR}-\ref{assump: nuisance estimation}, respectively.

Next, I show via an example that the min-max regret of $\UCB$ grows linearly with the number of rounds $T$.

\subsubsection{Classic $\UCB$ Algorithm}
\label{subsubsec: classic ucb}
It is not hard to find instances of bandits in $\mathcal{C}_2$ that make the regret of the standard $\UCB$ algorithm grow linearly with $T$. For example, suppose that $\calA=\{1,2\}, C_a \sim \Be(1/2), a\in\calA,$ and 
\[\begin{cases}
    R_1 \sim \mathsf{Unif}([0,1/2]), \quad &\text{if } C_1 = 1, \\
    R_1 \sim \mathsf{Unif}([1/2,1]), \quad &\text{if } C_1 = 0, 
\end{cases}, \qquad\qquad R_2 \sim \mathsf{Unif}([0,3/4]).\]
As $t$ grows large, the probability limit of $\widehat{R}^{\UCB}_a(t)$ is $\E_\nu[R_a \mid C_a=1]$. Under Assumption \ref{assump: oracle DR}, the probability limit of $\widehat{R}_2^{\ODR}(t)$ is $\theta_a$, whereas the same holds for $\widehat{R}_2^{\DR}(t)$ under Assumptions \ref{assump: non-oracle DR}-\ref{assump: nuisance estimation} (see Lemma \ref{lemma: concentration of estimated rewards under dependent missingness} and Lemma \ref{lemma: concentration of estimated DR rewards under dependent missingness} for formal arguments). For the second arm, $\E_\nu[R_2\mid C_2=1]=\theta_2$, thus $\widehat{R}_2(t)$ and $\widehat{R}_2^{\ODR}(t)$ share the same probability limit. However, for the first arm $\theta_1 = 1/2 > 1/4 = \E_\nu[R_1\mid C_1=1]$, thus the two mean reward estimators converge to different values. In this example, the optimal arm is $a^\star=1$ because $\theta_1=1/2>3/8=\theta_2$. The $\ODR$-$\UCB$ uses the right mean reward estimator and consistently chooses the first arm. On the contrary, the standard $\UCB$ algorithm will eventually end up stuck selecting the second arm because $\E[R_1\mid C_1=1] = 1/4 < 3/8 = \E[R_2\mid C_2=1]$. 

The example above belongs to the class of bandits $\mathcal{C}_2$, for which the standard \textsc{UCB} algorithm consistently selects a suboptimal arm, leading to regret that grows linearly with $T$. More generally, the standard $\UCB$ algorithm has linear regret in all those bandits in which the censoring is negatively correlated with the rewards, that is smaller rewards are observed with higher probability. When such censoring is not properly addressed, the estimated ranking of actions might be a scrambled version of the true one.

Now, I proceed showing that $\ODR$-$\UCB$ and $\DR$-$\UCB$ achieve nearly optimal regret rates.

\subsubsection{Oracle Doubly-Robust $\UCB$ Algorithm}

The next lemma justifies the particular choice of bonus term for the $\ODR$-$\UCB$ algorithm:
\[b_{a,t}^{\ODR}(\delta):= \Kodr\sqrt{\frac{2\ln(2AT/\delta)}{P_a(t)}}.\]
This result will also be useful when illustrating the properties of the feasible version of this algorithm.
\begin{lemma}
    \label{lemma: concentration of estimated rewards under dependent missingness}
    Let Assumptions \ref{assump: conditional ignorability} and \ref{assump: oracle DR} hold with $\underline{q} >0$, $\delta\in(0,1), a\in\calA,$ and $ t\in[T]$. Then, with probability at most $\delta$, it holds that
    \[\left| \widehat{R}^{\ODR}_a(t) - \theta_a\right|\geq  b_{a,t}^{\ODR}(AT\delta).\]
\end{lemma}
\begin{flushright}
[\hyperref[proof: concentration of estimated rewards under dependent missingness]{Proof}]
\end{flushright}

The following theorem provides a high-probability regret bound for the $\ODR$-$\UCB$ algorithm that holds uniformly over any horizon $T$ and for any bandit in the class $\mathcal{C}_2$. Since this result is a special case of Theorem~\ref{thm: regret UCB in MAB with dependent missingness}, I do not present a separate proof. Instead, I refer the reader to the proof of that more general result.

\begin{theorem}
\label{thm: regret oracle UCB in MAB with dependent missingness}
Let Assumptions \ref{assump: conditional ignorability} and \ref{assump: oracle DR} hold and $\delta \in(0,1)$. Then, for any horizon $T\in\N$ and bandit $\nu\in\mathcal{C}_2$ 
\begin{align*}
\regr(\piodrucb) \leq \frac{4\osigma}{\underline{q}_\lambda}\sqrt{AT\ln(2AT/\delta)} +  A\overline{K}
\end{align*}
with probability $1-\delta$.
\end{theorem}
\begin{flushright}
[\hyperref[proof: regret oracle UCB in MAB with dependent missingness]{Proof}]
\end{flushright}

The above result claims that, under Assumption \ref{assump: conditional ignorability}, the $\ODR$-$\UCB$ algorithm achieves a nearly optimal rate for the worst-case regret.

\subsubsection{Feasible Doubly-Robust $\UCB$ Algorithm}

The major drawback of the $\ODR$-$\UCB$ algorithm is that it is unfeasible to use in practice. Indeed, Assumption \ref{assump: oracle DR} is particularly stringent as it requires both the conditional probability of censoring $q_a(\cdot)$ and the conditional expected reward function $\theta_a(\cdot)$ to be known for each action $a\in\calA$. This assumption can be relaxed by relying on appropriate estimators $\hat{q}_a(\bx)$ and $\hat{\theta}_a(\bx)$ (in the sense of Assumption \ref{assump: nuisance estimation}), whose probability limits are denoted as $q_a(\bx)$ and $\theta_a(\bx)$, respectively, for each $\bx\in\calX$. It then suffices to assume that at least one of them is correctly specified, i.e., either $q_a(\bx) = q_a^\star(\bx)$ holds or $\theta_a(\bx)=\theta_a^\star(\bx)$; see Assumption \ref{assump: non-oracle DR} for a formalization of this concept.

Once appropriate $\hat{\theta}_a$ and $\hat{q}_a$ have been constructed for each $a\in\calA$, the feasible doubly-robust mean reward estimator is
\[ \widehat{R}^{\DR}_a(t) = \frac{1}{P_a(t)} \sum_{\ell=1}^{t-1}\indic[A_\ell=a] \left(\frac{C_{a,\ell}(R_{a,\ell} - \hat{\theta}_a(\bX_{a,\ell}))}{\hat{q}_a(\bX_{a,\ell})} + \hat{\theta}_a(\bX_{a,\ell})\right).\]

The next lemma shows that the bonus term for the $\DR$-$\UCB$ algorithm has been chosen appropriately in the sense that it controls the probability with which $\widehat{R}^{\DR}_a(t)$ deviates from $\theta_a$.

This result is of independent interest as it is the first one that provides high-probability bounds for a doubly-robust estimator under mild assumptions. The strategy of the proof is simple: it first decomposes $\widehat{R}^{\DR}_a(t)$ as the sum of $\widehat{R}^{\ODR}_a(t)$ and three residual terms $R_{a,j}(t),j\in\{1,2,3\}$. Then, it shows that each of the four terms in $b_{a,t}^\DR(\delta AT)$ serves a specific role in bounding in probability each of the terms mentioned above. In particular, $b_{a,t}^\ODR(\delta AT)$ controls $|\widehat{R}^{\ODR}_a(t) - \theta_a|$ (as already proven in Lemma \ref{lemma: concentration of estimated rewards under dependent missingness}), whilst each $b_{a,t}^{[j]}(\delta AT)$ controls $|R_{a,j}(t)$ for $j\in\{1,2,3\}$.

\begin{lemma}
    \label{lemma: concentration of estimated DR rewards under dependent missingness}
    Let Assumptions \ref{assump: conditional ignorability}, \ref{assump: non-oracle DR}, and \ref{assump: nuisance estimation} hold, $\delta\in(0,1), a\in\calA,$ and $t\in[T]$. Then, 
    \[\left| \widehat{R}^{\DR}_a(t) - \theta_a\right|\geq  b_{a,t}^\DR(\delta AT)\]
    with probability at most $\delta$.
\end{lemma}
\begin{flushright}
[\hyperref[proof: concentration of estimated DR rewards under dependent missingness]{Proof}]
\end{flushright}

Define the failure event
\[\calF^{\DR}(\delta) := \left\{\exists\,a\in\calA,t\in[T],\left| \widehat{R}^{\DR}_a(t) - \theta_a\right|\geq b_{a,t}^\DR(\delta)\right\}.\]
When $\calF^{\DR}(\delta)$ occurs the optimistic doubly-robust reward estimator $\widetilde{R}_a^{\DR}(t) = \widehat{R}_a^{\DR}(t) + b_{a,t}^{\DR}(t)$ does not cover the true mean reward $\theta_a$. The next lemma shows that such an event occurs with arbitrarily small probability.
\begin{lemma}
\label{lemma: probability of failure event with dependent missingness}
    Let Assumptions \ref{assump: conditional ignorability}, \ref{assump: non-oracle DR}, and \ref{assump: nuisance estimation} hold and $\delta\in(0,1)$. Then,
    $$\PP[\calF^{\DR}(\delta)]\leq\delta.$$
\end{lemma}
\begin{flushright}
[\hyperref[proof: probability of failure event with dependent missingness]{Proof}]
\end{flushright}

Similarly to Lemma \ref{lemma: UCB property}, the next lemma shows an implication of the event ${\calF}^\DR(\delta)$, which turns out to be useful when bounding the regret.
\begin{lemma}\label{lemma: UCB property with dependent missingness}
    Let Assumptions \ref{assump: conditional ignorability}, \ref{assump: non-oracle DR}, and \ref{assump: nuisance estimation} hold and $\delta\in(0,1)$. With probability at least $1-\delta$ or, equivalently, under the event, $\overline{\calF^{\DR}(\delta)}$ it holds that
    \[\forall\,a\in\calA,t\in[T],\qquad \widetilde{R}_a^{\DR}(t,\delta)= \widehat{R}^{\DR}_a(t)  + b_{a,t}^\DR(\delta) \geq \theta_a.\]
\end{lemma}
\begin{flushright}
[\hyperref[proof: UCB property with dependent missingness]{Proof}]
\end{flushright}

Finally, the next theorem shows that the regret of the $\DR$-$\UCB$ algorithm is nearly optimal (up to logarithmic factors) and provides an upper bound that holds with high probability uniformly over the horizon $T$ and the class of bandits $\mathcal{C}_2$.
\begin{theorem}
\label{thm: regret UCB in MAB with dependent missingness}
Let Assumptions \ref{assump: conditional ignorability}, \ref{assump: non-oracle DR}, and \ref{assump: nuisance estimation} hold with $\delta \in(0,1)$ and  $\delta_{\mathfrak{c}} \in(0,1)$. Then, for any horizon $T\in\N$ and bandit $\nu\in\mathcal{C}_2$ 
\begin{align*}
\regr(\pidrucb)\leq \frac{4\osigma}{\underline{q}}\sqrt{AT\ln(2AT/\delta)} + \widetilde{o}(\sqrt{T}) + A\overline{K}
\end{align*}
with probability $1-\delta-\delta_\mathfrak{c}$.
\end{theorem}
\begin{flushright}
[\hyperref[proof: regret UCB in MAB with dependent missingness]{Proof}]
\end{flushright}

\subsection{Lower Bound on Minimax Regret}
\label{subsec: lower bound minimax regret}
In this section, I show that the minimax regret
\[
\regr^\star(\mathcal{C}_j) := \inf_{\pi\in\Pi}\sup_{\nu\in\mathcal{C}_j} \regr(\pi;\nu), \quad j\in\{1,2\}
\]
is lower bounded by a constant times $\sqrt{T}$.

The proof follows standard arguments using Le Cam’s two-point method. Specifically, I analyze the regret of an arbitrary policy $\tilde{\pi}\in\Pi$ on two carefully chosen instances $\nu,\nu'\in\mathcal{C}$, and show that
\[
\sup_{\tilde{\nu}\in\mathcal{C}}\regr(\tilde{\pi};\tilde{\nu}) \geq \max\{\regr(\tilde{\pi},\nu), \regr(\tilde{\pi},\nu')\} \geq f(T)
\]
for some function $f(\cdot)$. Since $\tilde{\pi}$ is arbitrary, this implies
\[
\regr^\star(\mathcal{C}) \geq f(T).
\]

To lower bound minimax regret over $\mathcal{C}_1$ and $\mathcal{C}_2$, I construct the same lower bound for the Gaussian subclasses
\[
\mathcal{C}_1^{\mathtt{gau}} := \{(\nu_a)_{a\in\calA}: \nu_a^{[R]}=\mathcal{N}(\theta_a, 1), \; \nu_a^{[C]}=\mathrm{Be}(q_a), \; q_a\in(0,1]\} \subset \mathcal{C}_1
\]
and
\begin{align*}
\mathcal{C}_2^{\mathtt{gau}} := \big\{(\nu_a)_{a\in\calA} :\; & \nu_a^{[R|X]} = \mathcal{N}(\theta_a(X), 1),\; \nu_a^{[C]} = \mathrm{Be}(q_a),\; q_a \in (0,1], \\
& \text{and Assumption \ref{assump: conditional ignorability} holds} \big\} \subset \mathcal{C}_2.
\end{align*}

Because the supremum in the minimax regret is taken over a smaller class, the same lower bound extends to $\mathcal{C}_1$ and $\mathcal{C}_2$. The argument mirrors that of Theorem 15.2 in \cite{lattimore2020BanditAlgorithms}, and the bounds are identical.

\begin{theorem}
    \label{thm: lower bound minimax regret}
    Let $T\in\N, T\geq A-1$ and consider the classes of bandits $\mathcal{C}_1$ and $\mathcal{C}_2$. Then,
    \[\regr^\star(\mathcal{C}_j) = \inf_{\pi\in\Pi}\sup_{{\nu}\in\mathcal{C}_j}\regr(\pi;{\nu})\geq \frac{\sqrt{T(A-1)}}{16\sqrt{e}}.\] 
\end{theorem}
\begin{flushright}
[\hyperref[proof: lower bound minimax regret]{Proof}]
\end{flushright}

\clearpage
\section{Simulations}
\label{sec: simulation}
This section provides more details about the simulation study presented in the main paper.

\subsection{Setup}\label{subsec: sim setup}
In this subsection, I suppress the dependence on $a$ of each quantity and illustrate the data-generating process for a generic action. Let $X_j \iid \mathsf{N}(0,1), j=1,\ldots, d,$ and $u_j\iid \mathsf{N}(0,\sigma_j^2),j\in\{C,R\}.$ Define 
\[C = \indic\left(\sum_{\ell=1}^dX_\ell \beta_\ell+u_C>\tau(q)\right), \qquad R = \theta + \sum_{\ell=1}^d X_\ell\beta_\ell + u_R,\qquad \bbeta:=(\beta_1,\ldots,\beta_d)^\top\in\R_{++},\]
where $\tau(q): \PP[\sum_{\ell=1}^dX_\ell \beta_\ell+u_C>\tau(q)] = q.$ Note that
\[R\sim\mathsf{N}(\theta,\sigma^2_{\bbeta}+\sigma_R^2), \qquad C\sim \mathsf{Be}(p),\qquad C\independent R \mid \bX,\]
where $\sigma^2_{\bbeta}:=\|\bbeta\|_2^2$. Define $W:=\sum_{\ell=1}^dX_\ell\sim\mathsf{N}(0,\sigma^2_{\bbeta})$ and note that
\begin{align*}
    \Cov(R,C) &=\E[RC]-\E[R]\E[C]\\
    &=\theta\E[C]  +\E[WC]+ \E[u_R C]-\theta\E[C] \\
    &=\E[WC] = \PP[C=1]\E[W\mid C=1]=q\cdot \E[W\mid C=1].
\end{align*}
Define $V := W + u_C$, the quantity above can be rewritten as 
\[\E[W\mid C=1]=\E[W\mid V>\tau(q)].\]
Note that
\[
\begin{bmatrix}
    W\\V
\end{bmatrix} \sim \mathsf{N}_2 \left(
\begin{bmatrix}
    0 \\ 0
\end{bmatrix},
\begin{bmatrix}
    \sigma^2_{\bbeta} & \sigma^2_{\bbeta} \\
    \sigma^2_{\bbeta} & \sigma^2_{\bbeta} + \sigma^2_C
\end{bmatrix}
\right),\qquad \rho = \frac{\sigma^2_{\bbeta}}{\sigma_{\bbeta}\sqrt{\sigma^2_{\bbeta} + \sigma^2_C}} = \frac{\sigma_{\bbeta}}{\sqrt{\sigma^2_{\bbeta}+\sigma^2_{C}}},\]
thus, using the formulas for bivariate normal random variables,
\[W \mid  V \sim \mathsf{N}\left(\rho^2 V, \rho^2\sigma^2_{C} \right).\]
Hence,
\[\E[W\mid V>\tau(q)] = \E[\E[W\mid V]\mid V>\tau(q)]=\rho^2\E[V\mid V>\tau(q)].\]
Using formulas for the truncated expectation of a normal distribution, one gets
\[\E[V\mid V>\tau(q)] = \theta_V + \sigma_V\frac{\phi\left(\tilde{\tau}(q)\right)}{1-\Phi(\tilde{\tau}(q))}=\sqrt{\sigma^2_{\bbeta}+\sigma^2_{C}}\frac{\phi\left(\tilde{\tau}(q)\right)}{q}, \qquad \tilde{\tau}(q):=\frac{\tau(q)}{\sqrt{\sigma^2_{\bbeta}+\sigma^2_{C}}},\]
where the second equality also uses the fact that
\[1-\Phi(\tilde{\tau}(q)) = 1-\PP[Z\leq \tau(q)/\sigma_V] = \PP[V \geq \tau(q)] = q.\]
Therefore
\begin{align*}
    \Cov(R,C) &=q\cdot \E[W\mid C=1] \\
    &=q\cdot\rho^2\cdot \E[V\mid V>\tau(q)] \\
    &=\rho^2 \sqrt{\sigma^2_{\bbeta}+\sigma^2_{C}}\phi\left(\tilde{\tau}(q)\right),
\end{align*}
and so
\begin{align}
    \mathrm{Corr}(R,C) = \rho^2 \sqrt{\frac{\sigma^2_{\bbeta}+\sigma^2_{C}}{(\sigma^2_{\bbeta}+\sigma^2_{R})q(1-q)}}\phi\left(\tilde{\tau}(q)\right)= \rho \frac{\sigma_{\bbeta}\phi\left(\tilde{\tau}(q)\right)}{\sqrt{(\sigma^2_{\bbeta}+\sigma^2_{R})q(1-q)}}.
    \label{eq: corr beta}
\end{align}
The above equation makes it clear that once the unconditional probability of missingness $q$ and the variance of the noise terms $u_R$ and $u_C$ have been specified, it is possible to search for $\bbeta\in\R_{++}$ such that $\mathrm{Corr}(R,C)$ matches a desired value.

\subsection{Simulation Design}\label{subsec: sim design}
Throughout, I set $A=2, T=5000,$ and $d=1$. For each action and each simulation scenario, I parametrize the data-generating process with $\sigma_{a,R}^2=1, \sigma_{a,C}^2=2, \theta_1=0.5,$ and $\theta_2=1$. The values of $q_a$ and $\bbeta$ are scenario-specific and are made explicit in Table \ref{tab: simulation scenarios}.

\begin{table}[!ht]
\caption{Parametrization of various simulation scenarios.}
\label{tab: simulation scenarios}
\centering
\begin{tabular}{cccccc}
\toprule\toprule
      & Missingness & $\bbeta$ & $(\theta_1,\theta_2)$ & $(\tilde{\theta}_1,\tilde{\theta}_2)$ & $(q_1,q_2)$ \\
\midrule
\textit{1.}     & \xmark & $\mathbf{0}$ & $(0.5,1)$ & $(0.5,1)$ & $(1,1)$ \\
\textit{2.}     & $C\independent R$ & $\mathbf{0}$ & $(0.5,1)$ & $(0.5,1)$ & $(0.25,0.9)$ \\
\textit{3.}     & $C\independent R\mid\bX$ & s.t. $\mathrm{Corr}(C,R)=0.2$ & $(0.5,1)$  & $(1.16, 1.08)$ & $(0.25,0.9)$ \\
\bottomrule
\end{tabular}%
\end{table}
More in detail:
\begin{enumerate}
    \item Scenario 1 is no missing data, thus the data-generating process is that of a standard multi-armed bandit, which is akin to specifying $\bbeta=\mathbf{0}$ and $q_1=q_2=1$;
    \item Scenario 2 is reward-independent missingness, thus $\bbeta=\mathbf{0}$ and $q_1,q_2\in(0,1)$;
    \item Scenario 3 is reward-dependent missingness, thus $\bbeta$ is selected so that $\mathrm{Corr}(C_a,R_a)=0.2$ in Equation \eqref{eq: corr beta}. In this case, the probability limit of the sample average of observed rewards, $\widehat{R}_a(T)$, is biased and different from $\theta_a$.
\end{enumerate}

Finally, the oracle versions of the $\UCB$ and $\DR$-$\UCB$ algorithms are computed using knowledge of the underlying data-generating process. 

More in detail, under scenarios 1 and 2, $\pi_\star^\UCB$ uses the following bonus term
\[\check{b}_{a,t}(\delta) = \mathfrak{q}_{1-\delta} \frac{\sigma_{a,R}}{\sqrt{N_a(t) + \lambda}},\]
whereas $\pi_\star^\UCB$ under scenario 3 uses the following bonus term
\[\dot{b}_{a,t}(\delta) = \mathfrak{q}_{1-\delta} \sqrt{\frac{\sigma_{a,R} + \|\bbeta\|_2^2}{P_a(t)}},\]
where $\mathfrak{q}_{1-\delta}$ is the $(1-\delta)$th quantile of a standard normal distribution.

Finally, nuisance estimation is conducted using an auxiliary sample, $\hat{\theta}_a$ are estimated via least squares, and $\hat{q}_a$ using a probit model.

\clearpage
\section{Proofs}
\label{sec: proofs}

\subsection{Proof of Lemma \ref{lemma: freedmans inequality}}
\begin{proof}\label{proof: freedmans inequality}
   Fix $\delta\in(0,1)$. First of all, note that by assumption we get
    \[\mathbb{E}\left[e^{\kappa D_k- \kappa^2 \nu_k^2 / 2} \mid \mathcal{F}^{k-1}\right] \leq
    1\quad \text{a.s.}.\]
    Let $\kappa\in\R\setminus\{0\}$, then by iteratively applying the law of iterated expectations    
    \begin{align*}
        \E\left[e^{\sum_{k=1}^n(\kappa D_k- \kappa^2 \nu_k^2 / 2)}\right] &= \E\left[e^{\sum_{k=1}^{n-1}(\kappa D_k- \kappa^2 \nu_k^2 / 2)}\E\left[e^{\kappa D_n- \kappa^2 \nu_n^2 / 2}\mid \mathcal{F}_{n-1}\right]\right]\leq \cdots
        \leq 1.
    \end{align*}
    Using Markov's inequality
    \[\PP\left[e^{\sum_{k=1}^n(\kappa D_k- \kappa^2 \nu_k^2 / 2)}\geq 2\delta^{-1}\right]\leq \frac{\E\left[e^{\sum_{k=1}^n(\kappa D_k- \kappa^2 \nu_k^2 / 2)}\right]}{2\delta^{-1}}\leq \delta/2.\]
    Finally
    \[\sum_{k=1}^n(\kappa D_k- \kappa^2 \nu_k^2 / 2)\geq \log (2/\delta)\quad \text{w.p. $\delta/2$}\quad \implies\quad \sum_{k=1}^nD_k\leq  \frac{\kappa}{2} \sum_{k=1}^n\nu_k^2 + \frac{1}{\kappa}\log (2/\delta)\quad \text{w.p. $1-\delta/2$}.\]
    The same logic can be applied to $-D_k,$ by first noting that it is still a martingale difference sequence and then absorbing the minus sign into the $\kappa$. The two statements together give us 
    \[\left|\sum_{k=1}^n D_k\right| \leq \frac{\kappa}{2}\sum_{k=1}^n\nu_k^2 + \frac{1}{\kappa}\log(2/\delta).\]
    Minimizing the upper bound over $\kappa\in\R\setminus\{0\},$ we get $\kappa^\star = \sqrt{2\log(2/\delta)(\sum_{k=1}^n\nu_k^2)^{-1}}$ and plugging it in the bound yields
    \[\left|\sum_{k=1}^n D_k\right| \leq \sqrt{2\log(2/\delta)\sum_{k=1}^n\nu_k^2},\]
    which holds with probability at least $1-\delta$. The last statement of the lemma follows immediately.
\end{proof}

\subsection{Proof of Lemma \ref{lemma: bernoulli and sg product}}
\begin{proof}\label{proof: bernoulli and sg product}
    By Proposition 2.5.2, part $(iv)$ in
    \cite{vershynin2018HighDimensionalProbabilityIntroduction} we know that a
    random variable $Z\sim\sG(\sigma)$ if and only if
    \[\E\left[\exp\left(\frac{Z^2}{C\sigma^2}\right)\right] \leq 2, \]
    for some $C>0$. Therefore, in our case
    \[\E\left[\exp\left(\frac{Z^2}{C\sigma^2}\right)\right]=\E\left[\exp\left(\frac{X^2\cdot
    Y^2}{C\sigma^2}\right)\right]\overset{(i)}{\leq}
    \E\left[\exp\left(\frac{X^2}{C\sigma^2}\right)\right]\overset{(ii)}{\leq}
    2,\] where $(i)$ follows from the fact that $Y^2=Y\leq 1$ almost surely and
    $X\sim\sG(\sigma)$. Thus, we conclude that $Z\sim\sG(\sigma)$ which was
    to be shown.
\end{proof}

\subsection{Proof of Lemma \ref{lemma: deviation of conditional expectation}}

\begin{proof}\label{proof: deviation of conditional expectation}
    First, I prove an auxiliary fact that will turn out to be useful: conditional sub-Gaussianity implies unconditional sub-Gaussianity. As $Y\mid X \sim \sG(\sigma)$, by definition of sub-Gaussianity one gets
    \[\forall\,\lambda\in\R, \qquad \E\left[e^{\lambda(Y-\E[Y\mid X])}\mid X\right]\leq e^{\frac{\lambda^2\sigma^2}{2}}\quad \text{a.s.}\]
    It follows that $Y\sim \sG(\sigma)$. To see this, fix $\lambda\in\R$ and note that
    \begin{align*}
        \E[e^{\lambda(Y-\E[Y])}] &= \E_Y[e^{\lambda(Y-\E_X[\E_Y[Y\mid X]])}] \tag{LIE}\\
        &\leq \E_{X,Y}[e^{\lambda(Y-\E_Y[Y\mid X])}] \tag{Jensen's inequality} \\
        &= \E_{X}[\E_{Y|X}[e^{\lambda(Y-\E_Y[Y\mid X])}]\mid X]  \\
        &\leq \E_X\left[e^{\frac{\lambda^2\sigma^2}{2}}\right] \tag{$Y\mid X \sim \sG(\sigma)$} \\
        &\leq e^{\frac{\lambda^2\sigma^2}{2}},
    \end{align*}
    which was claimed.\footnote{Note that in principle one could make $\sigma^2$ a random variable that is $\sigma(X)$-measurable and integrable, and the result would also go through.}

    Now, consider $W:=\E[Y\mid X]-\E[Y]$. Note that $\E[W]=0$ and fix $\lambda\in\R$. Then,
    \begin{align*}
        \E[e^{\lambda W}] &= \E[e^{\lambda(\E[Y\mid X]-\E[Y])}] \\
        &=\E[e^{\lambda\E[Y\mid X]}]e^{-\lambda\E[Y]} \\
        &\leq \E[\E[e^{\lambda Y}\mid X]]e^{-\lambda\E[Y]} \tag{Jensen's inequality} \\
        &=\E[e^{\lambda Y}]e^{-\lambda\E[Y]} \\
        &=\E[e^{\lambda (Y-\E[Y])}] \\
        & \leq e^{\frac{\lambda^2\sigma^2}{2}} \tag{$Y\sim \sG(\sigma)$},
    \end{align*}
    where the last line follows because of the fact proven above. Thus, it follows that $W \sim \sG(\sigma)$, which was to be shown.
\end{proof}

\subsection{Proof of Lemma \ref{lemma: kl properties}}
\begin{proof}\label{proof: kl properties}
The first result follows from the fact that
\begin{align*}
\dkl(P,Q)
&=\int_\calX\int_\calY p(x,y)\,\ln\frac{p(x,y)}{q(x,y)}\de y \de x\\
&=\int_\calX\int_\calY p(x)\,p(y| x)\,
   \ln\frac{p(x)\,p(y| x)}{q(x)\,q(y| x)} \de y \de x\\
&=
\int_\calX p(x)\ln\frac{p(x)}{q(x)}\de x
+
\int_\calX\int_\calY p(x)\,p(y| x)\,\ln\frac{p(y| x)}{q(y| x)}\de y \de x\\
&=
\dkl(P_X,Q_X)
\;+\;
\E_{X\sim P_X}\Bigl[\dkl(P_{Y| X},Q_{Y| X})\Bigr].
\end{align*}
Furthermore, if $X\independent Y$, then $P_{Y|X}=P_Y$ and $Q_{Y|X}=Q_Y$, thus
\[\E_{X\sim P_X}\Bigl[\dkl(P_{Y| X},Q_{Y| X})\Bigr] = \dkl(P_Y,Q_Y),\]
which proves the second fact.
\end{proof}

\subsection{Proof of Lemma \ref{lemma: concentration of estimated rewards}}
\begin{proof}\label{proof: concentration of estimated rewards}
    Fix $a\in\calA, t\in[T],\lambda>0,$ and $\delta\in(0,1)$. First of all, for some $\lambda>0$ note that 
    \begin{align*}
        \widehat{R}^{\UCB}_a(t) - \theta_a &= \frac{1}{N_a(t)+\lambda}\sum_{\ell=1}^{t-1}\indic[A_\ell=a]C_{a,\ell}R_{a,\ell}  -\theta_a \\
        &=\frac{1}{N_a(t)+\lambda}\sum_{\ell=1}^{t-1}\indic[A_\ell=a]C_{a,\ell}(R_{a,\ell} -\theta_a) - \frac{\lambda\theta_a}{N_a(t)+\lambda}. 
    \end{align*}
    Define the auxiliary event
    \[\calE_{a,t}(\delta):= \left\{\left| \widehat{R}^{\UCB}_a(t) - \theta_a\right|\geq  b^{\UCB}_{a,t}(\delta/AT)\right\}.\]
    Via the triangular inequality, we have
    \begin{align*}
        \calE_{a,t}(\delta)&\subseteq 
        \left\{\left|\frac{1}{N_a(t)+\lambda}\sum_{\ell=1}^{t-1}\indic[A_\ell=a]C_{a,\ell}(R_{a,\ell} -\theta_a)\right| + \frac{\lambda\overline{K}}{N_a(t)+\lambda}\geq  b^{\UCB}_{a,t}(\delta/AT)\right\} \\
        &=\left\{\left|\frac{1}{N_a(t)+\lambda}\sum_{\ell=1}^{t-1}\indic[A_\ell=a]C_{a,\ell}(R_{a,\ell} -\theta_a)\right| + \frac{\lambda\overline{K}}{N_a(t)+\lambda}\geq  \frac{\osigma}{\underline{q}_\lambda}\sqrt{\frac{2\ln(2/\delta)}{P_a(t)+\lambda}}+\frac{\lambda\overline{K}}{N_a(t)+\lambda}\right\} \\
        &=\left\{\left| \frac{1}{N_a(t)+\lambda}\sum_{\ell=1}^{t-1}\indic[A_\ell=a]C_{a,\ell}(R_{a,\ell} -\theta_a)\right|\geq  \frac{\osigma}{\underline{q}_\lambda}\sqrt{\frac{2\ln(2/\delta)}{P_a(t)+\lambda}}\right\}\\
        &\subseteq\left\{\left| \frac{1}{P_a(t)+\lambda}\sum_{\ell=1}^{t-1}\indic[A_\ell=a]C_{a,\ell}(R_{a,\ell} -\theta_a)\right|\geq  \osigma\sqrt{\frac{2\ln(2/\delta)}{P_a(t)+\lambda}}\right\} \\
         &=\left\{\left| \sum_{\ell=1}^{t-1}\indic[A_\ell=a]C_{a,\ell}(R_{a,\ell} -\theta_a)\right|\geq  \osigma\sqrt{2\ln(2/\delta)(P_a(t)+\lambda)}\right\}.
    \end{align*}
    Then, we get
    \begin{align}
       \PP\left[\calE_{a,t}(\delta)\right] &\leq \PP\left[\left| \sum_{\ell=1}^{t-1}\indic[A_\ell=a]C_{a,\ell}(R_{a,\ell} -\theta_a)\right|\geq  \osigma\sqrt{2\ln(2/\delta)(P_a(t)+\lambda)}\right] 
       \leq \delta,\label{eq: aux freedman bound}
    \end{align}
    where the last inequality follows from Freedman's inequality (Lemma
    \ref{lemma: freedmans inequality}). To justify the use of such inequality, I
    show that $\{W_{a,\ell}\}_{\ell=1}^\tau$, where $W_{a,\ell}:=
    \indic[A_\ell=a]C_{a,\ell}(R_{a,\ell}-\theta_a)$ is a martingale difference
    sequence for an appropriately defined filtration. Let such filtration be
    defined as $\{\calF_\ell\}_{\ell=0}^t, \calF_\ell = \sigma(\{(R_{A_j,j},
    C_{A_j,j}), j=1,\ldots,\ell\})$.  It follows by construction that
    $\{W_{a,\ell}\}_{\ell=1}^{t-1}$ is $\{\calF_\ell\}_{\ell=0}^{t-1}$-adapted
    and integrable. Note that $\indic[A_\ell=a]$ is deterministic once we
    condition on $\calF_{\ell-1}$ as the $\UCB$ algorithm picks $A_\ell$ with
    all the information available at the beginning of round $\ell$ (see Protocol
    \ref{prot: stochastic mab with dependent missingness} and Algorithm
    \ref{alg: UCB algorithm}). Therefore, conditional on $\calF_{\ell-1}$ either
    $W_{a,\ell}=0$ a.s. or $W_{a,\ell} = C_{a,\ell}(R_{a,\ell}-\theta_a)$,
    hence, whenever $\{A_\ell\neq a\}$ realizes it follows immediately that
    $\E[W_{a,\ell}\mid \calF_{\ell-1}]=0$, whereas if $\{A_\ell= a\}$ occurs,
    then 
    \[\E[W_{a,\ell}\mid \calF_{\ell-1}] = \E[C_{a,\ell}(R_{a,\ell}-\theta_a)\mid \calF_{\ell-1}]= \E[C_{a,\ell}(R_{a,\ell}-\theta_a)]= \E[C_{a,\ell}]\E[(R_{a,\ell}-\theta_a)]=0,\]
    where the second equality follows because $(C_{a,\ell},R_{a,\ell})\iid \nu_a$ and the third equality from Assumption \ref{assump: independent missingness}. Finally, we have that 
    \[\forall\,\kappa\in\R\setminus\{0\},\: \mathbb{E}\left[e^{\kappa W_{a,\ell}} \mid \mathcal{F}^{\ell-1}\right] \leq
    e^{\kappa^2 \nu_\ell^2 / 2}\quad \text {a.s. }\]
    with
    \[\nu_\ell^2 = \sigma_a^2\indic[A_\ell=a]
    \implies \sum_{\ell=1}^{t-1}\nu_\ell^2 \leq \osigma^2P_{a}(t)<\osigma^2(P_{a}(t)+\lambda) \:\:\text{ a.s.},\]
    where the first inequality follows from the fact that: (i) $\indic[A_\ell=a]$ is $\calF_{\ell-1}$-measurable; (ii) $W_{a,\ell}\mid \calF_{\ell-1}\sim\sG(\sigma_a)$ by Lemma \ref{lemma: bernoulli and sg product} and the fact that $(R_{a,\ell},C_{a,\ell})\iid\nu_a$; and (iii) for a random variable $Z$ and sigma-algebra $\calF$, if $Z \mid \calF \sim \sG(\sigma),$ then $bZ\mid \calF\sim\sG(|b|\sigma)$ for a random variable $b$ that is $\calF$-measurable. The result in \eqref{eq: aux freedman bound} follows from Lemma \ref{lemma: freedmans inequality}.
\end{proof}

\subsection{Proof of Lemma \ref{lemma: probability of failure event}}
\begin{proof} \label{proof: probability of failure event}
 Fix some $\delta\in(0,1)$ to be chosen later and consider the failure event 
    $$\calF^{\UCB}(\delta)=\left\{\exists\,a\in\calA,t\in[T] : \left|\widehat{R}^{\UCB}_a(t) - \theta_a \right|\geq b^\UCB_{a,t}(\delta)\right\}.$$
    Then
    \begin{align*}
        \PP[\calF^{\UCB}(\delta)] &= \PP\left[\bigcup_{a\in\calA}\bigcup_{t\in[T]} \left\{\left|\widehat{R}^{\UCB}_a(t) - \theta_a \right|\geq b^\UCB_{a,t}(\delta)\right\}\right] \\
        &\leq \sum_{a\in\calA}\sum_{t\in[T]}\PP\left[\left|\widehat{R}^{\UCB}_a(t) - \theta_a \right|\geq b^\UCB_{a,t}(\delta)\right] \tag{union bound}\\
        &\leq \sum_{a\in\calA}\sum_{t\in[T]} \frac{\delta}{AT}=\delta \tag{Lemma \ref{lemma: concentration of estimated rewards}},
    \end{align*}
    which was to be shown.
\end{proof}

\subsection{Proof of Lemma \ref{lemma: probability of missingness event}}
\begin{proof}\label{proof: probability of missingness event}
Fix $a\in\calA, t, \kappa_a\in[T]$, and $\delta\in(0,1)$. Then
\begin{align}\label{eq: multiplicative chernoff bound}
\PP\left[\sum_{\ell=1}^{\kappa_a}\indic[C_{a,\ell}=1]\leq(1-\delta)q_a\kappa_a\right]\leq \exp\left\{-\frac{\delta^2q_a\kappa_a}{2}\right\},
\end{align}
where the inequality follows from the multiplicative version of a multiplicative Chernoff bound. 

Pick $\delta\in(0,1)$, then the probability of the missingness event is
\begin{align}
    \PP[\calF^{\MIS}(\delta)]&= \PP\left[\bigcup_{a\in\calA}\bigcup_{t\in[T]}\left\{N_a(t)\leq(1-\delta)q_aP_a(t),P_a(t) \geq \underline{T}_a\right\}\right] \nonumber\\
    &\leq \sum_{a\in\calA}\sum_{t\in[T]}\PP\left[\left\{N_a(t)\leq (1-\delta)q_aP_a(t),P_a(t) \geq \underline{T}_a\right\}\right] \tag{union bound} \nonumber\\
    &\leq \sum_{a\in\calA}\sum_{t\in[T]}\PP\left[\exists\,\kappa_a\in[T]:\left\{\sum_{\ell=1}^{\kappa_a}\indic[C_{a,\ell}=1]\leq(1-\delta)q_a\kappa_a,\kappa_a \geq \underline{T}_a\right\}\right]  \nonumber\tag{Assumption \ref{assump: independent missingness}}\\
    &\leq  \sum_{a\in\calA}\sum_{t=1}^T\sum_{\kappa_a=\underline{T}_a}^T\PP\left[\sum_{\ell=1}^{\kappa_a}\indic[C_{a,\ell}=1]\leq(1-\delta)q_a\kappa_a\right] \tag{union bound} \nonumber\\
    &\leq \sum_{a\in\calA}\sum_{t=1}^T\sum_{\kappa_a=\underline{T}_a}^T \exp\left\{-\frac{\delta^2q_a\kappa_a}{2}\right\} \tag{by \eqref{eq: multiplicative chernoff bound}}  \\
    &= T\cdot \sum_{a\in\calA}\sum_{\kappa_a=\underline{T}_a}^T \exp\left\{-\frac{\delta^2q_a\kappa_a}{2}\right\}.\label{eq: aux 1}
\end{align}
Now, note that via an integral comparison, one gets
\begin{align*}
    \sum_{\kappa_a=\underline{T}_a}^T \exp\left\{-\frac{\delta^2q_a\kappa_a}{2}\right\} &\leq \int\limits_{\underline{T}_a - 1}^{T}\exp\left\{-\frac{\delta^2q_au}{2}\right\}\de u \\
    &=\left[-\frac{2}{\delta^2q_a}\exp\left\{-\frac{\delta^2q_au}{2}\right\}\right]_{\underline{T}_a-1}^T \\
    &= \left[-\frac{2}{\delta^2q_a}\exp\left\{-\frac{\delta^2q_au}{2}\right\}\right]_{\underline{T}_a-1}^{\kappa_a} \tag{summands are negative and $\kappa_a\leq T$} \\
    &\leq \frac{2}{\delta^2q_a}\exp\left\{-\frac{\delta^2q_a(\underline{T}_a-1)}{2}\right\} \\
    &= \frac{2}{\delta^2q_a}T^{-12\delta^2} \tag{$\underline{T}_a = 1+\frac{24\ln(T)}{q_a}$}.
\end{align*}
Therefore, using the above result in \eqref{eq: aux 1}
\begin{align*}
    \PP[\calF^{\MIS}(\delta)]&\leq T\cdot \sum_{a\in\calA}\frac{2}{\delta^2q_a}T^{-12\delta^2} = \frac{2}{\delta^2}T^{1-12\delta^2} \acen,
\end{align*}
where $\acen=\sum_{a\in\calA}q_a^{-1}$, which was to be shown.
\end{proof}

\subsection{Proof of Lemma \ref{lemma: UCB property}}
\begin{proof}\label{proof: UCB property}
    Fix $\delta\in(0,1)$. Note that 
    \[\overline{\calF^\UCB(\delta)} = \left\{\forall\,a\in\calA,t\in[T],\left|\widehat{R}^{\UCB}_a(t) - \theta_a\right|< b^\UCB_{a,t}(\delta) \right\},\]
    thus for all $a\in\calA$ and $t\in[T]$ one has
    \begin{align*}
        \widetilde{R}^{\UCB}_a(t,\delta) &= \widehat{R}^{\UCB}_a(t) + b^\UCB_{a,t}(\delta) \tag{definition} \\
        &= \theta_a +\widehat{R}^{\UCB}_a(t) - \theta_a + b^\UCB_{a,t}(\delta) \\
        &> \theta_a ,
    \end{align*}
    where the last inequality follows because under the event $\overline{\calF^\UCB(\delta)}$ it occurs that
    \[\forall\, a\in\calA,t\in[T],\qquad -b^\UCB_{a,t}(\delta)< \widehat{R}^{\UCB}_a(t) - \theta_a < b^\UCB_{a,t}(\delta),\]
    where the first inequality implies
    \[\widehat{R}^{\UCB}_a(t) - \theta_a + b^\UCB_{a,t}(\delta) > 0.\]
\end{proof}

\subsection{Proof of Lemma \ref{lemma: bound on inverse of number of observed}}
\begin{proof}\label{proof: bound on inverse of number of observed}
    Fix $\delta\in(0,1),$ if $\overline{\calF^{\MIS}(\delta)}$ holds then for all $a\in\calA, t\in[T]$ we have that $ N_{a}(t)>(1-\delta)q_aP_{a}(t)$. Hence for each $a\in\calA$
    \begin{align}
        (1-\delta)\sum_{t=1}^T\frac{1}{{N_a(t)+\lambda}} 
    &\leq \sum_{t=1}^T\frac{1}{{q_aP_a(t) +\frac{\lambda}{(1-\delta)}}} \tag{$\overline{\calF^{\MIS}(\delta)}$ holds}\nonumber \\ 
    &= \sum_{a\in\calA}\sum_{\ell=1}^{t-1}\indic[A_\ell=a]\frac{1}{{q_a\cdot\ell +\frac{\lambda}{(1-\delta)}}} \tag{$T=\sum_{a\in\calA} P_a(T)$}\nonumber \\
    &\overset{(i)}{\leq} \sum_{a\in\calA}\int\limits_0^{t-1}\indic[A_\ell=a] \frac{1}{{q_a\cdot u +\frac{\lambda}{(1-\delta)}}}\de u\nonumber \\
   & \overset{(ii)}{\leq} \sum_{a\in\calA}\frac{1}{{q_a}}\ln\left(q_aP_a(T) + \frac{\lambda}{(1-\delta)}\right),
    \label{eq: aux 2}
    \end{align}
    where $(i)$ follows from an integral comparison and $(ii)$ by standard computations. 

    Now the goal is to construct a generic upper bound on $\sum_{a\in\calA}\frac{1}{{q_a}}\ln(q_aP_a(T) + \lambda)$ that holds for any process that causes missing data $\{q_a\}_{a\in\calA}$. To do so, one can solve the following constrained optimization problem
    \begin{align*}
        \max_{\bx \in\R^A} \sum_{a\in\calA}\frac{1}{{q_a}}\ln(q_ax_a + \alpha), \qquad \text{s.to} \quad \bx \succeq \mathbf{0}_A, \: \mathbf{1}_A^\top\bx = T,
    \end{align*}
    where $\alpha:=\lambda/(1-\delta).$ The problem above is a standard convex problem \citep[see the water-filling problem in][Example 5.2, p.245]{boyd2004ConvexOptimization} and has the following KKT conditions
    \begin{align*}
    \begin{gathered}
\bx^{\star} \succeq \mathbf{0}_A, \quad \mathbf{1}_A^\top \bx^{\star}=T, \quad \bmu^{\star} \succeq \mathbf{0}_A, \quad \theta_a^{\star} x_a^{\star}=0, \quad a=1, \ldots, A \\
-1 /\left(\alpha+q_ax_a^{\star}\right)-\theta_a^{\star}+\nu^{\star}=0, \quad a=1, \ldots, A,
\end{gathered}
    \end{align*}
    where $\bmu^\star$ are the Lagrange multipliers for the inequality constraints and $\nu^\star$ is the multiplier of the equality constraint. The unique solution of this problem is given by
    \[x_a^\star = \frac{1}{q_a}\left(\frac{1}{\nu^\star}- \frac{\lambda}{1-\delta}\right), \quad a=1, \ldots, A,\]
    where $\nu^\star$ is such that $\mathbf{1}_A^\top\bx = T$ and so $\nu^\star=\left(\frac{\lambda}{1-\delta}+\frac{T}{\acen}\right)^{-1}$ giving us
    \[x_a^\star = \frac{T}{q_a\acen}, \quad a=1, \ldots, A,\]
    which yields a maximum value of \eqref{eq: aux 2} equal to
    \[\acen\ln\left(\frac{T}{\acen} +\frac{\lambda}{1-\delta}\right).\]

    A similar logic can be used to show that 
    \[\sum_{t=1}^T\frac{1}{\sqrt{N_a(t)+\lambda}}\leq \frac{\acen}{\sqrt{1-\delta}}\sqrt{\frac{T}{\acen} +\frac{\lambda}{1-\delta}}.\]
\end{proof}

\subsection{Proof of Theorem \ref{thm: regret UCB in MAB}}
\begin{proof}\label{proof: regret UCB in MAB}

    Define the good event $\calG(\delta_1,\delta_2) = \overline{\calF^\UCB(\delta_1)} \cap \overline{\calF^{\MIS}(\delta_2)}$ for some $\delta_1,\delta_2\in(0,1)$ to be chosen later and consider the regret of the $\UCB$ algorithm. Recall that under the $\UCB$ policy, the action at round $t$ is chosen as $A_t:=\argmax_{a\in\mathcal{A}}\widetilde{R}^{\UCB}_a(t,\delta)$. Note that
    \begin{align*}
        \regr(\piucb) = \sum_{t=1}^T\left(\overline{\theta} - \theta_{A_t}\right) = \sum_{t=1}^T\Delta_t,
    \end{align*}
    with $\Delta_t:=\overline{\theta} - \theta_{A_t}$ is the sub-optimality gap at time $t.$ Furthermore, let $\Delta_t(\calE) := \E_\nu[R_{a^\star, t} -R_{A_t, t}\mid \calE]$ denote the sub-optimality gap conditional on the event $\calE$.
    
    First of all, via Lemma \ref{lemma: probability of failure event}, Lemma \ref{lemma: probability of missingness event}, and a union bound we get
    \begin{align}
    \label{eq: aux union bound}
        \PP[{\calG(\delta_1,\delta_2)}] \geq 1 -  \delta_1 - \frac{2}{\delta_2^2}\acen T^{1-12\delta_2^2}.
    \end{align}
    Assume $\calG(\delta_1,\delta_2)$ holds, then
\begin{align*}
    \Delta_t(\calG(\delta_1,\delta_2)) &= \overline{\theta} - \theta_{A_t} \\
    &\leq \widetilde{R}_{a^\star}(t,\delta_1) - \theta_{A_t} \tag{Lemma \ref{lemma: UCB property}}\\
    &\leq \widetilde{R}_{A_t}(t,\delta_1) - \theta_{A_t} \tag{by $\UCB$, $A_t:=\argmax_{a\in\mathcal{A}}\widetilde{R}_a(t,\delta_1)$}\\
    &= \widehat{R}_{A_t}(t) - \theta_{A_t} + b^\UCB_{A_t,t}(\delta_1) \tag{definition of $\widetilde{R}_{a}(t,\delta_1)$} \\
    &\leq 2b^\UCB_{A_t,t}(\delta_1) \tag{$\overline{\calF^\UCB(\delta_1)}$ holds}.
\end{align*}
Define $\widetilde{\lambda}:=\frac{\lambda}{1-\delta_2}$. Then, using the result above and the fact that $\lambda>0$
\begin{align}
\sum_{t=1}^T\Delta_t(\calG(\delta_1,\delta_2)) &\leq \frac{2\osigma}{\underline{q}_\lambda}\sqrt{2\ln(2AT/\delta_1)}\sum_{t=1}^T\frac{1}{\sqrt{P_{A_t}(t)}}+2\lambda\overline{K}\sum_{t=1}^T\frac{1}{N_{A_t}(t)+\lambda} \nonumber\\
    &\leq \frac{2\osigma}{\underline{q}_\lambda}\sqrt{2\ln(2AT/\delta_1)}\sum_{t=1}^T\frac{1}{\sqrt{P_{A_t}(t)}}+ 2\acen \overline{K}\widetilde{\lambda}\ln\left(\frac{T}{\acen} +\widetilde{\lambda}\right) \tag{Lemma \ref{lemma: bound on inverse of number of observed}} \nonumber\\
&=\frac{2\osigma}{\underline{q}_\lambda}\sqrt{2\ln(2AT/\delta_1)}\sum_{a\in\calA}\sum_{\ell=1}^{t-1}\indic[A_\ell=a]\frac{1}{\sqrt{\ell}} +  2\acen\overline{K} \widetilde{\lambda}\ln\left(\frac{T}{\acen} +\widetilde{\lambda}\right)\nonumber\\
&\leq\frac{4\osigma}{\underline{q}_\lambda}\sqrt{2\ln(2AT/\delta_1)}\sum_{a\in\calA}\sqrt{P_a(T)} +  2\acen \overline{K}\widetilde{\lambda}\ln\left(\frac{T}{\acen} +\widetilde{\lambda}\right) \tag{$\sum_{j=1}^k\frac{1}{\sqrt{j}}\leq 2\sqrt{k}$}\nonumber\\
&\leq \frac{4\osigma}{\underline{q}_\lambda}\sqrt{2\ln(2AT/\delta_1)}\sqrt{\sum_{a\in\calA} 1 \cdot \sum_{a\in\calA}P_a(T)} +  2\acen\overline{K} \widetilde{\lambda}\ln\left(\frac{T}{\acen} +\widetilde{\lambda}\right) \tag{Cauchy-Schwarz}\nonumber\\
&\leq \frac{4\osigma}{\underline{q}_\lambda}\sqrt{2AT\ln(2AT/\delta_1)} +  2\acen \overline{K}\widetilde{\lambda}\ln\left(\frac{T}{\acen} +\widetilde{\lambda}\right).\nonumber
\end{align}

Therefore, by \eqref{eq: aux union bound} 
\[\regr(\piucb) \leq \frac{4\osigma}{\underline{q}_\lambda}\sqrt{2AT\ln(2AT/\delta_1)} +  2\acen \overline{K}\widetilde{\lambda}\ln\left(\frac{T}{\acen} +\widetilde{\lambda}\right)\]
with probability at least $1-\delta_1-\frac{2}{\delta_2^2}\acen T^{1-12\delta_2^2}.$ Note that for $\delta_2 =\sqrt{\frac{1+\kappa}{12}} $, one gets that $\frac{2}{\delta_2^2}\acen T^{1-12\delta_2^2} = O(T^{-\kappa})$. Therefore, for $\kappa>0$ and $\lambda=o(T^{1/2}),$ it follows tht
    \[\regr(\piucb) \leq \frac{4\osigma}{\underline{q}_\lambda}\sqrt{2AT\ln(2AT/\delta_1)} + o(\sqrt{T}),\]
which was to be shown. 

Finally, recall that each arm has been pulled once during the 'burn-in" period; thus, an additional factor of $\overline{K}$ needs to be taken into account.
\end{proof}

\subsection{Proof of Lemma \ref{lemma: concentration of estimated rewards under dependent missingness}}
\begin{proof}\label{proof: concentration of estimated rewards under dependent missingness}

Recall that
\begin{align*}
     \widehat{R}^{\ODR}_a(t) = \frac{1}{P_a(t)} \sum_{\ell=1}^{t-1}\indic[A_\ell=a] \left(\frac{R_{a,\ell}C_{a,\ell}}{q_a(\bX_{a,\ell})} - \frac{\theta_a(\bX_{a,\ell})}{q_a(\bX_{a,\ell})}\left(C_{a,\ell}-q_a(\bX_{a,\ell})\right)\right).
\end{align*}
Define the auxiliary event
\[\calE_{a,t}(\delta):= \left\{\left| \widehat{R}^\ODR_a(t) - \theta_a\right|\geq  b^{\ODR}_{a,t}(\widetilde{\delta})\right\},\]
where
\[b^{\ODR}_{a,t}(\widetilde{\delta}) = \Kodr\sqrt{\frac{2\ln(2/\delta)}{P_a(t)}}.\]
Note that 
\[\widehat{R}^\ODR_a(t) = \frac{1}{P_a(t)}\sum_{\ell=1}^{t-1} (W_{a,\ell} + V_{a,\ell}),\]
with
\[W_{a,\ell}:=\indic[A_\ell=a]\frac{C_{a,\ell}}{q_a(\bX_{a,\ell})}(R_{a,\ell}-\theta_a(\bX_{a,\ell})), 
\qquad V_{a,\ell} = \indic[A_\ell=a](\theta_a(\bX_{a,\ell}) - \theta_a).\]
Using the fact that $\{|X|+|Y|\geq a + b\} \implies \{|X|\geq a\}\cup\{|Y|\geq b\}$ for any two random variables $X,Y$ and $a,b\in\R$,
one gets
\begin{align*}
    \calE_{a,t}(\delta) &=\left\{ \left|\sum_{\ell=1}^{t-1} (W_{a,\ell} + V_{a,\ell})\right| \geq \frac{\osigma}{\underline{q}}\sqrt{2\ln(2/\delta)P_a(t)} + \osigma \sqrt{2\ln(2/\delta)P_a(t)}\right\}\\
    &=\left\{ \left|\sum_{\ell=1}^{t-1} W_{a,\ell}\right| \geq \frac{\osigma}{\underline{q}}\sqrt{2\ln(2/\delta)P_a(t)}\right\}
    \bigcup \left\{ \left|\sum_{\ell=1}^{t-1} V_{a,\ell}\right| \geq \osigma\sqrt{2\ln(2/\delta)P_a(t)}\right\}.
\end{align*}
I now show that both $\{W_{a,\ell}\}_{\ell=1}^{t-1}$ and $\{V_{a,\ell}\}_{\ell=1}^{t-1}$ are martingale difference sequences with respect to appropriately defined filtrations.
Define the collections of sigma-algebras $\{\calF_\ell\}_{\ell=1}^{t}, \calF_\ell = \sigma(\{\bX_{a,\ell},a\in\calA\})\otimes\sigma(\{\bZ_j, j=1,\ldots,\ell-1\})$ and $\{\calG_\ell\}_{\ell=1}^{t}, \calG_\ell = \sigma(\{\bZ_j, j=1,\ldots,\ell-1\})$, where $\bZ_j =\{(R_{a,j}, C_{a,j}, \bX_{a,j}), a\in\calA\}$.  It follows by construction that $\{V_{a,\ell}\}_{\ell=1}^{t-1}$ is $\{\calG_\ell\}_{\ell=0}^{t-1}$-adapted and integrable and $\{W_{a,\ell}\}_{\ell=1}^{t-1}$ is $\{\calF_\ell\}_{\ell=0}^{t-1}$-adapted and integrable. Note that $\indic[A_\ell=a]$ is deterministic conditionally on either $\calF_{\ell}$ or $\calG_{\ell}$. Therefore, conditional on $\calF_{\ell}$ or on $\calG_{\ell}$, either $\{A_\ell \neq a\}$ realizes, and so $W_{a,\ell}=V_{a,\ell}=0$ almost surely and $\E[V_{a,\ell}\mid \calG_{\ell}]=\E[W_{a,\ell}\mid \calF_{\ell}]=0$ follow immediately. If instead $\{A_\ell=a\}$ realizes, then $\E[V_{a,\ell}\mid \calG_{\ell}]=\E[V_{a,\ell}]=0$ by the law of iterated expectations, whereas for $\E[W_{a,\ell\mid \calF_\ell}]$ two cases need to be considered.

First, suppose Assumption \ref{assump: oracle DR}\textcolor{ptonorange}{(a)} holds, i.e., $q_a(\bx)$ is the conditional probability of missingness. Then
\begin{align*}
    \E[W_{a,\ell}\mid \calF_{\ell}]&= \E[\indic[A_\ell=a](C_{a,\ell}R_{a,\ell}- \theta_a(\bX_{a,\ell})(C_{a,\ell}-q_a(\bX_{a,\ell})-\theta_aq_a(\bX_{a,\ell}))/q_a(\bX_{a,\ell})\mid \calF_{\ell}] \\
    &= \E[(C_{a,\ell}R_{a,\ell}- \theta_a(\bX_{a,\ell})(C_{a,\ell}-q_a(\bX_{a,\ell}))/q_a(\bX_{a,\ell})\mid\calF_{\ell}]-\theta_a \tag{$\{A_\ell=a\}$ occurs}\\
    &= \E[(C_{a,\ell}R_{a,\ell}- \theta_a(\bX_{a,\ell})(C_{a,\ell}-q_a(\bX_{a,\ell}))/q_a(\bX_{a,\ell})]-\theta_a \tag{$(R_{a,\ell},C_{a,\ell},\bX_{a,\ell})\iid\nu_a$} \\
    &=\E\left[\E\left[\frac{C_{a,\ell}R_{a,\ell}}{q_a(\bX_{a,\ell})}\mid \bX_{a,\ell}\right]\right] -\theta_a + \E\left[\E\left[\frac{C_{a,\ell}-q_a(\bX_{a,\ell})}{q_a(\bX_{a,\ell})}\mid \bX_{a,\ell}\right]\right] \tag{iterated expectations}\\
    &=\E\left[\E\left[\frac{C_{a,\ell}R_{a,\ell}}{q_a(\bX_{a,\ell})}\mid \bX_{a,\ell}\right]\right] -\theta_a \tag{Assumption \ref{assump: oracle DR}\textcolor{ptonorange}{(a)}}\\
    &= \E[q_a(\bX_{a,\ell})^{-1}\E[C_{a,\ell}\mid \bX_{a,\ell}]\E[R_{a,\ell}\mid \bX_{a,\ell}]] -\theta_a \tag{Assumption \ref{assump: conditional ignorability}}\\
    &=\E[\E[R_{a,\ell}\mid \bX_{a,\ell}]] -\theta_a= 0. \tag{Assumption \ref{assump: oracle DR}\textcolor{ptonorange}{(a)}}
\end{align*}

If instead Assumption \ref{assump: oracle DR}\textcolor{ptonorange}{(b)} holds, i.e., $\theta_a(\bx)$ is the conditional mean reward, then
\begin{align*}
    \E[W_{a,\ell}\mid\calF_{\ell}]
    &= \E[\indic[A_\ell=a](C_{a,\ell}R_{a,\ell}- \theta_a(\bX_{a,\ell})(C_{a,\ell}-q_a(\bX_{a,\ell})-\theta_aq_a(\bX_{a,\ell}))/q_a(\bX_{a,\ell})\mid \calF_{\ell}] \\
    &= \E[(C_{a,\ell}R_{a,\ell}- \theta_a(\bX_{a,\ell})(C_{a,\ell}-q_a(\bX_{a,\ell}))/q_a(\bX_{a,\ell})\mid\calF_{\ell}]-\theta_a \tag{$\{A_\ell=a\}$ occurs}\\
    &= \E[(C_{a,\ell}R_{a,\ell}- \theta_a(\bX_{a,\ell})(C_{a,\ell}-q_a(\bX_{a,\ell}))/q_a(\bX_{a,\ell})]-\theta_a \tag{$(R_{a,\ell},C_{a,\ell},\bX_{a,\ell})\iid\nu_a$} \\
    &= \E[\theta_a(\bX_{a,\ell})] -\theta_a + \E\left[q_a(\bX_{a,\ell})^{-1}\E\left[C_{a,\ell}(R_{a,\ell}-\theta_a(\bX_{a,\ell}))\mid \bX_{a,\ell}\right]\right] \tag{iterated expectations}\\
    &=\E\left[q_a(\bX_{a,\ell})^{-1}\E\left[C_{a,\ell}(R_{a,\ell}-\theta_a(\bX_{a,\ell}))\mid \bX_{a,\ell}\right]\right] \tag{Assumption \ref{assump: oracle DR}\textcolor{ptonorange}{(b)}}\\
    &= \E[q_a(\bX_{a,\ell})^{-1}\E[C_{a,\ell}\mid \bX_{a,\ell}](\E[R_{a,\ell}\mid \bX_{a,\ell}]-\theta_a(\bX_{a,\ell}))] -\theta_a \tag{Assumption \ref{assump: conditional ignorability}}\\
    &=0. \tag{Assumption \ref{assump: oracle DR}\textcolor{ptonorange}{(b)}}
\end{align*}

Moreover, it follows that 
    \[\forall\,\kappa\in\R,\: \mathbb{E}\left[e^{\kappa W_{a,\ell}} \mid \mathcal{F}_{\ell}\right] \leq
    e^{\kappa^2 \nu_\ell^2 / 2}\quad \text {a.s.,}\quad \text{with}\:\:
    \nu_\ell^2 = \frac{\sigma_a^2}{\underline{q}^2}\indic[A_\ell=a]
    \implies \sum_{\ell=1}^{t-1}\nu_\ell^2 \leq \frac{\osigma^2}{\underline{q}^2}P_{a}(t)\:\: \text{ a.s.},\]
where the first inequality follows from the fact that: (i) $\indic[A_\ell=a]$ is $\{\bZ_j, j=1,\ldots,\ell\}$-measurable; (ii) $\hat{q}_a(\bX_{a,\ell})-q_a(\bX_{a,\ell})$ is $\sigma(\{\bX_{a,\ell},a\in\calA\})\otimes\sigma(\mathcal{D})$-measurable; (iii) by Assumption \ref{assump: conditional ignorability} and Lemma \ref{lemma: bernoulli and sg product}, $C_{a,\ell}\epsilon_{a,\ell}\mid \calF_\ell\sim\sG(\sigma_a)$; and (iv) for a random variable $Z$ and sigma-algebra $\calF$, if $Z \mid \calF \sim \sG(\sigma),$ then $bZ\mid \calF\sim\sG(|b|\sigma)$ for a random variable $b$ that is $ \calF$-measurable.

With a similar argument and using Lemma \ref{lemma: deviation of conditional expectation}, it also follows that
    \[\forall\,\kappa\in\R,\: \mathbb{E}\left[e^{\kappa V_{a,\ell}} \mid \mathcal{G}_{\ell}\right] \leq
    e^{\kappa^2 \xi_\ell^2 / 2}\quad \text {a.s.,}\quad \text{with}\:\:
    \xi_\ell^2 = \sigma_a^2\indic[A_\ell=a]
    \implies \sum_{\ell=1}^{t-1}\xi_\ell^2 \leq \osigma^2 P_{a}(t)\:\: \text{ a.s.}.\]

Put differently, all the requirements of Freedman's inequality (Lemma \ref{lemma: freedmans inequality}) are satisfied, thus  for any fixed $a\in\calA$ and round $t\in[T]$
\[\PP\left[\calE_{a,t}(\delta)\right]\leq \PP\left[ \left|\sum_{\ell=1}^{t-1} W_{a,\ell}\right| \geq \frac{\osigma}{\underline{q}}\sqrt{2\ln(2/\delta)P_a(t)}\right] + \PP\left[\left|\sum_{\ell=1}^{t-1} V_{a,\ell}\right| \geq \osigma\sqrt{2\ln(2/\delta)P_a(t)}\right] \leq 2\delta,\]
where the inequality follows from a union bound. Reparametrizing $\delta$ yields the desired result.
\end{proof}

\subsection{Proof of Theorem \ref{thm: regret oracle UCB in MAB with dependent missingness}}
\begin{proof}\label{proof: regret oracle UCB in MAB with dependent missingness}
    The proof of this result is a particular case of the proof of Theorem \ref{thm: regret UCB in MAB with dependent missingness}, because
    \[{b}^\DR_{a,t}(\delta) = {b}_{a,t}^\ODR(\delta) + {b}_{a,t}^{[1]}(\delta)+ {b}_{a,t}^{[2]}(\delta).\]
    Thus, one can ignore the construction of the high-probability bounds on ${b}_{a,t}^{[1]}(\delta)$ and $ {b}_{a,t}^{[2]}(\delta)$ and obtain a proof for this theorem.
\end{proof}

\subsection{Proof of Lemma \ref{lemma: concentration of estimated DR rewards under dependent missingness}}
\begin{proof}\label{proof: concentration of estimated DR rewards under dependent missingness}

Before getting started with the actual proof, it is useful to think of the data-generating process as
\begin{align*}
    R_{a} = \theta_a^\star(\bX_{a}) + \epsilon_{a}, \qquad 
    C_{a} = q_a^\star(\bX_{a}) + \xi_{a}, \quad \forall\,a\in\calA.
\end{align*}
Note that the two equations above are \textit{definitional} and do not impose conditions other than Assumption \ref{assump: conditional ignorability} on the data-generating process. Furthermore, by Assumption \ref{assump: conditional ignorability} it follows that $\epsilon_{a}\mid\bX_{a}\sim\sG(\sigma_a)$ and $\xi_{a}\mid\bX_{a}\in[-1,1]$ a.s., and so
\[\epsilon_{a}\mid\bX_{a} \sim \sG(\sigma_a), \qquad \xi_{a}\mid\bX_{a}\sim \sG(1), \quad \forall\,a\in\calA.\]

Fix $a\in\calA,t\in[T], \delta\in(0,1),$ and let $\widetilde{\delta}:=AT\delta$. Consider the bonus term
\[{b}^\DR_{a,t}(\delta) = {b}_{a,t}^\ODR(\delta) + {b}_{a,t}^{[1]}(\delta)+ {b}_{a,t}^{[2]}(\delta) + b_{a,t}^{[3]}(\delta),\]
where
\begin{alignat*}{3}
   {b}_{a,t}^\ODR(\delta) &= \Kodr\sqrt{\frac{2\ln(2AT/\delta)}{P_a(t)}}, \qquad
   &{b}_{a,t}^{[1]}(\delta) &:= \frac{\osigma}{\underline{q}^2}\sqrt{\frac{2\ln(2AT/\delta)}{P_a(t)} }\Err_{t}(\hat{q}_a), \\
   {b}_{a,t}^{[2]}(\delta) &:=  \frac{1}{\underline{q}}\sqrt{\frac{2\ln(2AT/\delta)}{P_a(t)} }\Err_{t}(\hat{\theta}_a),\qquad & {b}_{a,t}^{[3]}(\delta)&=\Err_{t}(\hat{\theta}_a)\Err_{t}(\hat{q}_a), 
\end{alignat*}
with
\begin{align*}
    \Err_{t}(\hat{\theta}_a):= \sqrt{\frac{1}{P_a(t)}\sum_{\ell=1}^{t-1}\indic[A_\ell=a](\hat{\theta}_a(\bX_{a,\ell}) -\theta_a(\bX_{a,\ell}))^2} , \:
    \Err_{t}(\hat{q}_a):= \sqrt{\frac{1}{P_a(t)}\sum_{\ell=1}^{t-1}\indic[A_\ell=a](\hat{q}_a(\bX_{a,\ell}) -q_a(\bX_{a,\ell}))^2} .
\end{align*}

Consider the following decomposition:
\begin{align*}
   \widehat{R}^{\DR}_a(t)-\theta_a&= \frac{1}{P_a(t)}\sum_{\ell=1}^{t-1}\indic[A_\ell=a] \left(\frac{C_{a,\ell}(R_{a,\ell} - \hat{\theta}_a(\bX_{a,\ell}))}{\hat{q}_a(\bX_{a,\ell})} + \hat{\theta}_a(\bX_{a,\ell})\right) -\theta_a \\
   &=  \frac{1}{P_a(t)}\sum_{\ell=1}^{t-1}\indic[A_\ell=a] \left(\frac{C_{a,\ell}(R_{a,\ell} - {\theta}_a(\bX_{a,\ell}))}{{q}_a(\bX_{a,\ell})} + {\theta}_a(\bX_{a,\ell})\right) -\theta_a \tag{$:=R_{a,\mathsf{IF}}(t)$}\\
   &\quad +  \frac{1}{P_a(t)}\sum_{\ell=1}^{t-1}\indic[A_\ell=a] \frac{C_{a,\ell}(R_{a,\ell} - {\theta}_a(\bX_{a,\ell}))}{\hat{q}_a(\bX_{a,\ell})q_a(\bX_{a,\ell})}\left({q}_a(\bX_{a,\ell})-\hat{q}_a(\bX_{a,\ell})\right) \tag{$:=R_{a,1}(t)$} \\
   &\quad +  \frac{1}{P_a(t)}\sum_{\ell=1}^{t-1}\indic[A_\ell=a] \left(\hat{\theta}_a(\bX_{a,\ell})-{\theta}_a(\bX_{a,\ell})\right)\left(\frac{q_{a}(\bX_{a,\ell})- C_{a,\ell}}{\hat{q}_a(\bX_{a,\ell})}\right). \tag{$:=R_{a,2}(t)$} \\
   &\quad +  \frac{1}{P_a(t)}\sum_{\ell=1}^{t-1}\indic[A_\ell=a] \left(\hat{\theta}_a(\bX_{a,\ell})-{\theta}_a(\bX_{a,\ell})\right)\left(\frac{{q}_a(\bX_{a,\ell})-\hat{q}_a(\bX_{a,\ell})}{\hat{q}_a(\bX_{a,\ell})}\right). \tag{$:=R_{a,3}(t)$}
\end{align*}

First, Assumptions \ref{assump: non-oracle DR} and \ref{assump: nuisance estimation - truncation} ensure all quantities are well defined as $q_a$ and $\hat{q}_a$ are bounded away from zero. Then, note that 
\begin{align}
    &\PP\left[\left|\widehat{R}^{\DR}_a(t)-\theta_a\right| \geq b^\DR_{a,t}(\widetilde{\delta}) \right]\nonumber \\
    &\qquad \overset{(1)}{\leq} \PP\left[|R_{a,\mathsf{IF}}(t)| + |R_{a,1}(t)|+|R_{a,2}(t)|+|R_{a,3}(t)|\geq b_{a,t}^\ODR(\widetilde{\delta}) + b_{a,t}^{[1]}(\widetilde{\delta})+ b_{a,t}^{[2]}(\widetilde{\delta}) +b_{a,t}^{[3]}(\widetilde{\delta}) \right]\nonumber \\
    &\qquad \overset{(2)}{\leq} \PP\left[|R_{a,\mathsf{IF}}(t)| \geq b_{a,t}^\ODR(\widetilde{\delta}) \right] + 
    \PP\left[|R_{a,1}(t)|\geq b_{a,t}^{[1]}(\widetilde{\delta}) \right] + 
    \PP\left[|R_{a,2}(t)|\geq b_{a,t}^{[2]}(\widetilde{\delta}) \right]+ 
    \PP\left[|R_{a,3}(t)|\geq b_{a,t}^{[3]}(\widetilde{\delta}) \right],\label{eq: aux DR R_DR}
\end{align}
where (1) follows from the triangle inequality, (2) follows from the fact that $\{|X|+|Y|\geq a + b\} \implies \{|X|\geq a\}\cup\{|Y|\geq b\}$ for any two random variables $X,Y$ and $a,b\in\R$. Therefore, the goal is to provide bounds for the three terms in \eqref{eq: aux DR R_DR}. Towards this goal, 
recall that we defined $\epsilon_{a,\ell}:=R_{a,\ell} - \theta_a(\bX_{a,\ell})$ and 
$\xi_{a,\ell}:=C_{a,\ell} - q_a(\bX_{a,\ell})$, $\xi_a\in[-1,1]$ a.s., and, by the law of iterated expectations, $\E[\epsilon_{a,\ell}]=0=\E[\xi_{a,\ell}]$ for all $\ell\in[T]$ and $a\in\calA$.

\uline{\textbf{Bound on} $\PP[|R_{a,\mathsf{IF}}(t)|\geq b_{a,t}^{[\ODR]}(\widetilde{\delta})]$.}

Note that Assumption \ref{assump: non-oracle DR} is just Assumption \ref{assump: oracle DR} where $q_a(\cdot)$ and $\theta_a(\cdot)$ are the probability limits of the nuisance estimators. Thus, it follows immediately from Lemma \ref{lemma: concentration of estimated rewards under dependent missingness} that 
\begin{align}
    \PP\left[|R_{a,\mathsf{IF}}(t)| \geq b_{a,t}^\ODR(\widetilde{\delta}) \right]\leq \delta.
    \label{eq: aux DR R_IF}
\end{align}

\uline{\textbf{Bound on} $\PP[|R_{a,1}(t)|\geq b_{a,t}^{[1]}(\widetilde{\delta})]$.}
    
    Note that
    \begin{align*}
        \left\{|R_{a,1}(t)|\geq b_{a,t}^{[1]}(\widetilde{\delta})\right\} &=\left\{\left|\frac{1}{P_a(t)}\sum_{\ell=1}^{t-1}\indic[A_\ell=a] \frac{C_{a,\ell}(R_{a,\ell} - {\theta}_a(\bX_{a,\ell}))}{\hat{q}_a(\bX_{a,\ell})q_a(\bX_{a,\ell})}\left({q}_a(\bX_{a,\ell})-\hat{q}_a(\bX_{a,\ell})\right)\right| \geq \frac{\osigma}{\underline{q}^2}\sqrt{\frac{2\ln(2/\delta)}{P_a(t)} }\Err_{t}(\hat{q}_a)\right\} \\
        &\subseteq\left\{\frac{1}{\underline{q}^2}\left|\frac{1}{P_a(t)}\sum_{\ell=1}^{t-1}\indic[A_\ell=a] C_{a,\ell}\epsilon_{a,\ell}\left({q}_a(\bX_{a,\ell})-\hat{q}_a(\bX_{a,\ell})\right)\right|\geq \frac{\osigma}{\underline{q}^2}\sqrt{\frac{2\ln(2/\delta)}{P_a(t)} }\Err_{t}(\hat{q}_a)\right\} \ \\
        &=\left\{\left|\sum_{\ell=1}^{t-1}\indic[A_\ell=a] C_{a,\ell}\epsilon_{a,\ell}\left({q}_a(\bX_{a,\ell})-\hat{q}_a(\bX_{a,\ell})\right)\right|\geq \osigma\sqrt{ 2\ln(2/\delta)P_a(t)}\Err_{t}(\hat{q}_a)\right\}, 
    \end{align*}
    where the inclusion follows from Assumptions \ref{assump: non-oracle DR} and \ref{assump: nuisance estimation - truncation}.

    Denote with $\mathcal{D}$ the data (i.e., collection of random variables) used to estimate $\{\hat{q}_a, \hat{\theta}_a\}_{a\in\calA}$ such that Assumption \ref{assump: nuisance estimation - independence} is satisfied and define $W_{a,\ell}:=\indic[A_\ell=a]C_{a,\ell}\epsilon_{a,\ell}({q}_a(\bX_{a,\ell})-\hat{q}_a(\bX_{a,\ell}))$ and the collection of sigma-algebras $\{\calF_\ell\}_{\ell=1}^{t}, \calF_\ell = \sigma(\{\bX_{a,\ell}, a\in\calA\})\otimes\sigma(\{\bZ_j, j=1,\ldots,\ell-1\})\otimes\sigma(\mathcal{D})$, where $\bZ_j =\{(R_{a,j}, C_{a,j}, \bX_{a,j}), a\in\calA\}$.  It follows by construction that $\{W_{a,\ell}\}_{\ell=1}^{t-1}$ is $\{\calF_\ell\}_{\ell=1}^{t-1}$-adapted and integrable. Again, if $\{A_\ell\neq a\}$ occurs, then $\E[W_{a,\ell}\mid \calF_\ell]=0$ a.s. follows immediately. If $\{A_\ell= a\}$ realizes, note that 
    \begin{align*}
        \E[W_{a,\ell}\mid \calF_{\ell-1}] &= \E[C_{a,\ell}\epsilon_{a,\ell}({q}_a(\bX_{a,\ell})-\hat{q}_a(\bX_{a,\ell}))\mid \calF_\ell]\tag{$\{A_\ell= a\}$ occurs}\\
        &=\E[C_{a,\ell}\epsilon_{a,\ell}\mid \calF_\ell]({q}_a(\bX_{a,\ell})-\hat{q}_a(\bX_{a,\ell})) \tag{Assumption \ref{assump: nuisance estimation - l2 error}} \\
        &=\E[C_{a,\ell}\epsilon_{a,\ell}\mid \bX_{a,\ell}]({q}_a(\bX_{a,\ell})-\hat{q}_a(\bX_{a,\ell})) 
        \tag{$(C_{a,\ell},\epsilon_{a,\ell})\mid\bX_{a,\ell}$ are i.i.d.}\\
        &=\E[C_{a,\ell}\mid \bX_{a,\ell}]\E[\epsilon_{a,\ell}\mid \bX_{a,\ell}]({q}_a(\bX_{a,\ell})-\hat{q}_a(\bX_{a,\ell}))\tag{Assumption \ref{assump: conditional ignorability}}\\
        &=0\tag{definition of $\epsilon_{a,\ell}$}.
    \end{align*}
    Moreover, for $\kappa\in\R$ we have
    \[\forall\,\kappa\in\R,\: \mathbb{E}\left[e^{\kappa W_{a,\ell}} \mid \calF_\ell\right] \leq
    e^{\kappa^2 \nu_\ell^2 / 2}\quad \text {a.s.},\]
    with
    \[\nu_\ell^2 = \osigma^2\indic[A_\ell=a](\hat{q}_a(\bX_{a,\ell})-{q}_a(\bX_{a,\ell}))^2
    \implies \sum_{\ell=1}^{t-1}\nu_\ell^2 \leq \osigma^2P_a(t)\Err_t(\hat{q}_a)^2\:\: \text{ a.s.},\]
    where the first inequality follows from the fact that: (i)  $\indic[A_\ell=a]$ is $\sigma(\{\bZ_j, j=1,\ldots,\ell-1\})$-measurable; (ii) $\hat{q}_a(\bX_{a,\ell})-q_a(\bX_{a,\ell})$ is $\sigma(\{\bX_{a,\ell}, a\in\calA\})\otimes\sigma(\mathcal{D})$-measurable; (iii) by Assumption \ref{assump: conditional ignorability} and Lemma \ref{lemma: bernoulli and sg product}, $C_{a,\ell}\epsilon_{a,\ell}\mid \calF_\ell\sim\sG(\sigma_a)$; and (iv) for a random variable $Z$ and sigma-algebra $\calF$, if $Z \mid \calF \sim \sG(\sigma),$ then $bZ\mid \calF\sim\sG(|b|\sigma)$ for a random variable $b$ that is $ \calF$-measurable. 
    
     All the conditions of Freedman's inequality (Lemma \ref{lemma: freedmans inequality}) are satisfied by $\{W_{a,\ell}\}_{\ell=1}^t$ and get 
    \begin{align*}
        \PP\left[\left|\sum_{\ell=1}^{t-1}\indic[A_\ell=a] C_{a,\ell}\epsilon_{a,\ell}\left({q}_a(\bX_{a,\ell})-\hat{q}_a(\bX_{a,\ell})\right)\right|\geq \osigma\sqrt{ 2\ln(2/\delta)P_a(t)}\Err_{t}(\hat{q}_a)\right] \leq \delta,
    \end{align*}
    which implies
    \begin{align}
        \PP[|R_{a,1}(t)|\geq b_{a,t}^{[1]}(\widetilde{\delta})]\leq \delta.
    \label{eq: aux DR R_1}
    \end{align}

    \uline{\textbf{Bound on} $\PP[|R_{a,2}(t)|\geq b_{a,t}^{[2]}(\delta)]$.}
    
    Regarding $R_{a,2}(t)$, note that
    \begin{align*}
        |R_{a,2}(t)| &= \left|\frac{1}{P_a(t)}\sum_{\ell=1}^{t-1}\indic[A_\ell=a] \left(\hat{\theta}_a(\bX_{a,\ell})-{\theta}_a(\bX_{a,\ell})\right)\left(\frac{q_a(\bX_{a,\ell})-C_{a,\ell}}{\hat{q}_a(\bX_{a,\ell})}\right)\right| \\
        &\leq \frac{1}{\underline{q}}\left|\frac{1}{P_a(t)}\sum_{\ell=1}^{t-1}\indic[A_\ell=a] \left(\hat{\theta}_a(\bX_{a,\ell})-{\theta}_a(\bX_{a,\ell})\right)\xi_{a,\ell}\right|.
    \end{align*}
    Using a symmetric argument to the one used above to bound $R_{a,1}(t)$ in probability, one gets
    \begin{align}
        \PP[|R_{a,2}(t)|\geq b_{a,t}^{[2]}(\widetilde{\delta})] = \PP\left[|R_{a,2}(t)|\geq \frac{1}{\underline{q}}\sqrt{2\ln(2/\delta)}\frac{\Err_{t}(\hat{\theta}_a)}{P_a(t)}\right]\leq \delta.
        \label{eq: aux DR R_2}
    \end{align}

    \uline{\textbf{Bound on} $\PP[|R_{a,3}(t)|\geq b_{a,t}^{[3]}(\delta)]$.}
    
    By the Cauchy-Schwarz inequality and Assumption \ref{assump: nuisance estimation - truncation}
    \begin{align*}
        |R_{a,3}(t)|&\leq\sqrt{\frac{1}{P_a(t)}\sum_{\ell=1}^{t-1}\indic[A_\ell=a](\hat{\theta}_a(\bX_{a,\ell})-\theta_a(\bX_{a,\ell}))^2}\cdot \sqrt{\frac{1}{P_a(t)}\sum_{\ell=1}^{t-1}\indic[A_\ell=a](\hat{q}_a(\bX_{a,\ell})-q_a(\bX_{a,\ell}))^2} \\
        &= \frac{1}{\underline{q}}\Err_{t}(\hat{\theta}_a)\Err_{t}(\hat{q}_a),
    \end{align*}
    almost surely, which yields
    \begin{align}
        \PP[|R_{a,3}(t)|\geq b_{a,t}^{[3]}(\widetilde{\delta})]\leq \delta.
    \label{eq: aux DR R_3}
    \end{align}

\uline{\textbf{Final bound}.}

Finally, by using \eqref{eq: aux DR R_IF}, \eqref{eq: aux DR R_1}, \eqref{eq: aux DR R_2}, and \eqref{eq: aux DR R_3} in \eqref{eq: aux DR R_DR}, it follows that
\[\PP\left[\left|\widehat{R}^{\DR}_a(t)-\theta_a\right| \geq b^\DR_{a,t}(\widetilde{\delta}) \right]\leq 3\delta.\]
Reparametrizing $\delta$ yields the desired result.
  \end{proof}

\subsection{Proof of Lemma \ref{lemma: probability of failure event with dependent missingness}}
\begin{proof}\label{proof: probability of failure event with dependent missingness}

 Fix some ${\delta}\in(0,1)$ and consider the failure event 
    $$\calF^{\DR}({\delta})=\left\{\exists\,a\in\calA,t\in[T] : \left| \widehat{R}^{\DR}_a(t) - \theta_a \right|\geq b_{a,t}^\DR(\delta)\right\}.$$
    Then, 
    \begin{align*}
        \PP[\calF^{\DR}({\delta})] &= \PP\left[\bigcup_{a\in\calA}\bigcup_{t\in[T]} \left\{\left| \widehat{R}^{\DR}_a(t) - \theta_a \right|\geq b_{a,t}^\DR(\delta)\right\}\right] \\
        &\leq \sum_{a\in\calA}\sum_{t\in[T]}\PP\left[\left| \widehat{R}^{\DR}_a(t) - \theta_a \right|\geq b_{a,t}^\DR(\delta)\right] \tag{union bound}\\
        &\leq \frac{\delta}{AT} \cdot A T = \delta. \tag{Lemma \ref{lemma: concentration of estimated DR rewards under dependent missingness}}
    \end{align*}
\end{proof}

\subsection{Proof of Lemma \ref{lemma: UCB property with dependent missingness}}
\begin{proof}\label{proof: UCB property with dependent missingness}
    Fix $\delta\in(0,1)$. Note that 
    \[\overline{\calF^{\DR}(\delta)} = \left\{\forall\,a\in\calA,t\in[T],\left| \widehat{R}^{\DR}_a(t) - \theta_a\right|\leq b_{a,t}^\DR(\delta)\right\},\]
    thus for all $a\in\calA$ and $t\in[T]$ we have
    \begin{align*}
        \widetilde{R}_{a,t}^{\DR}(t,\delta) &=  \widehat{R}^{\DR}_a(t) + b_{a,t}^\DR(\delta) \\
        &= \theta_a + \underbrace{ \widehat{R}^{\DR}_a(t) - \theta_a + b_{a,t}^\DR(\delta)}_{\geq 0} \\
        &\geq  \theta_a,
    \end{align*}
    where the last line follows because under $\overline{\calF^{\DR}(\delta)}$ we have
    \[\forall\, a\in\calA,t\in[T],\qquad -b_{a,t}^\DR(\delta)\leq  \widehat{R}^{\DR}_a(t) - \theta_a \leq b_{a,t}^\DR(\delta),\]
    where the first inequality gives us
    \[ \widehat{R}^{\DR}_a(t) - \theta_a + b_{a,t}^\DR(\delta) \geq 0.\]
\end{proof}

\subsection{Proof of Theorem \ref{thm: regret UCB in MAB with dependent missingness}}
\begin{proof}\label{proof: regret UCB in MAB with dependent missingness}

    Define the good event $\calG(\delta) = \overline{\calF^{\DR}(\delta)}$ for some $\delta\in(0,1)$ to be chosen later. Consider the regret of the $\DR$-$\UCB$ algorithm that uses $ \widehat{R}^{\DR}_a(t)$ as an estimator for mean rewards. Moreover, recall that under the $\UCB$ policy, the action at round $t$ is chosen as $A_t:=\argmax_{a\in\mathcal{A}}\widetilde{R}^{\DR}_a(t,\delta)$. Note that
    \begin{align*}
        \regr(\pidrucb) = \sum_{t=1}^T\left(\overline{\theta} - \theta_{A_t}\right) = \sum_{t=1}^T\Delta_t,
    \end{align*}
    with $\Delta_t:=\overline{\theta} - \theta_{A_t}$ is the sub-optimality gap at time $t.$ Furthermore, let $\Delta_t(\calE) := \E_\nu[R_{a^\star, t} -R_{A_t, t}\mid \calE]$ denote the sub-optimality gap conditional on the event $\calE$.

    Suppose $\calG(\delta)$ holds. By Lemma \ref{lemma: probability of failure event with dependent missingness}, this occurs with probability at least $1-\delta$. Then,
\begin{align*}
    \Delta_t(\calG(\delta)) &= \overline{\theta} - \theta_{A_t} \\
    &\leq \widetilde{R}^{\DR}_{a^\star}(t,\delta) - \theta_{A_t} \tag{Lemma \ref{lemma: UCB property with dependent missingness}}\\
    &\leq \widetilde{R}^{\DR}_{A_t}(t,\delta) - \theta_{A_t} \tag{by $\DR$-$\UCB$, $A_t:=\argmax_{a\in\mathcal{A}}\widetilde{R}^{\DR}_a(t,\delta)$}\\
    &= \widehat{R}^\DR_{A_t}(t) - \theta_{A_t} + b_{A_t,t}^\DR(\delta) \tag{definition of $\widetilde{R}^{\DR}_{A_t}(t,\delta)$} \\
    &\leq 2b_{A_t,t}^\DR(\delta) \tag{Lemma \ref{lemma: probability of failure event with dependent missingness}}.
\end{align*}

Therefore, decomposing $b_{A_t,t}^\DR(\delta)$ and summing over rounds
\begin{align*}
    2\sum_{t=1}^T{b}_{A_t,t}^\ODR(\delta) &= \frac{2\osigma}{\underline{q}}\sqrt{\ln(2AT/\delta)}\sum_{t=1}^T\frac{1}{\sqrt{P_{A_t}(t)}}, \\
    &=\frac{2\osigma}{\underline{q}}\sqrt{2\ln(2AT/\delta)}\sum_{a\in\calA}\sum_{\ell=1}^{P_{a}(T)}\frac{1}{\sqrt{\ell}}  \\
    &\leq\frac{4\osigma}{\underline{q}}\sqrt{2\ln(2AT/\delta)}\sum_{a\in\calA}\sqrt{P_a(T)}  \tag{$\sum_{j=1}^k\frac{1}{\sqrt{j}}\leq 2\sqrt{k}$}\\
    &\leq \frac{4\osigma}{\underline{q}}\sqrt{2\ln(2AT/\delta)}\sqrt{\sum_{a\in\calA} 1 \cdot \sum_{a\in\calA}P_a(T)}   \tag{Cauchy-Schwarz}\\
    &\leq \frac{4\osigma}{\underline{q}}\sqrt{2AT\ln(2AT/\delta)}. 
\end{align*}
Moreover, under Assumption \ref{assump: nuisance estimation - l2 error} with $\delta_{\mathfrak{c}}\in(0,1)$, it follows that $\mathfrak{c}_q(P_a(t))\lesssim P_a(t)^{-\alpha_q}$ for some $\alpha_q>0$. Then,
\begin{align*}
   2\sum_{t=1}^T{b}_{A_t,t}^{[1]}(\delta) &= \frac{2\osigma}{\underline{q}^2}\sqrt{ 2\ln(2AT/\delta)}\sum_{t=1}^T\sqrt{\frac{\Err_{P_{A_t}(t)}(\hat{q}_{A_t})}{P_{A_t}(t)}}\\
   &\leq \frac{2\osigma}{\underline{q}^2}\sqrt{ 2\ln(2AT/\delta)}\sum_{t=1}^T\frac{\mathfrak{c}_t^q}{\sqrt{P_{A_t}(t)}}\\
   &\lesssim \frac{2\osigma}{\underline{q}^2}\sqrt{ 2\ln(2AT/\delta)}\sum_{t=1}^T\frac{1}{P_{A_t}(t)^{1/2+\alpha_q}}\\
   &=\widetilde{o}\left(\sqrt{T}\right),
\end{align*}
with probability $1-\delta-\delta_\mathfrak{c}$ and where the last line follows from a comparison with $\sum_{t=1}^TP_{A_t}(t)^{-1/2}$. Similarly, using Assumption \ref{assump: nuisance estimation - l2 error} again, for some $\alpha_\theta>0$ and $\alpha>1/2$ we get
\begin{align*}
    2\sum_{t=1}^T{b}_{A_t,t}^{[2]}(\delta) &\leq 2\sum_{t=1}^T\Err_{t}(\hat{\theta}_a)\Err_{t}(\hat{\theta}_a) + \frac{2}{\underline{q}}\sqrt{2\ln(2AT/\delta) }\sum_{t=1}^T\frac{\Err_{t}(\hat{q}_a)}{\sqrt{P_a(t)}} \\
    &\lesssim T^{1-\alpha} + \frac{2}{\underline{q}}\sqrt{2\ln(2AT/\delta) }\sum_{t=1}^T\frac{1}{P_{A_t}(t)^{1/2+\alpha_\theta}} = \widetilde{o}\left(\sqrt{T}\right)
\end{align*}
with probability $1-\delta-\delta_\mathfrak{c}$. Thus, one can conclude that 
\begin{align*}
\regr(\pidrucb)=\sum_{t=1}^T\Delta_t(\calG(\delta)) \leq \frac{4\osigma}{\underline{q}}\sqrt{AT\ln(2AT/\delta)} + \widetilde{o}(\sqrt{T})
\end{align*}
with probability $1-\delta-\delta_\mathfrak{c}$. Finally, recall that each arm has been pulled once during the ``burn-in" period; thus, an additional factor of $A\overline{\theta}$ needs to be taken into account.
\end{proof}

\subsection{Proof of Theorem \ref{thm: lower bound minimax regret}}
\begin{proof}\label{proof: lower bound minimax regret}
Fix $T\in\N$ and choose a generic policy $\pi\in\Pi$ and two Gaussian-Bernoulli bandits $\nu$ and $\nu'$ in $\mathcal{C}_2^{\mathtt{gau}},$ defined as follows:
\begin{enumerate}
    \item $\forall\,a\in\calA$, $\nu=\{\nu_a\}_{a\in\calA}$ is such that $\nu_a^{[R]} = \normal(\theta_a, 1)$ with $\theta_a = \Delta\indic[a=1]$ and $\nu_a^{[C]}=\Be(q_a), q_a\in [\underline{q},1]$; 
    \item $\forall\,a\in\calA$, $\nu'=\{\nu'_a\}_{a\in\calA}$ is such that $\nu_a^{[X]}=\nu_a^{'[X]}, \nu_a^{'[R|X]} = \normal(\theta_a', 1)$ with $\theta_a' = \Delta\indic[a=1]+2\Delta\indic[a=i^\star]$, $\nu_a^{'[C|X]}=\Be(q_a)$ and where
    \[i^\star:=\argmin_{a\in\calA\setminus\{1\}}\E_\nu[P_a(T)].\]
\end{enumerate}
The rationale for choosing these two bandits is that they are sufficiently hard to distinguish from each other, but they induce different strategies. To summarize, the two mean vectors are
\[(\Delta, 0 ,\ldots, 0) \quad \text{and}\quad (\Delta,0,\ldots,0,2\Delta,0,\ldots,0).\]
This strategy exactly matches what an adversarial nature would play, making it the natural benchmark when focusing on minimax regret. Furthermore, by the definition of $i^\star$ and the fact that $P_a(T)\geq 0$ almost surely for each $a\in\calA$ 
\[\sum_{a\in\calA}\E_\nu[P_a(T)] = \E_\nu[P_1(T)] + \sum_{a\in\calA\setminus\{1\}}\E_\nu[P_a(T)] \geq \sum_{a\in\calA\setminus\{1\}}\E_\nu[P_a(T)] \geq (A-1)\E_\nu[P_{i^\star}(T)].\]
Then, because $\sum_{a\in\calA}\E_\nu[P_a(T)] =T$ it follows that 
\begin{align}\label{eq: aux pull istar}
    \E_\nu[P_{i^\star}(T)]\leq \frac{T}{A-1}.
\end{align}
By the classical decomposition of regret see Lemma 4.2 in \cite{lattimore2020BanditAlgorithms}, it follows that
\[\regr(\nu) = \sum_{a\in\calA}\Delta_a\E_\nu[P_a(T)] = \Delta \sum_{a\in\calA\setminus\{1\}}\E_\nu[P_a(T)]=\Delta(T-\E_\nu[P_1(T)])\]
and
\[\regr(\nu') = \sum_{a\in\calA}\Delta_a\E_{\nu'}[P_a(T)] = \Delta\E_{\nu'}[P_1(T)] + 2\Delta\sum_{a\in\calA\setminus\{1,i^\star\}}\E_{\nu'}[P_a(T)]\geq \Delta\E_{\nu'}[P_1(T)].\]
Then, define the event $\calA:=\{P_1(T)\leq T/2\}$ and note that
\begin{align}\label{eq: aux regret nu}
    \regr(\nu) = \Delta\E_\nu[T-P_1(T)]\geq \Delta \E_\nu[T-P_1(T)\mid \calA]\PP_\nu[\calA] \geq \frac{\Delta T}{2}\PP_\nu[P_1(T)\leq T/2].  
\end{align}
Similarly,
\begin{align}\label{eq: aux regret nu'}
    \regr(\nu') = \Delta\E_{\nu'}[P_1(T)] \geq \Delta\E_{\nu'}[P_1(T)\mid \overline{\calA}]\PP_{\nu'}[\overline{\calA}] \geq \frac{\Delta T}{2}\PP_{\nu'}[P_1(T)> T/2].
\end{align}
Thus, it follows that
\begin{align*}
    \regr(\nu) + \regr(\nu') &\geq \frac{\Delta T}{2}\left(\PP_\nu[P_1(T)\leq T/2]+\PP_{\nu'}[P_1(T)> T/2]\right) \tag{\eqref{eq: aux regret nu} and \eqref{eq: aux regret nu'}} \\
    &\geq \frac{\Delta T}{4}\exp\left\{-\dkl(\PP_\nu,\PP_{\nu'})\right\} \tag{Bretagnole-Huber inequality} \\
    &=\frac{\Delta T}{4} \exp\left\{-\sum_{a\in\calA}\E_\nu[P_a(T)]\dkl(\nu,\nu')\right\},\tag{Lemma 15.1, $\mathsf{LS}$}
\end{align*}
where $\mathsf{LS}$ is short for \cite{lattimore2020BanditAlgorithms}. Then, by Lemma \ref{lemma: kl properties} and the fact that $\nu$ and $\nu'$ differ only in action $i^\star$
\begin{align*}
    \regr(\nu) + \regr(\nu') &\geq \frac{\Delta T}{4} \exp\left\{-\E_\nu[P_{i^\star}(T)]\dkl(\nu_{i^\star},\nu_{i^\star}') \right\} 
\end{align*}
In both cases, $\nu,\nu'\in\mathcal{C}_2^{\mathtt{gau}}$, the Kullback-Leibler divergence between $\nu_{i^\star}$ and $\nu_{i^\star}'$ is $2\Delta^2/2$. Hence,
\begin{align*}
    \regr(\nu) + \regr(\nu') &\geq \frac{\Delta T}{4} \exp\left\{-\E_\nu[P_{i^\star}(T)]\frac{2\Delta^2}{2}\right\}\\
    &\geq \frac{\Delta T}{4} \exp\left\{-\frac{2T\Delta^2}{2(K-1)}\right\} \tag{\eqref{eq: aux pull istar}} \\
    &\geq \frac{T}{4} \sqrt{\frac{A-1}{4 T}} e^{-1 / 2} \tag{$\Delta=\sqrt{\frac{A-1}{4 T}} \leq \frac{1}{2}$} \\
    &=\frac{\sqrt{T(A-1)}}{8\sqrt{e}}.
\end{align*}
Finally, note that
\[\sup_{\tilde{\nu}\in\mathcal{C}_j}\regr(\pi;\tilde{\nu}) \geq \max\{\regr(\pi;\nu);\regr(\pi;\nu')\} \geq \frac{1}{2}\left(\regr(\nu) + \regr(\nu')\right) \geq \frac{\sqrt{T(A-1)}}{16\sqrt{e}}.\]
Because $\pi\in\Pi$ was generically chosen, it follows that 
\[\regr^\star(\mathcal{C}^\mathtt{gau}_2) = \inf_{\pi\in\Pi}\sup_{{\nu}\in\mathcal{C}^\mathtt{gau}_2}\regr(\pi;{\nu})\geq \frac{\sqrt{T(A-1)}}{16\sqrt{e}},\]
which was to be shown. The proof for $\mathcal{C}^\mathtt{gau}_1$ is identical.
\end{proof}

\clearpage
\putbib
\end{bibunit}

\end{document}